\documentclass[aps,prd,nofootinbib,superscriptaddress,preprint]{revtex4}
\usepackage{amssymb,amsmath,amsbsy}
\usepackage{mathrsfs}

\newlength{\captsize}           \let\captsize=\small
\newlength{\captwidth}          
\newlength{\beforetableskip}    
\newcommand{\capt}[1]{\begin{minipage}{\captwidth}
              \let\normalsize=\captsize
              \caption[0]{#1}
              \end{minipage}\\ \vspace{\beforetableskip}}

\newenvironment{Eqnarray}%
     {\arraycolsep 0.14em\begin{eqnarray}}{\end{eqnarray}}
\let\Re\relax
\let\Im\relax
\DeclareMathOperator{\Re}{Re}
\DeclareMathOperator{\Im}{Im}
\def\beqa{\begin{Eqnarray}}
\def\eeqa{\end{Eqnarray}}

\def\ddel{\!\!\mathrel{\raise1.5ex\hbox{$\leftrightarrow$\kern-.85em
\lower1.7ex\hbox{$\partial$}}}}

\def\Tr{{\rm Tr}}
\def\T{{\mathsf T}}
\def\abar{{\bar a}}
\def\bbar{{\bar b}}
\def\cbar{{\bar c}}
\def\dbar{{\bar d}}
\def\ebar{{\bar e}}
\def\fbar{{\bar f}}
\def\gbar{{\bar g}}
\def\hb{{\bar h}}
\def\Lama{\lambda_1^\prime}
\def\Lamb{\lambda_2^\prime}
\def\Lamc{\lambda_3^\prime}
\def\Lamd{\lambda_4^\prime}
\def\Lame{\lambda_5^\prime}
\def\Lamf{\lambda_6^\prime}
\def\Lamg{\lambda_7^\prime}

\def\lam{\lambda}

\def\ben{\begin{enumerate}}
\def\een{\end{enumerate}}
\def\beq{\begin{equation}}
\def\eeq{\end{equation}}

\def\ifmath#1{\relax\ifmmode #1\else $#1$\fi}
\def\lsim{\mathrel{\raise.3ex\hbox{$<$\kern-.75em\lower1ex\hbox{$\sim$}}}}
\def\gsim{\mathrel{\raise.3ex\hbox{$>$\kern-.75em\lower1ex\hbox{$\sim$}}}}
\def\sect#1{Section~\ref{#1}}

\def\eq#1{Eq.~\!(\ref{#1})}
\def\Ref#1{Ref.~\cite{#1}}
\def\Refs#1#2{Refs.~\cite{#1} and \cite{#2}}

\def\eqs#1#2{Eqs.~(\ref{#1}) and (\ref{#2})}

\def\eqst#1#2{Eqs.~(\ref{#1})--(\ref{#2})}
\def\eqthree#1#2#3{Eqs.~(\ref{#1}), (\ref{#2}), and (\ref{#3})}
\def\Eq#1{Eq.~(\ref{#1})}
\def\Eqst#1#2{Eqs.~(\ref{#1})--(\ref{#2})}
\def\Eqs#1#2{Eqs.~(\ref{#1}) and (\ref{#2})}
\def\Eqst#1#2{Eqs.~(\ref{#1})--(\ref{#2})}

\def\anti{\overline}

\def\Maa{m_{11}^{\prime\,2}}
\def\Mbb{m_{22}^{\prime\,2}}
\def\Mab{m_{12}^{\prime\,2}}
\def\st  {s_{\beta}}
\def\ct  {c_{\beta}}
\def\cosbb{c_{\beta}^2}
\def\sinbb{s_{\beta}^2}
\def\stwob  {s_{2\beta}}
\def\ctwob  {c_{2\beta}}

\def\ctwot  {c_{2\beta}}
\def\cthreet{c_{3\beta}}
\def\sthreet{s_{3\beta}}
\def\mud{M_U}
\def\mdd{M_D}
\def\vev#1{\langle #1 \rangle}

\def\ur{U_R}
\def\dr{D_R}

\def\ctwob{c_{2\beta}}
\def\stwob{s_{2\beta}}
\def\sthreeb{s_{3\beta}}

\def\cthreeb{c_{3\beta}}

\def\lamtil{\lam\ls{345}}

\def\phm{\phantom{-}}
\def\phaa{\phantom{AA}}

\def\beq{\begin{equation}}
\def\eeq{\end{equation}}
\def\ifmath#1{\relax\ifmmode #1\else $#1$\fi}

\def\stwob  {s_{2\beta}}
\def\ctwob  {c_{2\beta}}

\def\lamtil{\lam_{345}}

\def\mw{m_W}

\def\ls#1{\ifmath{_{\lower1.5pt\hbox{$\scriptstyle #1$}}}}
\def\lss#1{\ifmath{^{\,\lower2.5pt\hbox{$\scriptstyle #1$}}}}
\def\lsup#1{^{\lower 6pt\hbox{$\scriptstyle#1$}}}
\def\llsup#1{^{\lower 3pt\hbox{$\scriptstyle#1$}}}
\def\lasup#1{^{\lower 2pt\hbox{$\scriptstyle#1$}}}
\def\nicefrac#1#2{\hbox{$\frac{#1}{#2}$}}

\def\half{\ifmath{{\textstyle{\frac{1}{2}}}}}

\def\quarter{\ifmath{{\textstyle{\frac{1}{4}}}}}

\def\nn{\nonumber}

\begin{document}
%
\preprint{CFTP/19-032\phantom{J} \cr
SCIPP-19/01\phantom{X}  \cr
December, 2019}

\vspace*{-2cm}

\title{Basis-independent treatment of the complex 2HDM}
\author{Rafael Boto}
 \email[E-mail: ]{rafael.boto@tecnico.ulisboa.pt}
 \affiliation{CFTP, Departamento de F\'{\i}sica, Instituto Superior T\'{e}cnico, Universidade de Lisboa, Avenida Rovisco Pais 1, Lisboa 1049, Portugal}
\author{Tiago V.~Fernandes}
 \email[E-mail: ]{tvfernandes1@hotmail.com}
 \affiliation{CFTP, Departamento de F\'{\i}sica, Instituto Superior T\'{e}cnico, Universidade de Lisboa, Avenida Rovisco Pais 1, Lisboa 1049, Portugal}
\author{\mbox{Howard E.~Haber}}
 \email[E-mail: ]{haber@scipp.ucsc.edu}
 \affiliation{Santa Cruz Institute for Particle Physics,
   University of California, Santa Cruz, California 95064, USA}
 \author{Jorge C.~Rom\~ao}
 \email[E-mail: ]{jorge.romao@tecnico.ulisboa.pt}
 \affiliation{CFTP, Departamento de F\'{\i}sica, Instituto Superior T\'{e}cnico, Universidade de Lisboa, Avenida Rovisco Pais 1, Lisboa 1049, Portugal}
 \author{Jo\~ao P.~Silva}
  \email[E-mail: ]{jpsilva@cftp.ist.utl.pt}
 \affiliation{CFTP, Departamento de F\'{\i}sica, Instituto Superior T\'{e}cnico, Universidade de Lisboa, Avenida Rovisco Pais 1, Lisboa 1049, Portugal}

\vspace*{4cm}

\begin{abstract}
  
The complex 2HDM (C2HDM) is the most general CP-violating two-Higgs-doublet model that possesses a softly broken $\mathbb{Z}_2$ symmetry. 
However, the physical consequences of the model cannot depend on the basis of scalar fields used to define it.
Thus, to get a better sense of the significance of the C2HDM parameters, we have analyzed this model by employing a basis-independent formalism.
This formalism involves transforming to the Higgs basis  (which is defined up to an arbitrary complex phase) and identifying quantities that are invariant
with respect to this phase degree of freedom.  Using this method, we have obtained the constraints that enforce the softly broken $\mathbb{Z}_2$ symmetry.
One can then relate the C2HDM parameters to basis-independent quantities up to a twofold ambiguity.  We then show how this remaining
ambiguity is resolved.  We also examine the possibility of spontaneous CP violation when the scalar potential of the C2HDM is explicitly CP conserving.
Basis-independent constraints are presented that govern the presence of spontaneous CP violation.

\end{abstract}
\maketitle

\section{Introduction}  \label{sec:intro}

The two-Higgs-doublet model (2HDM) is one of the most well-studied
extensions of the Standard Model (SM).  Various motivations for adding a second
hypercharge-one complex Higgs doublet to the Standard Model
have been advocated in the literature%
~\cite{Lee:1973iz,Peccei:1977hh,Fayet:1974fj,Inoue:1982ej,Flores:1982pr,ghhiggs,hhg,Botella:1994cs,branco,Carena:2002es,Djouadi:2005gj,Branco:2011iw}.
In most cases, the structure of the 2HDM scalar potential is constrained in some way.
For example, many papers assume a CP-conserving scalar potential and vacuum
in order to simplify the resulting Higgs phenomenology.  In such models,
the three neutral Higgs bosons are states of definite CP, consisting of two CP-even
scalars and one CP-odd scalar.   

The assumption of CP conservation in the bosonic sector of the 2HDM may not be tenable in light of the
CP-violating effects that necessarily exist in the Higgs-fermion Yukawa couplings [which are the source
of the phase of the Cabibbo-Kobayashi-Maskawa (CKM) matrix that governs flavor physics].    However,
the most general 2HDM scalar potential and Yukawa couplings generically yield 
Higgs-mediated flavor-changing neutral currents \mbox{(FCNCs)} at tree level in conflict with experimental observations (which
imply that FCNCs are significantly suppressed).   The simplest way to avoid tree-level Higgs-mediated FCNCs
is to impose a discrete $\mathbb{Z}_2$ symmetry on the Higgs Lagrangian~\cite{Weinberg,Paschos,Georgi}.   Remarkably, such a symmetry, if exact,
removes tree-level Higgs-mediated FCNCs in the Yukawa sector while eliminating all CP-violating phases in the bosonic
sector of the theory.  However, the imposition of an exact $\mathbb{Z}_2$ symmetry is too restrictive.  For example,
no decoupling limit exists in the $\mathbb{Z}_2$-symmetric 2HDM~\cite{Gunion:2002zf}.   Since the LHC Higgs data imply that the
observed Higgs boson is SM-like in its properties, one can only achieve approximate Higgs alignment without decoupling 
by a fine-tuning of the Higgs scalar potential parameters~\cite{Gunion:2002zf,Craig:2013hca,Haber:2013mia,Asner:2013psa,Carena:2013ooa,Dev:2014yca,Das:2015mwa,Grzadkowski:2018ohf}.  

It is possible to satisfy the phenomenological constraint of suppressed Higgs-mediated FCNCs by introducing a soft breaking
of the $\mathbb{Z}_2$ symmetry.   Having introduced such a symmetry breaking term in the Higgs Lagrangian, it is now
possible that unremovable complex phases in the scalar potential exist, in which case Higgs-mediated CP-violating effects 
will be present.  The 2HDM with a softly broken $\mathbb{Z}_2$ symmetry and unremovable complex
phases in the scalar potential is called the complex 2HDM (often denoted as the C2HDM)~\cite{Ginzburg:2002wt,Ginzburg:2004vp,ElKaffas:2006gdt,Arhrib:2010ju,Barroso:2012wz,Inoue:2014nva,Fontes:2014xva,Grzadkowski:2014ada,Fontes:2017zfn}.
 
The C2HDM is typically exhibited in a scalar field basis in which the  $\mathbb{Z}_2$ symmetry of the dimension-four terms of
the Higgs Lagrangian is manifest.   Nevertheless, the physical consequences of the C2HDM are independent of the choice
of basis.  It is often convenient to employ a basis-independent formalism~\cite{Davidson:2005cw}, in which the relevant parameters of the model
are manifestly independent of the basis choice.  Indeed, basis-independent couplings (in principle) can always be directly related to
physical observables.
Thus, it is useful to express the parameters of the C2HDM, defined in the basis in which the $\mathbb{Z}_2$ symmetry is manifestly realized,
in terms of basis-independent quantities.  

To see utility of the basis-independent approach, consider the 
well-known quantity,
\beq \label{tanbdef}
\tan\beta\equiv\frac{|\vev{\Phi_2^0}|}{|\vev{\Phi_1^0}|}\,,
\eeq
given by the ratio of the absolute values 
of the two neutral Higgs field vacuum expectation values defined in some basis of the scalar fields.
In the most general 2HDM, this quantity is basis dependent and thus no physical observable can depend on it.
In the C2HDM, $\tan\beta$ is defined via the Higgs-fermion Yukawa couplings
in a basis where the $\mathbb{Z}_2$ 
symmetry of the dimension-four terms of the Higgs Lagrangian is manifestly realized.  
However, even given such a definition, some residual basis dependence remains.
Moreover, no coupling
in the bosonic sector of the C2HDM depends on $\tan\beta$~\cite{Haber:2006ue}.  

In this paper, we follow the basis-independent formalism of
\Refs{Davidson:2005cw}{Haber:2006ue}, which was inspired by an elegant formulation
of the 2HDM in \Ref{Botella:1994cs} that was subsequently described in more detail in \Ref{branco}.
An alternative approach to basis-independent methods in the 2HDM based on employing 
a set of independent physical couplings is given in Refs.~\cite{Grzadkowski:2014ada,Grzadkowski:2016szj}.
The translation between these two approaches can be found in Appendix~D of Ref.~\cite{Grzadkowski:2018ohf}.
The bilinear formalism of the 2HDM employed
in Refs.~\cite{Ivanov:2005hg,Nishi:2006tg,Maniatis:2007vn,Ferreira:2010hy,Ferreira:2010yh,Ivanov:2019kyh} 
also provides a powerful framework for establishing basis-independent results that 
can be applied in numerous applications.

In order to make this paper self-contained, we
recapitulate in \sect{sec:two} the ingredients of the basis-independent treatment of the 2HDM developed in
\Refs{Davidson:2005cw}{Haber:2006ue} in full detail.  In particular,
we emphasize the singular importance of 
the Higgs basis (defined to be a basis
in which one of the two neutral scalar fields has zero vacuum
expectation value), which possesses some important invariant
features.  In this regard, we tweak the formalism of Ref.~\cite{Haber:2006ue} to emphasize the significance of
the complex phase degree of freedom associated with the definition of the Higgs basis.  This allows us to define
\textit{invariant} Higgs basis scalar fields, which simplifies the subsequent analysis.  

In \sect{sec:three}, we obtain expressions for the charged and neutral Higgs mass-eigenstate fields
in terms of the invariant Higgs basis fields, which can then be expressed in terms of the scalar fields of the
original basis.   The neutral Higgs mass eigenstates arise after the diagonalization of a $3\times 3$
squared-mass matrix, which yields three invariant mixing angles.  Although we have slightly modified
the formalism of Ref.~\cite{Haber:2006ue}, we can explicitly show that one invariant mixing angle 
combines additively with a parameter that represents the phase dependence implicit in the definition of
the Higgs basis.  Hence, only two of the three invariant mixing angles can be related to physical observables.  

In \sect{sec:four}, we introduce a basis-invariant description of the Higgs-fermion Yukawa interactions.
We again tweak the formalism of Ref.~\cite{Haber:2006ue} in order to construct matrix invariant Yukawa
couplings.  We then introduce the Type-I and Type-II Yukawa Higgs-quark couplings~\cite{Haber:1978jt,Donoghue:1978cj,Hall:1981bc}
by imposing a (softly broken) $\mathbb{Z}_2$ symmetry that defines the parameter $\tan\beta$ and
guarantees the absence of tree-level Higgs-mediated FCNCs.   Although the physics literature treats $\tan\beta$ as a physical
parameter of the 2HDM,\footnote{The definition of the term ``physical parameter'' requires some care.  In this paper, we identify a Lagrangian
parameter as a physical parameter if it can be uniquely related to 
quantities that can be obtained (in principle) from direct experimental measurements.    Note that parameters that cannot be defined in terms of quantities that are 
invariant with respect to field redefinitions are not physical parameters.}
we emphasize that a residual basis dependence is still present and associated 
with the freedom to interchange the
two Higgs fields in a basis where the softly broken $\mathbb{Z}_2$ 
symmetry is manifestly realized.    

In \sect{sec:five}, a basis-independent treatment of the softly broken $\mathbb{Z}_2$ symmetry (which is needed in the 
construction of the Type-I and Type-II Yukawa interactions) is presented.   Formal basis-independent
expressions were originally given in Ref.~\cite{Davidson:2005cw}, and explicit results in the case of the CP-conserving
2HDM were presented in Ref.~\cite{Haber:2015pua}.  In this paper, we provide the corresponding results that are
applicable if CP violation is present in the 2HDM, with a careful analysis of all possible special cases.  
We subsequently noticed that some equivalent results can also be found in a paper by Lavoura~\cite{Lavoura:1994yu},
although the basis-independent nature of Lavoura's results was not initially appreciated.

In \sect{sec:six}, we are finally ready to carry out the basis-independent treatment of the C2HDM.  In the literature,
the parameters of the C2HDM are typically defined in the basis where the softly broken $\mathbb{Z}_2$ symmetry is manifest and where the two scalar field vacuum
expectation values are real and positive. 
Our goal was to
provide a translation between these parameters and the corresponding parameters of the basis-independent formalism.
In doing so, one gains insight into the nature of the original C2HDM parameters and their relations to physical quantities.
We again emphasize the significance of the residual basis dependence associated with the interchange of the two scalar fields.

In \sect{compare}, we return to the paper of Lavoura~\cite{Lavoura:1994yu}. 
We provide 
the necessary detail to derive Lavoura's results and indicate where his results fall short (i.e., special cases in which
Lavoura's results do not apply).   Lavoura attempted to find two invariant conditions for identifying the presence of spontaneous
CP violation in the 2HDM.  He was able to find one of the conditions but unable to find the second one.   We complete his search
and discuss various special cases in which only one invariant condition is required.

We briefly summarize our conclusions in \sect{conclusions}.   Additional details are relegated to five appendices.   
Appendix~\ref{appA} provides the necessary formulae for transforming between two scalar field bases.   In particular, we exhibit
how the parameters of the original basis of the 2HDM are expressed in terms of the parameters of the Higgs basis.
Appendix~\ref{erps} treats the so-called exceptional region of the 2HDM parameter space (the nomenclature was introduced in Ref.~\cite{Ferreira:2009wh}).
Indeed, in this parameter regime special attention is mandated as some of our derivations of basis-independent conditions provided in the main text are not applicable in this case.
Appendix~\ref{appC} demonstrates that the formal basis-independent conditions for a (softly broken) $\mathbb{Z}_2$ symmetry given in
 Ref.~\cite{Davidson:2005cw} are equivalent to the results of the explicit derivation given in \sect{sec:five}.   Appendix~\ref{appD} provides a simple proof for the
 existence of a particular basis of scalar field in which the CP-odd invariants employed in \sect{compare} take on especially convenient forms.  Finally, Appendix~\ref{appE}
 examines the mixing of the three neutral physical scalars of the 2HDM in a generic basis of the two scalar fields.

\section{Basis-independent formalism of the 2HDM}
\label{sec:two}

The fields of the two-Higgs-doublet model (2HDM) consist of two
identical complex hypercharge one, SU(2) doublet scalar fields
$\Phi_a(x)\equiv (\Phi^+_a(x)\,,\,\Phi^0_a(x))$, 
where the ``Higgs flavor'' index $a=1,2$ labels the two-Higgs-doublet fields.
The most general renormalizable SU(2)$_L\times$U(1)$_Y$ invariant scalar potential is given
by
\beqa  \label{pot}
\mathcal{V}&=& m_{11}^2\Phi_1^\dagger\Phi_1+m_{22}^2\Phi_2^\dagger\Phi_2
-[m_{12}^2\Phi_1^\dagger\Phi_2+{\rm H.c.}]+\half\lambda_1(\Phi_1^\dagger\Phi_1)^2
+\half\lambda_2(\Phi_2^\dagger\Phi_2)^2
+\lambda_3(\Phi_1^\dagger\Phi_1)(\Phi_2^\dagger\Phi_2)\nonumber\\[8pt]
&&\quad 
+\lambda_4(\Phi_1^\dagger\Phi_2)(\Phi_2^\dagger\Phi_1)
+\left\{\half\lambda_5(\Phi_1^\dagger\Phi_2)^2
+\big[\lambda_6(\Phi_1^\dagger\Phi_1)
+\lambda_7(\Phi_2^\dagger\Phi_2)\big]
\Phi_1^\dagger\Phi_2+{\rm H.c.}\right\}\,,
\eeqa
where $m_{11}^2$, $m_{22}^2$, and $\lam_1,\cdots,\lam_4$ are real parameters
and $m_{12}^2$, $\lambda_5$, $\lambda_6$ and $\lambda_7$ are
potentially complex parameters.  We assume that the
parameters of the scalar potential are chosen such that
the minimum of the scalar potential respects the
U(1)$\ls{\rm EM}$ gauge symmetry.  Then, the scalar field
vacuum expectations values (vevs) are of the form
\beq \label{potmin}
\langle \Phi_1 \rangle={\frac{1}{\sqrt{2}}} \left(
\begin{array}{c} 0\\ v_1\end{array}\right), \qquad \langle
\Phi_2\rangle=
{\frac{1}{\sqrt{2}}}\left(\begin{array}{c}0\\ v_2\, e^{i\xi}
\end{array}\right)\,,
\eeq
where $v_1$ and $v_2$ are real and non-negative, $0\leq \xi< 2\pi$, and $v$ is determined by the Fermi constant,
\beq \label{v246}
v\equiv (v_1^2+v_2^2)^{1/2}=\frac{2\mw}{g}=(\sqrt{2}G_F)^{-1/2}=246~{\rm GeV}\,.
\eeq
In writing \eq{potmin},
we have used a global U(1)$_Y$ hypercharge transformation to eliminate the phase
of $v_1$.  The bosonic part of the Higgs Lagrangian consists of a sum of the 
scalar potential [\eq{pot}] and the gauge invariant kinetic energy term,
\beq \label{KE}
\mathscr{L}_{KE}=(D_\mu\Phi)^\dagger_{\abar} (D^\mu\Phi)_a\,.
\eeq
In \eq{KE}, the covariant derivative of the electroweak gauge group acting on the scalar fields yields
\beq \label{covder}
D_\mu\Phi_a=\left(\begin{array}{c} \displaystyle
\partial_\mu\Phi^+_a+\left[\frac{ig}{c_W}\left(\half-s_W^2\right)Z_\mu
+ieA_\mu\right]\Phi^+_a+\frac{ig}{\sqrt{2}}W_\mu^+\Phi^0_a \\[8pt]
\displaystyle \partial_\mu\Phi^0_a-\frac{ig}{2c_W}Z_\mu\Phi_a^0+
\frac{ig}{\sqrt{2}}W_\mu^-\Phi^+_a\end{array}\right)\,,
\eeq
where $s_W\equiv\sin\theta_W$ and $c_W\equiv\cos\theta_W$. 

Since the scalar doublets $\Phi_1$ and $\Phi_2$
have identical SU(2)$\times$U(1) quantum numbers, one
is free to express the scalar potential in terms of two orthonormal linear combinations of the original
scalar fields.  The parameters appearing in \eq{pot} depend on a
particular \textit{basis choice} of the two scalar fields (denoted henceforth as the $\Phi$ basis).
The most general redefinition of the scalar fields that leaves $\mathscr{L}_{KE}$ invariant
corresponds to a global U(2) transformation,
$\Phi_a\to U_{a\bbar}\Phi_b$ [and $\Phi_\abar^\dagger\to\Phi_\bbar^\dagger
U^\dagger_{b\abar}$], where the $2\times 2$ unitary matrix $U$ satisfies
$U^\dagger_{b\abar}U_{a\cbar}=\delta_{b\cbar}$.   In our convention of employing unbarred and barred indices,
there is an implicit sum over unbarred--barred index pairs such as $a$ and $\abar$.\footnote{Note that replacing an unbarred index with a barred index is
equivalent to complex conjugation.  An alternative but equivalent convention
makes use of lower and upper Higgs flavor indices in place of barred and unbarred indices, in which case there is an implicit 
sum over a repeated upper-lower index pair.}

Following Refs.~\cite{Botella:1994cs,branco,Davidson:2005cw}, the scalar potential can be written
in U(2)-covariant form:
\beq \label{genericpot}
\mathcal{V}=Y_{a\bbar}\Phi_\abar^\dagger\Phi_b
+\half Z_{a\bbar c\dbar}(\Phi_\abar^\dagger\Phi_b)
(\Phi_\cbar^\dagger\Phi_d)\,,
\eeq
where the quartic couplings satisfy
$Z_{a\bbar c\dbar}=Z_{c\dbar a\bbar}$.
The hermiticity of the scalar potential implies that
$Y_{a \bbar}= (Y_{b \abar})^\ast$ and
$Z_{a\bbar c\dbar}= (Z_{b\abar d\cbar})^\ast$.
Under a flavor-U(2) transformation, the tensors $Y_{a\bbar}$ and
$Z_{a\bbar c\dbar}$ transform covariantly:
$Y_{a\bbar}\to U_{a\cbar}Y_{c\dbar}U^\dagger_{d\bbar}$
and $Z_{a\bbar c\dbar}\to U_{a\ebar}U^\dagger_{f\bbar}U_{c\gbar}
U^\dagger_{h\dbar} Z_{e\fbar g\hb}$. The coefficients of
the scalar potential depend on the choice of basis.  The
transformation of these coefficients under a U(2) basis change, exhibited explicitly in \eqst{maa}{Lam7def}, are
precisely the transformation laws of $Y$ and $Z$ given above.

For the convenience of the reader, we
recapitulate the ingredients of the basis-independent approach employed in
Ref.~\cite{Haber:2006ue}, in order to make this paper self-contained.
In an arbitrary scalar basis, the vevs of the two-Higgs-doublet fields [cf.~\eq{potmin}] 
can be written compactly~as %
\beq \label{vhat}
\langle\Phi_a\rangle=
\frac{v}{\sqrt{2}}\begin{pmatrix} 0\\ \widehat{v}_a\end{pmatrix}\, ,
\eeq
where $\widehat{v}=(\widehat{v}_1,\widehat{v}_2)$ is a complex vector of unit norm.
The $\widehat{v}_a$ are the nonzero solutions to the equation obtained by minimizing the scalar potential,
\beq \label{potmingeneric}
\widehat v_\abar^{\,\ast}\,
[Y_{a\bar b}+\half v^2 Z_{a\bbar c\dbar}\, \widehat v_\cbar^{\,\ast}\,
\widehat v_d]=0 \,.
\eeq
A second unit vector $\widehat{w}$ can be defined that is orthogonal to $\widehat{v}$,
\beq \label{what}
\widehat{w}_b=\widehat{v}_{\abar}^{\,\ast}\epsilon_{ab}\,,
\eeq
where $\epsilon_{12}=-\epsilon_{21}=1$ and $\epsilon_{11}=\epsilon_{22}=0$.
Indeed, $\widehat{v}$ and $\widehat{w}$ are orthogonal due to the vanishing of the complex dot product,
$\widehat{v}_{\bbar}^{\,\ast}\widehat{w}_b=0$.  
Note that under a U(2) transformation, 
\beq \label{wtrans}
\widehat{v}_a\to U_{a\bbar}\,\widehat{v}_b, \quad \text{which implies that \quad
$\widehat{w}_a\to (\det U)^{-1}U_{a\bbar}\,\widehat{w}_b$}.
\eeq

Since the
tensors $Y_{a\bbar}$ and $Z_{a\bbar c\dbar}$ exhibit tensorial
properties with respect to global U(2) transformations in the Higgs flavor
space, one can easily construct invariants with respect to the U(2) by
forming U(2)-scalar quantities.   
It is convenient to define two Hermitian projection operators,
\beq
V_{a\bbar}\equiv \widehat{v}_a \widehat{v}_\bbar^{\,\ast}\,,\qquad\quad W_{a\bbar}\equiv \widehat{w}_a\widehat{w}_\bbar^{\,\ast}=\delta_{a\bbar}-V_{a\bbar}\,.
\eeq
The matrices $V$ and $W$ can be used to define the following manifestly
basis-invariant real quantities that depend on the scalar potential parameters [cf.~\eq{genericpot}],
\beqa
Y_1 &\equiv & \Tr(Y V)\,,\qquad\qquad \quad
Y_2\equiv \Tr(Y W)\,,\label{syvv}\\
Z_1&\equiv& Z_{a\bbar c\dbar}\,V_{b\abar}V_{d\cbar}\,,\qquad\quad\,\,\,
Z_2\equiv  Z_{a\bbar c\dbar}\,W_{b\abar}W_{d\cbar}\,,\label{szvv12}\\
Z_3 &\equiv&  Z_{a\bbar c\dbar}\,V_{b\abar}W_{d\cbar}\,,\qquad\quad\,
Z_4 \equiv   Z_{a\bbar c\dbar}\,V_{b\cbar}W_{d\abar}\,.\label{szvv34}
\eeqa
In addition, we shall define the following pseudoinvariant
(potentially complex) quantities,
\beqa
Y_3&\equiv & Y_{a\bbar}\,\widehat{v}_{\abar}^{\,\ast} \widehat{w}_b\,, \label{syvv3} \\
Z_5&\equiv & Z_{a\bbar c\dbar} \,\widehat{v}_{\abar}^{\,\ast} \widehat{w}_b \widehat{v}_{\cbar}^{\,\ast} \widehat{w}_d\,,\label{szvv5}\\
Z_6&\equiv & Z_{a\bbar c\dbar}\, \widehat{v}_{\abar}^{\,\ast} \widehat{v}_b \widehat{v}_{\cbar}^{\,\ast} \widehat{w}_d\,,\label{szvv6}\\
Z_7&\equiv & Z_{a\bbar c\dbar}\, \widehat{v}_{\abar}^{\,\ast} \widehat{w}_b \widehat{w}_{\cbar}^{\,\ast} \widehat{w}_d\,.\label{szvv7}
\eeqa
In particular, \eq{wtrans} implies that under a basis transformation,
$\Phi_a\to U_{a\bbar}\Phi_b$,
\beq \label{rephasing}
 [Y_3, Z_6, Z_7]\to (\det~U)^{-1}[Y_3, Z_6, Z_7] \quad{\rm and}\quad
Z_5\to  (\det~U)^{-2} Z_5\,.
\eeq
Note that $Z_5^* Z_6^2$, $Z_5^* Z_7^2$ and $Z_6^* Z_7$ are basis-invariant quantities that can be obtained from the pseudoinvariants $Z_5$, $Z_6$ and $Z_7$.

Once the scalar potential minimum is determined, which defines
$\widehat v_a$, one can introduce new Higgs-doublet fields that define the Higgs basis,
\beq \label{hbasisdef}
H_1=(H_1^+\,,\,H_1^0)\equiv \widehat v_{\abar}^{\,\ast}\Phi_a\,,\qquad\qquad
H_2=(H_2^+\,,\,H_2^0)\equiv \widehat w_{\abar}^{\,\ast}\Phi_a\,.
\eeq
The definitions of $H_1$ and $H_2$ imply that
\beq \label{higgsvevs}
\vev{H_1^0}=\frac{v}{\sqrt{2}}\,,\qquad\qquad \vev{H_2^0}=0\,,
\eeq
where we have used \eq{vhat} and the fact that $\widehat{v}$ and $\widehat{w}$ are complex orthogonal unit vectors.
Note that the definition of the scalar field $H_1$ is basis-independent, whereas the scalar field $H_2$ is a pseudoinvariant field due to the transformation properties of $\widehat{w}$ given in \eq{wtrans}.  That is, $H_2\to (\det U)H_2$ under $\Phi_a\to U_{a\bbar}\Phi_b$, where $\det U$ is a pure phase.  The pseudoinvariant nature of $H_2$ is ultimately due to the fact that
one can rephase $H_2$ while maintaining \eq{higgsvevs} which defines the Higgs basis.   Thus, one should really speak of a class of Higgs bases that is characterized by an arbitrary phase angle.

The significance of the quantities defined by \eqst{syvv}{szvv7}
becomes clearer after rewriting the scalar potential in terms of the Higgs basis fields, 
\beqa
 \mathcal{V}&=& Y_1 H_1^\dagger H_1+ Y_2 H_2^\dagger H_2 +[Y_3 
H_1^\dagger H_2+{\rm H.c.}]
\nn\\
&&\quad 
+\half Z_1(H_1^\dagger H_1)^2+\half Z_2(H_2^\dagger H_2)^2
+Z_3(H_1^\dagger H_1)(H_2^\dagger H_2)
+Z_4( H_1^\dagger H_2)(H_2^\dagger H_1) \nn \\
&&\quad
+\left\{\half Z_5(H_1^\dagger H_2)^2 +\big[Z_6  (H_1^\dagger
H_1) +Z_7 (H_2^\dagger H_2)\big] H_1^\dagger H_2+{\rm
H.c.}\right\}\,.\label{higgsbasispot}
\eeqa
The minimization of the scalar potential in the Higgs basis yields 
\beq \label{minconds}
Y_1=-\half Z_1 v^2\,,\qquad\quad Y_3=-\half Z_6 v^2\,.
\eeq

In light of \eq{rephasing}, the form of the scalar potential is invariant under the rephasing of the pseudoinvariant Higgs basis field $H_2$.   However, one can make the basis invariance of the scalar potential even more explicit by introducing
invariant Higgs basis fields,
\beq \label{hbasisdef2}
\mathcal{H}_1\equiv H_1\,,\qquad\qquad \mathcal{H}_2\equiv e^{i\eta} H_2\,,
\eeq
where  $e^{i\eta}$ is a pseudoinvariant quantity that transforms under the basis transformation,
$\Phi_a\to U_{a\bbar}\Phi_b$, as
\beq \label{etatrans}
e^{-i\eta}\to (\det~U) e^{-i\eta}\,.
\eeq
\Eq{hbasisdef2} provides a new way of exhibiting explicitly the existence of the class of Higgs bases parametrized by the phase angle $\eta$.
Equivalently, one can write,
\beq \label{inverting}
\Phi_a=\mathcal{H}_1 \widehat v_a+ e^{-i\eta}\mathcal{H}_2 \widehat w_a \,.
\eeq
In terms of the invariant Higgs basis fields, the scalar potential is given by, 
 \beqa
 \mathcal{V}&=& Y_1 \mathcal{H}_1^\dagger \mathcal{H}_1+ Y_2 \mathcal{H}_2^\dagger \mathcal{H}_2 +[Y_3 e^{-i\eta}
\mathcal{H}_1^\dagger \mathcal{H}_2+{\rm H.c.}]
\nn\\
&&\quad 
+\half Z_1(\mathcal{H}_1^\dagger \mathcal{H}_1)^2+\half Z_2(\mathcal{H}_2^\dagger \mathcal{H}_2)^2
+Z_3(\mathcal{H}_1^\dagger \mathcal{H}_1)(\mathcal{H}_2^\dagger \mathcal{H}_2)
+Z_4( \mathcal{H}_1^\dagger \mathcal{H}_2)(\mathcal{H}_2^\dagger \mathcal{H}_1) \nn \\
&&\quad
+\left\{\half Z_5 e^{-2i\eta}(\mathcal{H}_1^\dagger \mathcal{H}_2)^2 +\big[Z_6 e^{-i\eta} (\mathcal{H}_1^\dagger
\mathcal{H}_1) +Z_7 e^{-i\eta} (\mathcal{H}_2^\dagger \mathcal{H}_2)\big] \mathcal{H}_1^\dagger \mathcal{H}_2+{\rm
H.c.}\right\}\,.\label{higgspot}
\eeqa
Due to \eqs{rephasing}{etatrans}, all the coefficients of the scalar potential given in \eq{higgspot} are  manifestly basis invariant.

It is instructive to see what happens if one transforms between two Higgs bases.   That is, suppose that $\vev{\Phi_1^0}=v/\sqrt{2}$ and $\vev{\Phi_2^0}=0$.   To transform to another Higgs basis, one can employ the U(2) transformation $\Phi_a\to U_{a\bbar}\Phi_b$, where $U={\rm diag}(1,e^{i\chi})$.   Then, \eq{etatrans} implies that $\eta\to\eta-\chi$.  
It then follows that
\beq \label{rephasing2}
[Y_3, Z_6, Z_7]\to e^{-i\chi}[Y_3, Z_6, Z_7] \quad \text{and} \quad
Z_5\to  e^{-2i\chi} Z_5\,.
\eeq
In contrast, $Y_1$, $Y_2$ and $Z_{1,2,3,4}$ are invariant when transforming between two Higgs bases.

To summarize, the class of Higgs bases corresponds to $\widehat{v}=(1,0)$ and $\widehat{w}=(0,1)$; different Higgs basis choices are parametrized by the phase angle $\eta$ via $\mathcal{H}_2= e^{i\eta}\Phi_2$ after inserting $\widehat{w}=(0,1)$ into \eq{hbasisdef}.  Indeed, inserting the Higgs basis values of $\widehat{v}$ and $\widehat{w}$ into \eqst{syvv}{szvv7} and then rewriting the scalar potential [\eq{genericpot}] in terms of the invariant Higgs basis fields defined in \eq{hbasisdef} yields \eq{higgspot} as expected.

Finally, we note that the 2HDM scalar potential and vacuum are CP invariant if one can find a choice of $\eta$ such that all the coefficients of the scalar potential in \eq{higgspot} are real after imposing the scalar potential minimum conditions given in \eq{minconds}.   This condition is satisfied if and only if~\cite{cpx} (see also Refs.~\cite{Davidson:2005cw,Haber:2006ue})
\beq \label{cpconds}
\Im(Z_5^* Z_6^2)=\Im(Z_5^* Z_7^2)=\Im(Z_6^* Z_7)=0\,.
\eeq

\section{The charged and neutral Higgs mass eigenstates}
\label{sec:three}

To determine the Higgs mass eigenstates, one must examine the terms of
the scalar potential that are quadratic in the scalar fields (after
imposing the scalar potential minimum conditions and defining shifted
fields with zero vevs).  We have slightly tweaked the procedure that was
carried out in \Ref{Haber:2006ue}, and we summarize the results here.  

We parametrize the invariant Higgs basis fields $\mathcal{H}_1$ and $\mathcal{H}_2$ as follows,
\beq
\label{hbasisfields}
\mathcal{H}_1=\left(\begin{array}{c}
G^+ \\ {\frac{1}{\sqrt{2}}}\left(v+\varphi_1^0+iG^0\right)\end{array}
\right)\,,\qquad
\mathcal{H}_2=\left(\begin{array}{c}
H^+ \\ {\frac{1}{\sqrt{2}}}\left(\varphi_2^0+ia^0\right)\end{array}
\right)\,,
\eeq
where $G^+$ (and its Hermitian conjugate) are the charged Goldstone bosons 
and $G^0$ is the neutral Goldstone boson.
The three remaining neutral fields mix, and
the resulting neutral Higgs
squared-mass matrix in the $\varphi_1^0$--$\varphi_2^0$--$a^0$ basis is:\
\beq  \label{matrix33}
\mathcal{M}^2=v^2\left( \begin{array}{ccc}
Z_1&\quad \Re(Z_6 e^{-i\eta}) &\quad -\Im(Z_6 e^{-i\eta})\\
\Re(Z_6 e^{-i\eta})  &\quad \half\bigl[Z_{34}+\Re(Z_5 e^{-2i\eta})\bigr]+Y_2/v^2 & \quad
- \half \Im(Z_5  e^{-2i\eta})\\ -\Im(Z_6  e^{-i\eta}) &\quad - \half \Im(Z_5  e^{-2i\eta}) &\quad
\half\bigl[Z_{34}-\Re(Z_5 e^{-2i\eta})\bigr]+Y_2/v^2\end{array}\right), 
\eeq
where
$Z_{34}\equiv Z_3+Z_4$.

The squared-mass matrix $\mathcal{M}^2$
is real symmetric; hence it can be diagonalized by
a special real orthogonal transformation
\beq \label{rmrt}
R\mathcal{M}^2 R^{\T}=\mathcal{M}^2_D\equiv {\rm diag}~(m_1^2\,,\,m_2^2\,,\,m_3^2)\,,
\eeq
where $R$ is a real matrix such that $RR^{\T}=I$, $\det~R=1$ and the $m_i^2$ are the eigenvalues of~$\mathcal{M}^2$.
A convenient form for $R$ is:
\beqa \label{rmatrix}
R=R_{12}R_{13}\overline{R}_{23} &=&\left( \begin{array}{ccc}
c_{12}\,\, &-s_{12}\quad &0\\
s_{12}\,\, &\phm c_{12}\quad &0\\
0\,\, &\phm 0\quad &1\end{array}\right)\left( \begin{array}{ccc}
c_{13}\quad &0\,\, &-s_{13}\\
0\quad & 1\,\,&\phm 0\\
s_{13}\quad &0\,\, &\phm c_{13}\end{array}\right) \left( \begin{array}{ccc}
1\quad &0\,\, &\phm 0\\
0\quad &\overline{c}_{23}\,\, &-\overline{s}_{23}\\
0\quad &\overline{s}_{23}\,\, &\phm \overline{c}_{23}\end{array}\right) \nonumber \\[10pt]
&=&
\left( \begin{array}{ccc}
c_{13}c_{12}\quad &-s_{12}\overline{c}_{23}-c_{12}s_{13}\overline{s}_{23}\quad &-c_{12}s_{13}\overline{c}_{23}
+s_{12}\overline{s}_{23}\\[6pt]
c_{13}s_{12}\quad &c_{12}\overline{c}_{23}-s_{12}s_{13}\overline{s}_{23}\quad
& -s_{12}s_{13}\overline{c}_{23}-c_{12}\overline{s}_{23}\\
s_{13}\quad &c_{13}\overline{s}_{23}\quad &c_{13}\overline{c}_{23}\end{array}\right)\,,
\eeqa
where $c_{ij}\equiv \cos\theta_{ij}$ and $s_{ij}\equiv\sin\theta_{ij}$.  We have written $\overline{c}_{23}\equiv\cos\bar{\theta}_{23}$ and
$\overline{s}_{23}\equiv\sin\bar{\theta}_{23}$ to distinguish between the angle $\theta_{23}$ defined in Ref.~\cite{Haber:2006ue} and the angle $\bar{\theta}_{23}$ defined above.
Indeed, the angles $\theta_{12}$, $\theta_{13}$ and $\bar{\theta}_{23}$  defined above are all invariant quantities since they are obtained by diagonalizing $\mathcal{M}^2$ whose matrix elements are manifestly basis invariant.

The neutral physical Higgs mass eigenstates are denoted by $h_1$, $h_2$ and
$h_3$,
\beq \label{rotated}
\left( \begin{array}{c}
h_1\\ h_2\\h_3 \end{array}\right)=R \left(\begin{array}{c} \varphi_1^0\\
\varphi_2^0\\ a^0 \end{array}\right)=RW \left(\begin{array}{c}
\sqrt{2}\,\Re~\mathcal{H}_1^0-v\\ \mathcal{H}_2^0\\  \,\,\mathcal{H}_2^{0\,\dagger} \end{array}\right)\,,
\eeq
which defines the unitary matrix $W$.
A straightforward calculation yields~\cite{Haber:2006ue}
\beq
RW=\left(\begin{array}{ccc}q_{11} 
&\qquad  \nicefrac{1}{\sqrt{2}}q^*_{12}\,e^{i\bar{\theta}_{23}}
&\qquad \nicefrac{1}{\sqrt{2}} q_{12}\,e^{-i\bar{\theta}_{23}}\\[4pt]
q_{21} &\qquad  \nicefrac{1}{\sqrt{2}}q^*_{22}\,e^{i\bar{\theta}_{23}}
&\qquad \nicefrac{1}{\sqrt{2}}q_{22}\,e^{-i\bar{\theta}_{23}} \\[4pt]
q_{31} &\qquad  \nicefrac{1}{\sqrt{2}}q^*_{32}\,e^{i\bar{\theta}_{23}}
&\qquad \nicefrac{1}{\sqrt{2}} q_{32}\,e^{-i\bar{\theta}_{23}}
\end{array}\right)\,,\label{RW}
\eeq
where the $q_{k\ell}$ are listed in Table~\ref{tabqij}.
 \begin{table}[t!]
\centering
\caption{The U(2)-invariant quantities $q_{k\ell}$ are functions of 
the neutral Higgs mixing angles $\theta_{12}$ and $\theta_{13}$, where
$c_{ij}\equiv\cos\theta_{ij}$ and $s_{ij}\equiv\sin\theta_{ij}$.  The neutral Goldstone boson
corresponds to $k=0$. \\
\label{tabqij}}
\begin{tabular}{|c||c|c|}\hline
$\phaa k\phaa $ &\phaa $q_{k1}\phaa $ & \phaa $q_{k2} \phaa $ \\ \hline
$0$ & $i$ & $0$ \\ 
$1$ & $c_{12} c_{13}$ & $-s_{12}-ic_{12}s_{13}$ \\
$2$ & $s_{12} c_{13}$ & $c_{12}-is_{12}s_{13}$ \\
$3$ & $s_{13}$ & $ic_{13}$ \\
\hline
\end{tabular}
\end{table}
Employing \eqs{hbasisdef}{rotated}, it follows that
\beq \label{hmassinv}
h_k=\frac{1}{\sqrt{2}}\left[\overline\Phi_{\abar}\lsup{0\,\dagger}
(q_{k1} \widehat v_a+q_{k2}\widehat w_a e^{-i\theta_{23}})
+(q^*_{k1}\widehat v^*_{\abar}+q^*_{k2}\widehat w^*_{\abar}e^{i\theta_{23}})
\overline\Phi_a\lsup{0}\right]\,,
\eeq
for $k=1,2,3$, 
where the shifted neutral fields are defined
by $\overline\Phi_a\lsup{0}\equiv \Phi_a^0-v\widehat v_a/\sqrt{2}$.  It is straightforward to verify that \eq{hmassinv} also applies to the neutral Goldstone
boson if we denote $h_0\equiv G^0$ and define $q_{01}=i$ and $q_{02}=0$ as indicated in Table~\ref{tabqij}.

We have also introduced the quantity,\footnote{Note that $\theta_{23}$ corresponds precisely to the angle of the same name employed in Ref.~\cite{Haber:2006ue}.}
\beq \label{twothree}
\theta_{23}\equiv\bar{\theta}_{23}+\eta\,.
\eeq
Note that $e^{-i\theta_{23}}$ is a pseudoinvariant quantity.  In particular, in light of \eq{etatrans} it follows that 
\beq \label{shift}
e^{-i\theta_{23}}\to (\det~U)e^{-i\theta_{23}}
\eeq
under a U(2) basis transformation, $\Phi_a\to U_{a\bbar}\Phi_b$. 
This transformation law is consistent with \eq{wtrans} and the fact that the neutral Higgs mass-eigenstates $h_k$ are invariant fields.\footnote{The remaining freedom to define  
the overall sign of $h_k$ is associated with the convention adopted for the domains of the mixing angles $\theta_{ij}$, as discussed in Ref.~\cite{Haber:2006ue}, and is independent
of scalar field basis transformations.}

For completeness, we note that \eqs{hbasisdef}{hbasisfields} yield expressions for the massless charged Goldstone field, $G^+=\widehat{v}^{\,\ast}_{\abar}\Phi^+_a$ and the charged Higgs field, $H^+=e^{i\eta}\widehat{w}^{\,\ast}_{\abar}\Phi^+_a$, with corresponding squared mass,
\beq \label{plusmass}
m_{H^\pm}^2=Y_2+\half Z_3 v^2\,.
\eeq
Nevertheless, one is always free to rephase the charged Higgs field without affecting any observable of the model.
It is convenient to rephase, $H^+\to e^{-i\bar{\theta}_{23}}H^+$, which yields
\beq \label{chhiggs}
H^+=e^{i\bar{\theta}_{23}}\mathcal{H}_2^+=e^{i\theta_{23}}\widehat{w}^{\,\ast}_{\abar}\Phi^+_a\,.
\eeq
Note that this rephasing is conventional and does not alter the fact that $H^+$ is an invariant field with respect to scalar field basis transformations.  

Finally, one can invert \eq{hmassinv} and include the charged scalars to obtain,\footnote{Here we differ slightly from Ref.~\cite{Haber:2006ue} where a noninvariant charged Higgs field, $H^+=\widehat{w}^{\,\ast}_{\abar}\Phi^+_a$, is employed.}
\beq \label{master}
\Phi_a=\left(\begin{array}{c}G^+\widehat v_a+H^+e^{-i\theta_{23}} \widehat w_a\\[6pt]
\displaystyle
\frac{v}{\sqrt{2}}\widehat v_a+\frac{1}{\sqrt{2}}\sum_{k=0}^3
\left(q_{k1}\widehat v_a+q_{k2}e^{-i\theta_{23}}\widehat w_a\right)h_k
\end{array}\right)\,.
\eeq

Although $\bar{\theta}_{23}$ is an invariant parameter, it has no physical significance, since 
it only appears in \eq{master} in the combination defined in \eq{twothree}.
Indeed, if we now insert \eq{master} into the expression for the scalar potential given in \eq{genericpot} to derive the bosonic couplings of the 2HDM, one sees that $\bar{\theta}_{23}$ never appears explicitly in any observable.   Consequently,  one can simply set $\bar{\theta}_{23}=0$ without loss of generality, which would identify $\eta=\theta_{23}$ as the pseudoinvariant phase angle that specifies the choice of Higgs basis. 

It is useful to rewrite the neutral Higgs mass diagonalization equation
[\eq{rmrt}] as follows.  With $R\equiv R_{12}R_{13}\overline{R}_{23}$ given by
\eq{rmatrix}, we define 
\beq \label{mtilmatrix}
\widetilde{\mathcal{M}}^2\equiv \overline{R}_{23}\mathcal{M}^2\overline{R}_{23}^{\T}=
v^2\left( \begin{array}{ccc}
Z_1&\,\, \Re(Z_6 \, e^{-i\theta_{23}}) &\,\, -\Im(Z_6 \, e^{-i\theta_{23}})\\
\Re(Z_6 e^{-i\theta_{23}}) &\,\,\Re(Z_5 \,e^{-2i\theta_{23}})+ A^2/v^2 & \,\,
- \half \Im(Z_5 \,e^{-2i\theta_{23}})\\ -\Im(Z_6 \,e^{-i\theta_{23}})
 &\,\, - \half \Im(Z_5\, e^{-2i\theta_{23}}) &\,\, A^2/v^2\end{array}\right),
\eeq
where 
${A}^2$ is the auxiliary quantity,
\beq \label{madef}
{A}^2\equiv Y_2+\half[Z_3+Z_4-\Re(Z_5 e^{-2i\theta_{23}})]v^2\,.
\eeq
Note that we have employed \eq{twothree}, which results in the appearance of $e^{-i\theta_{23}}$ in the appropriate places given that the matrix elements of $\widetilde{\mathcal{M}}^2$ are invariant quantities (but with no separate dependence on the invariant angle $\bar{\theta}_{23}$).
The diagonal neutral Higgs squared-mass matrix is then given by:
\beq \label{diagtil}
\widetilde{R}\,\widetilde{\mathcal{M}}^2\,\widetilde{R}^{\T}=\mathcal{M}^2_D={\rm
 diag}(m_1^2\,,\,m_2^2\,,\,m_3^2)\,, 
\eeq
where the diagonalizing matrix $\widetilde{R}\equiv R_{12}R_{13}$ 
depends only on the invariant angles $\theta_{12}$ and~$\theta_{13}$,
\beq \label{rtil}
\widetilde R=\left(\begin{array}{ccc}c_{12}c_{13} & \quad -s_{12} &
\quad -c_{12}s_{13} \\ c_{13}s_{12} & \quad \phm c_{12} & \quad-s_{12}s_{13}\\
s_{13} & \quad \phm 0 & \quad c_{13}\end{array}\right)=\begin{pmatrix} q_{11} & \quad \Re~q_{12} & \quad  \Im~q_{12} \\
 q_{21} & \quad \Re~q_{22} & \quad  \Im~q_{22} \\
 q_{31} & \quad \Re~q_{32} & \quad  \Im~q_{32}\end{pmatrix}\,.
\eeq

Explicit expressions for the neutral Higgs boson squared masses requires one to solve a cubic characteristic equation that yields the eigenvalues of $\widetilde{\mathcal{M}}^2$.
The resulting expressions are unwieldy and impractical.   Nevertheless, one can derive useful relations by rewriting \eq{diagtil} as 
$\widetilde{\mathcal{M}}^2=\widetilde{R}^{\T}\mathcal{M}^2_D\widetilde{R}$ and employing \eq{rtil}.  It then follows that
\beqa
Z_1  &=& \frac{1}{v^2}\sum_{k=1}^3 m_k^2 (q_{k1})^2\,,\label{zee1id} \\
Z_4  &=& \frac{1}{v^2}\left[\sum_{k=1}^3 m_k^2 |q_{k2}|^2 -2m_{H^\pm}^2\right]\,, \label{zeefour} 
\eeqa
after making use of \eq{plusmass} in the evaluation of \eq{zeefour}, and
\beqa
Z_5  e^{-2i\theta_{23}} &=& \frac{1}{v^2}\sum_{k=1}^3 m_k^2 (q_{k2}^*)^2\,, \label{zee5id}\\
Z_6  e^{-i\theta_{23}} &=& \frac{1}{v^2}\sum_{k=1}^3 m_k^2 \,q_{k1} q_{k2}^*\,.\label{zee6id}
\eeqa

The conditions for a CP-invariant scalar potential and vacuum were given in \eq{cpconds}.   These conditions are satisfied in the following two cases:
\beqa
&& 1.~~\Im(Z_5 e^{-2i\theta_{23}})=\Im(Z_6 e^{-i\theta_{23}})=\Im(Z_7 e^{-i\theta_{23}})=0\,,\label{one}  \\
&&\hspace{1.75in}  \text{or} \nonumber \\
&& 2.~~\Im(Z_5 e^{-2i\theta_{23}})=\Re(Z_6 e^{-i\theta_{23}})=\Re(Z_7 e^{-i\theta_{23}})=0\,.\label{two} 
\eeqa
In both cases the neutral scalar squared-mass matrix given in \eq{mtilmatrix} assumes a block diagonal form consisting of a $2\times 2$ mass matrix that yields the squared masses of two neutral CP-even Higgs bosons and a $1\times 1$ mass matrix corresponding to the squared mass of a neutral CP-odd Higgs boson.   
In this paper, our primary focus is the 2HDM with a scalar sector that exhibits either explicit or spontaneous CP violation,
in which case neither \eq{cpconds} nor \eqs{one}{two} are satisfied.

\section{Higgs-fermion Yukawa interactions}
\label{sec:four}

The Higgs boson couplings to the fermions arise from the Yukawa Lagrangian.
We shall slightly tweak the results that were initially presented in Ref.~\cite{Haber:2006ue} (with some corrections subsequently noted in Ref.~\cite{Haber:2010bw}). 
In terms of the quark mass-eigenstate fields, the Yukawa Lagrangian in the $\Phi$ basis is given by
\beq \label{yuklag}
-\mathscr{L}_{\rm Y}=\anti U_L \Phi_{\abar}^{0\,*}{{h^U_a}} \ur -\anti
D_L K^\dagger\Phi_{\abar}^- {{h^U_a}}\ur
+\anti U_L K\Phi_a^+{{h^{D\,\dagger}_{\abar}}} \dr
+\anti D_L\Phi_a^0 {{h^{D\,\dagger}_{\abar}}}\dr+{\rm H.c.}\,,
\eeq
where $Q_{R,L}\equiv P_{R,L}Q$, with $P_{R,L}\equiv\half(1\pm\gamma\ls{5})$ [for $Q=U,D$],
$K$ is the CKM mixing matrix, and the $h^{U,D}$
are $3\times 3$ Yukawa coupling matrices.   
We can construct invariant matrix Yukawa couplings $\kappa^Q$ and $\rho^Q$ by defining,\footnote{We have modified the definition of $\rho^Q$ as compared to the one employed in Refs.~\cite{Davidson:2005cw,Haber:2006ue,Haber:2010bw} by including a factor of $e^{i\theta_{23}}$.   This new definition has been adopted as a matter of convenience since $\rho^Q$ defined as in \eq{kapparho} is invariant with respect to basis transformations of the scalar fields.}
\beq \label{kapparho}
\kappa^{Q}\equiv \widehat v^*_\abar h^{Q}_a\,,\qquad\qquad
\rho^{Q}\equiv e^{i\theta_{23}} \widehat w^*_\abar h^{Q}_a\,.
\eeq
Inverting these equations yields
\beq \label{hexpand}
h^Q_a=\kappa^Q\widehat v_a+e^{-i\theta_{23}}\rho^Q\widehat w_a\,.
\eeq

Inserting the above result into \eq{yuklag} and employing \eqthree{hbasisdef}{hbasisdef2}{twothree}, we end up with the Yukawa Lagrangian in terms of the invariant Higgs basis fields,
\beqa 
&& -\mathscr{L}_{\rm Y}=\anti U_L (\kappa^U \mathcal{H}_1^{0\,\dagger}
+e^{-i\bar{\theta}_{23}}\rho^U \mathcal{H}_2^{0\,\dagger})\ur
-\anti D_L K^\dagger(\kappa^U \mathcal{H}_1^{-}+e^{-i\bar{\theta}_{23}}\rho^U \mathcal{H}_2^{-})\ur \nonumber \\
&& \quad +\anti U_L K (\kappa^{D\,\dagger}\mathcal{H}_1^+ +e^{i\bar{\theta}_{23}}\rho^{D\,\dagger}\mathcal{H}_2^+)\dr
+\anti D_L (\kappa^{D\,\dagger}\mathcal{H}_1^0+e^{i\bar{\theta}_{23}}\rho^{D\,\dagger}\mathcal{H}_2^0)\dr+{\rm H.c.} \label{yukhbasis}
\eeqa
In light of \eq{higgsvevs}, $\kappa^U$ and $\kappa^D$ are proportional to the
(real non-negative) diagonal quark mass matrices $M_U$ and $M_D$,
respectively.  In particular,
\beq \label{MQ}
M_U=\frac{v}{\sqrt{2}}\kappa^U={\rm diag}(m_u\,,\,m_c\,,\,m_t)\,,\qquad
M_D=\frac{v}{\sqrt{2}}\kappa^{D\,\dagger}={\rm
diag}(m_d\,,\,m_s\,,\,m_b) \,.
\eeq
In contrast, the matrices $\rho^U$ and $\rho^D$ are independent complex
$3\times 3$ matrices.  

One can now reexpress the Higgs basis fields in terms of mass-eigenstate charged and neutral Higgs fields by inverting \eq{rotated} and
employing \eq{chhiggs}
to obtain the Yukawa couplings of the quarks to the physical scalars and to the Goldstone bosons.
Of course, the same result can be obtained directly
by inserting \eq{master} into \eq{yuklag}.   The end result is,  
\beqa
 && \hspace{-0.2in} -\mathscr{L}_Y = \frac{1}{v}\overline D
\biggl\{M_D (q_{k1} P_R + q^*_{k1} P_L)+\frac{v}{\sqrt{2}}
\left[q_{k2}\,\rho^{D^\dagger} P_R+
q^*_{k2}\,\rho^D P_L\right]\biggr\}Dh_k \nonumber \\
&&\qquad \,\, +\frac{1}{v}\overline U \biggl\{M_U (q_{k1}
P_L + q^*_{k1} P_R)+\frac{v}{\sqrt{2}}\left[
q^*_{k2}\,\rho^U P_R+
q_{k2}\,\rho^{U\dagger} P_L\right]\biggr\}U h_k
\nonumber \\
&&\quad +\biggl\{\overline U\left[K\rho^{D\dagger}
P_R-\rho^{U\dagger} KP_L\right] DH^+  +\frac{\sqrt{2}}{v}\,\overline
U\left[K\mdd P_R-\mud KP_L\right] DG^+ +{\rm
H.c.}\biggr\}\,, \label{Yukawas}
\eeqa
where there is an implicit sum over $k=0,1,2,3$ (and $h_0\equiv G^0$). 
\clearpage

As expected, 
the Higgs-quark Yukawa couplings depend only on invariant quantities, namely, 
$M_Q$ and $\rho^Q$ (for $Q=U$, $D$) and the
invariant angles $\theta_{12}$, $\theta_{13}$, while all dependence on $\bar{\theta}_{23}$ has canceled. 
Since $\rho^Q$ is in general a complex matrix,
\eq{Yukawas} exhibits CP-violating neutral-Higgs--fermion interactions.
Moreover, Higgs-mediated 
FCNCs are present at tree level in cases where the $\rho^Q$ are
not flavor diagonal. 

To avoid tree-level Higgs-mediated FCNCs, we shall impose a $\boldsymbol{\mathbb{Z}_2}$ symmetry
on the Higgs Lagrangian specified by \eqthree{pot}{KE}{yuklag}.   If the scalar potential respects the discrete symmetry 
$\Phi_1\to\Phi_1$
and $\Phi_2\to -\Phi_2$, 
then it follows that $m_{12}^2=\lambda_6=\lambda_7=0$.   
However, phenomenological considerations allow for the presence
of a soft $\boldsymbol{\mathbb{Z}_2}$-breaking term, $m_{12}^2\neq 0$.  Consequently, we shall henceforth apply the 
$\boldsymbol{\mathbb{Z}_2}$ symmetry exclusively to the dimension-four terms of the Higgs Lagrangian.
Note that the action of the $\boldsymbol{\mathbb{Z}_2}$  symmetry on the scalar fields is basis dependent. In Section~\ref{sec:five},
we shall recast this action in a basis-independent form.  

\begin{table}[t!]
 \begin{center}
 \caption{Four possible $\mathbb{Z}_2$ charge assignments that forbid
tree-level Higgs-mediated FCNC effects in the 2HDM Higgs-quark Yukawa interactions, and the corresponding invariant Yukawa coupling matrix parameters.  The Type Ia and Ib cases (collectively referred to as Type I) and the Type IIa and IIb cases (collectively referred to as Type II) differ respectively by the interchange of $\Phi_1\to\Phi_2$ or equivalently by the interchange of $\cot\beta\to\tan\beta$. The presence of the $\mathbb{Z}_2$ symmetry fixes $\rho^U$ and $\rho^D$ to be diagonal matrices as exhibited below.}
\vskip -0.2in
\label{Tab:type}
\vskip 0.1in
\begin{tabular}{|cl||c|c|c|c|c|c|c|}
 \hline &  &\, $\Phi_1$\, &\,  $\Phi_2$ \, &\, $U_R^{}$ \, &\,  $D_R^{}$ \,  & $U_L$, $D_L$  & $\rho^U$ & $\rho^D$ \\  \hline
Type Ia  && $+$ & $-$ & $-$ & $-$ & $+$  & $\,\phm e^{i(\xi+\theta_{23})}(\sqrt{2}M_U/v)\cot\beta\,$ & $\,\phm e^{i(\xi+\theta_{23})}(\sqrt{2}M_D/v)\cot\beta\,$  \\
Type Ib  && $+$ & $-$ & $+$ & $+$ & $+$  & $\, -e^{i(\xi+\theta_{23})}(\sqrt{2}M_U/v)\tan\beta\,$ & $\,-e^{i(\xi+\theta_{23})}(\sqrt{2}M_D/v)\tan\beta\,$  \\
Type IIa && $+$ & $-$ & $-$ & $+$ & $+$  & $\,\phm e^{i(\xi+\theta_{23})}(\sqrt{2}M_U/v)\cot\beta\,$ & $\, -e^{i(\xi+\theta_{23})}(\sqrt{2}M_D/v)\tan\beta\,$  \\
Type IIb && $+$ & $-$ & $+$ & $-$ & $+$  & $\,-e^{i(\xi+\theta_{23})}(\sqrt{2}M_U/v)\tan\beta\,$ & $\,\phm e^{i(\xi+\theta_{23})}(\sqrt{2}M_D/v)\cot\beta\,$  \\
\hline
\end{tabular}
\end{center}
\vskip -0.2in
\end{table}

One must also impose the  $\boldsymbol{\mathbb{Z}_2}$ symmetry on the Yukawa Lagrangian, which defines the so-called $\mathbb{Z}_2$ basis.   
Four possible $\mathbb{Z}_2$ charge assignments are exhibited in Table~\ref{Tab:type},
\beqa
&& \text{Type Ia:}~~~h^U_1=h^D_1=0\,,\qquad   \text{Type Ib:}~~~h^U_2=h^D_2=0\,, \label{typeoneab}\\
&& \text{Type IIa:}~~h^U_1=h^D_2=0\,, \qquad  \text{Type IIb:}~~h^U_2=h^D_1=0\,.\label{typetwoab}
\eeqa
Of course, the above conditions are basis dependent.  
Types Ia and Ib (collectively denoted by Type I) and Types IIa and IIb (collectively denoted by Type II) 
are essentially equivalent, respectively, differing only in which scalar is denoted by $\Phi_1$ and which is denoted by $\Phi_2$.
In \Ref{Haber:2006ue}, the following basis-independent conditions were given,
\beqa
&& \text{Type I:}~~~\epsilon_{\abar\bbar}h_a^Dh_b^U=\epsilon_{ab}h^{D\dagger}_{\abar}h^{U\dagger}_{\bbar}=0\,,\\
&& \text{Type II:}~~\delta_{a\bbar}h^{D\dagger}_{\abar}h^U_b=0\,,
\eeqa
which are clearly satisfied in the $\mathbb{Z}_2$ basis.   Employing \eq{hexpand} yields the invariant conditions,
\beqa
&& \text{Type I:}~~~\kappa^D\rho^U-\kappa^U\rho^D=0\,,\\
&& \text{Type II:}~~\kappa^D\kappa^U+\rho^{D\dagger}\rho^U=0\,,
\eeqa
where we have used the fact that $\kappa^Q$ is a real matrix [cf.~\eq{MQ}].

In the $\mathbb{Z}_2$ basis, \eq{potmin} yields 
$\widehat{v}=(\cos\beta\,,\,e^{i\xi}\sin\beta)$ and $\widehat{w}=(-e^{-i\xi}\sin\beta\,,\,\cos\beta)$,
where $\tan\beta\equiv |v_2|/|v_1|$.
Hence using \eqs{kapparho}{MQ}, one obtains
\beqa
&& \text{Type Ia:}~~~\rho^U=\frac{e^{i(\xi+\theta_{23})}\sqrt{2}M_U\cot\beta}{v}\,,\qquad\quad  \rho^D=\frac{e^{i(\xi+\theta_{23})}\sqrt{2}M_D\cot\beta}{v}\,, \label{rhoType1a}\\
&& \text{Type Ib:}~~~\rho^U=-\frac{e^{i(\xi+\theta_{23})}\sqrt{2}M_U\tan\beta}{v}\,,\qquad  \rho^D=-\frac{e^{i(\xi+\theta_{23})}\sqrt{2}M_D\tan\beta}{v}\,, \label{rhoType1b}\\
&& \text{Type IIa:}~~\rho^U=\frac{e^{i(\xi+\theta_{23})}\sqrt{2}M_U\cot\beta}{v}\,,\qquad\quad  \rho^D=-\frac{e^{i(\xi+\theta_{23})}\sqrt{2}M_D\tan\beta}{v}\,,\label{rhoType2a} \\
&& \text{Type IIb:}~~\rho^U=-\frac{e^{i(\xi+\theta_{23})}\sqrt{2}M_U\tan\beta}{v}\,,\qquad\,  \rho^D=\frac{e^{i(\xi+\theta_{23})}\sqrt{2}M_D\cot\beta}{v}\,,\label{rhoType2b}
\eeqa
which we have also recorded in Table~\ref{Tab:type}.  Indeed $\rho^U$ and $\rho^D$ are proportional to the diagonal quark matrices $M_U$ and $M_D$, respectively, indicating that the tree-level Higgs-quark couplings are flavor diagonal.
Since the $\rho^Q$ are basis invariants, the quantity, $e^{i(\xi+\theta_{23})}\tan\beta$, is a physical parameter in the 2HDM with Type-I or Type-II Yukawa couplings.  

In particular, note that one still has the freedom to make a transformation that interchanges $\Phi_1\leftrightarrow\Phi_2$ in the $\mathbb{Z}_2$ basis.  In performing such
a basis transformation, one must also  interchange $\tan\beta\leftrightarrow\cot\beta$ while changing the sign of the quantity $e^{i(\xi+\theta_{23})}$ [as we shall demonstrate in \eq{xi23}]. 
These two parameter transformations simply result in the interchange the $a$ and $b$ versions of the Type-I and Type-II Yukawa couplings.  
Once a specific discrete symmetry is chosen (among the four specified in Table~\ref{Tab:type}), $\tan\beta$ is promoted to a physical parameter of the model.  It then follows that
$e^{i(\xi+\theta_{23})}$ is also physical.  However, the parameters $\xi$ and $\theta_{23}$ separately retain their basis-dependent nature.

In contrast, the parameter $\tan\beta$ does not appear in the bosonic couplings of the 2HDM. 
This statement is easily checked by inserting \eq{master} into \eqs{pot}{KE}, which yields the Higgs self-couplings and the Higgs couplings to vector bosons~\cite{Haber:2006ue}.  The couplings of the Higgs bosons to the gauge bosons 
depend only on the gauge couplings and the invariant mixing angles $\theta_{12}$ and $\theta_{13}$ by virtue of \eqs{KE}{master}.\footnote{Note that the Type-I or Type-II conditions remove two of the four gauge invariant Yukawa couplings [cf.~\eqs{typeoneab}{typetwoab}], which ultimately provide meaning for the parameter $\tan\beta$.  In contrast, the imposition of the (softly broken) $\mathbb{Z}_2$ symmetry does \textit{not} remove any of the Higgs boson--gauge boson couplings, whose forms are fixed by gauge invariance.}    
The Higgs self-couplings will additionally depend on invariant combinations of the $Z_i$ and $e^{-i\theta_{23}}$.    If there exists a scalar field basis in which $\lambda_6=\lambda_7=0$, then this basis is related to the Higgs basis by a rotation by the angle~$\beta$.   The existence of such a basis will yield an invariant relation among the $Z_i$ that will be derived in the next section.   It is only through this relation [cf.~\eqs{sincos}{tanbeta}] that $\tan\beta$ can be indirectly probed via the Higgs self-couplings.

\section{Basis-independent treatment of the $\boldsymbol{\mathbb{Z}_2}$ symmetry}
\label{sec:five}

The  $\mathbb{Z}_2$ symmetry of the 2HDM scalar potential is manifestly realized in a scalar field basis where $m_{12}^2=\lambda_6=\lambda_7=0$, and is softly broken if $m_{12}^2\neq 0$ in a basis where \mbox{$\lambda_6=\lambda_7=0$.}  Of~course, such a description is basis-dependent.  In this section, we explore a basis-independent characterization of the $\boldsymbol{\mathbb{Z}_2}$ symmetry, where the symmetry is either exact or softly broken.  We obtain conditions in terms of Higgs basis parameters that are independent of the initial choice of scalar field basis.  Our analysis generalizes results previously obtained in Refs.~\cite{Lavoura:1994yu,Ginzburg:2004vp,Haber:2015pua}.  The connection of the results obtained in this section with the basis-independent conditions that are independent of the vacuum, derived in Ref.~\cite{Davidson:2005cw}, is discussed in Appendix~\ref{appC}.   An alternative basis-independent treatment of the $\mathbb{Z}_2$ symmetry based on the bilinear formalism of the 2HDM scalar potential can be found in Refs.~\cite{Ivanov:2005hg,Ferreira:2010hy,Ferreira:2010yh}. 

\subsection{The inert doublet model}
\label{idm}

A very special case of the 2HDM is the so-called inert doublet model (IDM).   In this model, the Higgs basis exhibits an exact $\mathbb{Z}_2$ symmetry, $\mathcal{H}_1\to \mathcal{H}_1$ and $\mathcal{H}_2\to -\mathcal{H}_2$.
Imposing this symmetry on the scalar potential given in \eq{higgspot} yields
\beq \label{idmcond}
Y_3=Z_6=Z_7=0\,.
\eeq
The conditions given in \eq{idmcond} are basis independent given that $Y_3$, $Z_6$ and $Z_7$ are 
pseudoinvariant  quantities.  
Note that it is sufficient to impose the $\mathbb{Z}_2$ symmetry on the dimension-four terms of \eq{higgspot}, since if $Z_6=0$ then $Y_3=0$ due to the scalar potential minimum conditions of \eq{minconds}.   Thus in this case, it is not possible to softly break the 
$\mathbb{Z}_2$ symmetry.

To complete the definition of the IDM, the 
Higgs-fermion Yukawa couplings are fixed by imposing the condition that  
all fermion fields are even under the $\mathbb{Z}_2$ symmetry.   This corresponds to Type-Ib Yukawa couplings as specified in Table~\ref{Tab:type} with $\tan\beta=0$.   In this case
$\rho^U=\rho^D=0$, which implies that the doublet $\mathcal{H}_2$ does not couple to the fermions.  Consequently, $\mathcal{H}_2$ is called an inert doublet.  
Due to the fact that 
$Z_6=0$, the tree-level couplings of the neutral scalar that resides in the doublet $\mathcal{H}_1$ are precisely those of the SM Higgs boson.
Moreover, in the bosonic sector of the theory, the scalar fields that reside in the doublet $\mathcal{H}_2$ can only couple in pairs to the gauge bosons and to the SM Higgs boson.

In light of \eq{idmcond}, $Z_5$ is the only potentially complex parameter of the IDM scalar potential.   This means that one is free to rephase the pseudoinvariant Higgs basis field $H_2$ such that
all Higgs basis scalar potential parameters are real.  Hence, the IDM scalar potential and vacuum are CP conserving.   Since the main interest of this paper is the 2HDM with
a softly broken $\mathbb{Z}_2$ symmetry and CP violation, we shall henceforth assume that the $\mathbb{Z}_2$ symmetry of the dimension-four terms of the scalar potential is manifestly realized  in a basis that is not the Higgs basis. That is, $Z_6$ and $Z_7$ are not both simultaneously equal to zero.
This assumption will allow for the possibility of a 2HDM scalar sector that exhibits either explicit or spontaneous CP violation.

\subsection{A softly broken $\mathbb{Z}_2$ symmetry}
\label{soft}

Suppose that the $\mathbb{Z}_2$ symmetry of the dimension-four terms of the scalar potential is manifestly realized in some scalar field basis (henceforth denoted as the $\mathbb{Z}_2$ basis), which implies that $\lambda_6=\lambda_7=0$ in this basis.  In light of \eqs{plus}{minus}, it follows that the $\mathbb{Z}_2$ basis exists if and only if values of $\beta$ and $\xi$ can be found such that,
 \beqa
&&\hspace{-0.5in}
\half\stwob\left(Z_1-Z_2\right)+c_{2\beta}\Re\bigl(Z_{67}e^{i\xi}\bigr)+i\Im\bigl(Z_{67}e^{i\xi}\bigr)= 0\,, \label{plusa}\\
&&\hspace{-0.5in}
\half\stwob\ctwob\left[Z_1+Z_2-2Z_{34}-2\Re(Z_5e^{2i\xi})\right]-is_{2\beta}\Im(Z_5 e^{2i\xi})+c_{4\beta}\Re\bigl[(Z_6-Z_7)e^{i\xi}\bigr]
\nonumber \\
&& \qquad\qquad\qquad\qquad\qquad\qquad\qquad\qquad
+ic_{2\beta}\Im\bigl[(Z_6-Z_7)e^{i\xi}\bigr]=0\,, \label{minusa}
\eeqa
where $Z_{34}\equiv Z_3+Z_4$ and $Z_{67}\equiv Z_6+Z_7$.
The real and imaginary parts of \eqs{plusa}{minusa} yield four independent real equations.

The $\mathbb{Z}_2$ basis is not unique.  Suppose we choose a $\Phi$ basis in which $\lambda_6=\lambda_7=0$.   To maintain the conditions, $\lambda_6=\lambda_7=0$, it is still possible to transform to a new $\Phi^\prime$ basis that is related to the $\Phi$ basis according to $\Phi^\prime_a=U_{a\bbar}\Phi_b$, where
\beq \label{U}
U=\begin{pmatrix} 0 & \quad e^{-i\xi} \\ e^{i\zeta} & \quad 0\end{pmatrix}\,.
\eeq
In particular, by noting that 
\beq
\begin{pmatrix} s_\beta \\ c_\beta e^{i\zeta}\end{pmatrix}=U\begin{pmatrix} c_\beta \\ s_\beta e^{i\xi}\end{pmatrix}\,,
\eeq
it immediately follows that $\beta^\prime=\half\pi-\beta$ and $\xi^\prime=\zeta$.  Moreover, after employing \eq{rephasing} where $\det U=-e^{i(\zeta-\xi)}$ it follows that if $\Phi_a\to U_{a\bbar}\Phi_b$ with $U$ given by \eq{U}, then
\beq \label{phitophiprime}
Z_5 e^{2i\xi}\to Z_5 e^{2i\xi}\,,\quad Z_6 e^{i\xi}\to -Z_6 e^{i\xi}\,,\quad Z_7 e^{i\xi}\to -Z_7 e^{i\xi}\,,\quad s_{2\beta}\to s_{2\beta}\,,\quad c_{2\beta}\to -c_{2\beta}\,.
\eeq
That is, the left-hand side of \eq{plusa} [\eq{minusa}] is transformed into [the negative of] its complex conjugate, and the 
four real equations obtained from \eqs{plusa}{minusa} are unchanged.
Likewise, using \eq{shift} it follows that if $\Phi_a\to U_{a\bbar}\Phi_b$ with $U$ given by \eq{U}, then
\beq \label{xi23}
e^{i(\xi+\theta_{23})}\to -e^{i(\xi+\theta_{23})}\,,
\eeq
which shows that the phase factor, $e^{i(\xi+\theta_{23})}$, appearing in the expressions for $\rho^Q$ exhibited in \eqst{rhoType1a}{rhoType2b}, changes sign when transforming from the $\Phi$ basis to the $\Phi^\prime$ basis.  Consequently, the effect of this scalar field transformation is to interchange the $a$ and $b$ versions of the Type-I and Type-II Yukawa couplings as asserted below \eq{rhoType2b}.

Returning to \eqs{plusa}{minusa}, we first take
the imaginary part of \eq{plusa} to obtain,
\beq \label{z67}
\Im(Z_{67}e^{i\xi})=0\,.
\eeq
Assuming that $Z_{67}\neq 0$ (we will return to the case of $Z_{67}=0$ later), we shall denote,
\beq \label{z67p}
Z_{67}=|Z_{67}|e^{i\theta_{67}}\,.
\eeq
Then, \eq{z67} implies that $\xi+\theta_{67}=n\pi$, for some integer $n$, or equivalently 
\beq \label{signchoice}
e^{i\xi}=\pm e^{-i\theta_{67}}\,.
\eeq
The two possible sign choices in \eq{signchoice} correspond to the $\Phi$ and $\Phi^\prime$ basis choices identified above in which $\lambda_6=\lambda_7=0$ is satisfied.
Employing 
\eq{signchoice} in \eqs{plusa}{minusa} yields,
\beqa
&&\hspace{-0.3in}
\half\stwob\left(Z_1-Z_2\right)\pm c_{2\beta}|Z_{67}|= 0\,, \label{plusb}\\
&&\hspace{-0.3in}
\half\stwob\ctwob\left[Z_1+Z_2-2Z_{34}-2\Re(Z_5e^{-2i\theta_{67}})\right]-is_{2\beta}\Im(Z_5 e^{-2i\theta_{67}})\pm c_{4\beta}\Re\bigl[(Z_6-Z_7)e^{-i\theta_{67}}\bigr]
\nonumber \\
&& \qquad\qquad\qquad\qquad\qquad\qquad\qquad\qquad
\pm ic_{2\beta}\Im\bigl[(Z_6-Z_7)e^{-i\theta_{67}}\bigr]=0\,. \label{minusb}
\eeqa

Assuming $Z_1\neq Z_2$ (we will return to the case of $Z_1=Z_2$ below), \eq{plusb} yields
\beq \label{tan2b}
\frac{s_{2\beta}}{c_{2\beta}}=\pm\,\frac{2|Z_{67}|}{Z_2-Z_1}\,.
\eeq
Since $0\leq\beta\leq\half\pi$, it follows that
\beq \label{sincos}
s_{2\beta}=\frac{2|Z_{67}|}{\sqrt{(Z_2-Z_1)^2+4|Z_{67}|^2}}\,,\qquad\quad
c_{2\beta}=\frac{\pm(Z_2-Z_1)}{\sqrt{(Z_2-Z_1)^2+4|Z_{67}|^2}}\,,
\eeq
In particular,
\beq \label{tanbeta}
\tan\beta=\sqrt{\frac{1-c_{2\beta}}{1+c_{2\beta}}}\,,
\eeq
which demonstrates that $\tan\beta$ in the $\Phi$ basis corresponds to $\cot\beta$ in the $\Phi^\prime$ basis. 
Moreover,
\beq \label{xit23}
e^{i(\xi+\theta_{23})}=\pm e^{i(\theta_{23}-\theta_{67})}=\pm\frac{|Z_{67}|}{Z_{67}e^{-i\theta_{23}}}=\left(\frac{Z_2-Z_1}{2Z_{67}e^{-i\theta_{23}}}\right)\frac{s_{2\beta}}{c_{2\beta}}\,.
\eeq
Note that \eq{xit23} is consistent with the result of \eq{xi23}.

Plugging the results of \eq{sincos} back into \eq{minusb},
\beqa
&&\hspace{-0.3in} |Z_{67}|(Z_2-Z_1)\left[Z_1+Z_2-2Z_{34}-2\Re(Z_5e^{-2i\theta_{67}})\right]
+\bigl[(Z_2-Z_1)^2-4|Z_{67}|^2\bigr]\Re\bigl[(Z_6-Z_7)e^{-i\theta_{67}}\bigr] \nonumber \\
&& \quad\qquad
\pm iD\left\{(Z_2-Z_1)\Im\bigl[(Z_6-Z_7)e^{-i\theta_{67}}\bigr]
-2|Z_{67}|\Im(Z_5 e^{-2i\theta_{67}})\right\}=0\,,
\eeqa
where $D\equiv \sqrt{(Z_2-Z_1)^2+4|Z_{67}|^2}$.
We can use \eq{z67p} to write $e^{-i\theta_{67}}=Z_{67}^*/|Z_{67}|$.   It then follows that
\beqa
&&\hspace{-0.3in} (Z_2-Z_1)\left[|Z_{67}|^2(Z_1+Z_2-2Z_{34})-2\Re(Z^*_5 Z_{67}^2)\right]
+\bigl[(Z_2-Z_1)^2-4|Z_{67}|^2\bigr]\bigl[|Z_6|^2-|Z_7|^2\bigr] \nonumber \\
&& \qquad\qquad
\pm 2iD\left\{(Z_1-Z_2)\Im(Z_6^*Z_7)+
\Im(Z^*_5 Z_{67}^2)\right\}=0\,.
\label{complexeq}
\eeqa
Taking the real and imaginary parts of \eq{complexeq} and massaging the real part yields 
\beqa
&& (Z_1-Z_2)\bigl[Z_{34} |Z_{67}|^2-Z_2 |Z_6|^2-Z_1 |Z_7|^2-(Z_1+Z_2)\Re(Z_6^* Z_7)+\Re(Z_5^* Z^2_{67})\bigr]\nonumber \\
&& \qquad\qquad\qquad\qquad\qquad -2|Z_{67}|^2\bigl(|Z_6|^2-|Z_7|^2\bigr)=0\,,\label{cond5}\\
&& (Z_1-Z_2)\Im(Z_6^* Z_7)+\Im\bigl(Z_5^* Z_{67}^2\bigr)=0\,.\label{cond6}
\eeqa
It is convenient to multiply \eq{cond6} by $-i$ and add the result to \eq{cond5}.   This yields a single 
complex equation.
Finally, since $Z_{67}\neq 0$ by assumption, one can divide 
this complex equation by $Z^*_{67}$ and take the complex conjugate of the result to obtain
\beq \label{finalcond}
(Z_1-Z_2)\bigl[Z_{34}Z^*_{67}-Z_1 Z^*_7-Z_2 Z^*_6+Z^*_5 Z_{67}\bigr]-2Z^*_{67}\bigl(|Z_6|^2-|Z_7|^2\bigr)=0\,.
\eeq

The cases where $Z_1=Z_2$ and/or $Z_{67}=0$ are easily treated.  First, if $Z_1=Z_2$ and $Z_{67}\neq 0$, then 
\eqs{plusb}{minusb} imply that $s_{2\beta}=1$ and $c_{2\beta}=0$, and it follows that $\Im(Z_5^* Z^2_{67})=0$ and $|Z_6|=|Z_7|$.   
Second, if $Z_{67}=0$ and $Z_1\neq Z_2$, then 
\eq{plusa} yields $s_{2\beta}=0$, which when inserted into \eq{minusa} implies that $Z_6 e^{i\xi}=0$.   That is, if $Z_{67}=0$ then $Z_6=Z_7=0$, and the $\mathbb{Z}_2$ symmetry is manifest in the Higgs basis, as noted in \sect{idm}.
In this latter case, one must employ the Type-Ib Yukawa interactions, which yield $\rho^U=\rho^D=0$.   This corresponds to the case of $\tan\beta=0$ in \eq{rhoType1b}.\footnote{If one were to employ Type-Ia Yukawa couplings, then one would find that $M_U=M_D=0$, while $\rho^U$ and $\rho^D$ are arbitrary complex matrices.}   Likewise, in the case of Type-II couplings, $M_U=\rho^D=0$ and $\rho^U$ is a arbitrary complex matrix.  In the IDM (corresponding to a Type-Ib Yukawa sector with $Z_6=Z_7=0$), the fermions couple only to the $\mathbb{Z}_2$-even scalar doublet~$\mathcal{H}_1$, whose tree-level interactions exactly coincide with those of the SM Higgs doublet.

Finally, the case of $Z_1=Z_2$ and $Z_{67}=0$ requires special treatment; this case has been dubbed the ``exceptional region'' of the 2HDM parameter space in Ref.~\cite{Ferreira:2009wh}.
The analysis of Appendix~\ref{erps} shows that in this exceptional case, 
there always exists a scalar field basis in which the softly broken $\mathbb{Z}_2$ symmetry is manifestly realized.
Furthermore, \eqs{cond6}{finalcond} are trivially satisfied in the exceptional region of the 2HDM parameter space.

In conclusion, \eq{finalcond} is a necessary condition for the presence of a softly broken $\mathbb{Z}_2$ symmetry.   It is also a sufficient condition in all cases with one exception.   Namely, if $Z_1=Z_2$, $Z_5\neq 0$ and $Z_{67}\neq 0$, then \eq{finalcond} must be supplemented with the additional constraint of $\Im(Z_5^* Z^2_{67})=0$ .

In the case of the CP-conserving 2HDM, it is possible to rephase the pseudoinvariant Higgs basis field $H_2$ 
such that all of the $Z_i$ are real.  In this real basis, \eq{finalcond} reduces to
\beq \label{real1}
(Z_1-Z_2)\bigl[(Z_{34}+Z_5)Z_{67}-Z_2 Z_6-Z_1 Z_7\bigr]-2Z^2_{67}\bigl(Z_6-Z_7\bigr)=0\,,
\eeq
a result previously given in eq.~(54) of Ref.~\cite{Haber:2015pua}.  The scalar basis in which $\lambda_6=\lambda_7=0$ is obtained from the Higgs basis by a rotation by an angle $\beta$, which is determined by \eq{tan2b},
\beq
\cot 2\beta=\frac{Z_2-Z_1}{2Z_{67}}\,,
\eeq
in a convention where $v_1$ and $v_2$ are non-negative [in which case $\xi=0$ so that ${\rm sgn}~\!Z_{67}=\pm 1$ in light of \eq{signchoice}].
Once again, the exceptional region of parameter space where $Z_1=Z_2$ and $Z_{67}=0$ must be treated separately.  
Using \eqs{except1}{except2} with $\xi=0$ and real $Z_i$, it follows that  
$\cot 2\beta$ is a solution of \eq{tantwobeta}, where $Z_5$ and $Z_6$ are real and $\pm$ is identified with ${\rm sgn}~\!Z_6$ (or equivalently, replace $|Z_6|$ with $Z_6$ and replace $\pm$ with a plus sign). 

\subsection{Softly broken $\boldsymbol{\mathbb{Z}_2}$ symmetry and spontaneously broken CP symmetry}
\label{scpv}

Suppose that the conditions for a softly broken $\boldsymbol{\mathbb{Z}_2}$-symmetric scalar potential obtained in Section~\ref{soft} are satisfied.  
Then a $\mathbb{Z}_2$ basis exists (which is not unique) in which $\lambda_6=\lambda_7=0$.  If in addition, 
\beq \label{imlam5}
\Im\bigl(\lambda_5^*[m^2_{12}]^2\bigr)=0\,,
\eeq
then one can rephase one of the scalar fields such that $m_{12}^2$ and $\lambda_5$ are simultaneously real.  In this case, the scalar potential is explicitly CP invariant.
In addition, if in this so-called real $\mathbb{Z}_2$ basis there is an unremovable complex phase in the vevs; that is,
\beq
\Im(v_1^* v_2)=\half v^2 s_{2\beta}\sin\xi\neq 0\,, 
\eeq
then the CP symmetry of the scalar potential is spontaneously broken.

Using \eqs{m12}{lam5def},  
\beqa \label{imexp}
\Im\bigl(\lambda_5^*[m^2_{12}]^2\bigr)&=& \biggl\{\tfrac14 s_{2\beta}^2\bigl[Z_1+Z_2-2Z_{345}\bigr]+\Re(Z_5 e^{2i\xi})+s_{2\beta}c_{2\beta}\Re\bigl[(Z_6-Z_7)e^{i\xi}\bigr]\biggr\} \nonumber \\
&& \qquad\qquad \times
\bigl[(Y_1-Y_2)s_{2\beta}+2\Re(Y_3 e^{i\xi})c_{2\beta}\bigr]\Im(Y_3 e^{i\xi})
\nonumber \\[5pt]
&& \!\!\!\!  -\biggl\{\tfrac14\bigr[(Y_1-Y_2)s_{2\beta}+2\Re(Y_3 e^{i\xi})c_{2\beta}\bigr]^2-\bigl[\Im(Y_3 e^{i\xi})\bigr]^2\biggr\}
\nonumber \\
&& \qquad\qquad \times
\bigl[c_{2\beta}\Im(Z_5 e^{2i\xi})+s_{2\beta}\Im[(Z_6-Z_7)e^{i\xi}]\bigr]\,,
\eeqa
where $Z_{345}\equiv Z_{34}+\Re(Z_5 e^{2i\xi})$.  Next, we employ the potential minimum conditions [\eq{minconds}], $Y_1=-\half Z_1 v^2$ and $Y_3=-\half Z_6 v^2$,
and we make use of \eq{sincos} for $s_{2\beta}$ and $c_{2\beta}$.  
To make further progress, we first assume that $Z_1\neq Z_2$ and $Z_{67}\neq 0$.   In this case, we can
use \eqs{z67p}{signchoice} to write $e^{i\xi}=\pm Z_{67}^*/|Z_{67}|$.  It is convenient to introduce the following notation
\beq \label{effs}
f_1\equiv |Z_{67}|^2\,,\qquad\quad f_2\equiv |Z_7|^2-|Z_6|^2\,,\qquad\quad f_3\equiv \Im(Z_6 Z_7^*)\,.
\eeq
It then follows that,
\beqa
\Re(Z_6 e^{i\xi})&=&\pm\frac{\Re(Z_6 Z_7^*)+|Z_6|^2}{|Z_{67}|}=\pm\half(f_1-f_2)f_1^{-1/2}\,,\label{e3} \\[5pt]
\Im(Z_6 e^{i\xi})&=&\pm\frac{\Im(Z_6 Z_7^*)}{|Z_{67}|}=\pm f_3 f_1^{-1/2}\,,\label{e4} \\[5pt]
\Re\bigl[(Z_6-Z_7)e^{i\xi}\bigr] &=& \pm\left(\frac{|Z_6|^2-|Z_7|^2}{|Z_{67}|}\right)=\mp f_2 f_1^{-1/2}\,,\label{e5} \\[5pt]
\Im\bigl[(Z_6-Z_7)e^{i\xi}\bigr] &=& \pm\frac{2\Im(Z_6 Z_7^*)}{|Z_{67}|}=\pm2f_3 f_1^{-1/2}\,.\label{e6} 
\eeqa
Finally, we employ \eqs{cond5}{cond6} to obtain,
\beqa
\Re(Z_5 e^{2i\xi})&=&\frac{\Re(Z_5^* Z^2_{67})}{|Z_{67}|^2}=\frac{2f_2}{Z_2-Z_1}+\half(Z_1+Z_2)-Z_{34}+\frac{(Z_1-Z_2)f_2}{2f_1}\,, \label{e1} \\[5pt]
\Im(Z_5 e^{2i\xi})&=& -\frac{\Im(Z_5^* Z^2_{67})}{|Z_{67}|^2}=\frac{(Z_2-Z_1)f_3}{f_1}\,.\label{e2}
\eeqa

Plugging the above results into \eq{imexp}, we end up with
\beq \label{imm12sq}
\Im\bigl(\lambda_5^*[m^2_{12}]^2\bigr)=\mp\frac{v^4 f_3 \mathcal{F}}{16f_1^2(Z_1-Z_2)\sqrt{(Z_2-Z_1)^2+4f_1}}\,,
\eeq
where the function $\mathcal{F}$ is given by,\footnote{An expression for $\mathcal{F}$ was first derived by Lavoura in Ref.~\cite{Lavoura:1994yu}, although his eq.~(22) contains a misprint in which the factor of $f_2$ in the coefficient of $(Z_1-Z_2)^2(Z_1+Z_2-2Z_{34})$ in \eq{fancy} was inadvertently dropped. \label{fnlav}}
\beqa
\mathcal{F}&=&f_1^2\left[16(Z_1-Z_2)\left(\frac{Y_2}{v^2}\right)^2
+16\left[f_2+(Z_1-Z_2)Z_{34}\right]\left(\frac{Y_2}{v^2}\right)+4f_2(Z_1+Z_2)\right. \nonumber \\
&& \qquad\qquad -(Z_1^2-Z_2^2)(Z_1+Z_2-4Z_{34})\biggr]-(f_2^2+4f_3^2)(Z_1-Z_2)^3\nonumber \\[6pt]
&& -2f_1 f_2(Z_1-Z_2)^2(Z_1+Z_2-2Z_{34})+4f_1(f_2^2-4f_3^2)(Z_1-Z_2)\,.\label{fancy}
\eeqa

Thus, $\Im\bigl(\lambda_5^*[m^2_{12}]^2\bigr)=0$ if one of two conditions are satisfied: $f_3=0$ and/or $\mathcal{F}=0$.   If $f_3=0$, then it follows that
$\Im(Z_5 e^{2i\xi})=  \Im(Z_6 e^{i\xi})= \Im(Z_7 e^{i\xi})=0$.   This implies that one can rephase the Higgs basis field $H_2$ such that $Z_5$, $Z_6$ and $Z_7$ are
simultaneously real [which also implies that $Y_3$ is real by \eq{minconds}].  That is, all the coefficients of the scalar potential in the Higgs basis and the
corresponding vevs are real, implying that the scalar potential and the vacuum are CP conserving.  In contrast, if $f_3\neq 0$ and $\mathcal{F}=0$,
then the scalar potential is explicitly CP conserving as noted below \eq{imlam5}.  However, in the $\mathbb{Z}_2$ basis in which all scalar potential parameters are real,
the vevs exhibit a complex phase that cannot be removed by a basis transformation while maintaining real coefficients in the scalar potential.
In particular, $\Im(Z_6 Z_7^*)\neq 0$ implies that no real Higgs basis exists, which is a signal of CP violation.\footnote{We define a real Higgs basis to be the basis in which the potentially complex parameters $Z_5$, $Z_6$ and $Z_7$ are simultaneously real.  In this case, $Y_3$ is also real in light of \eq{minconds}.  Note that a real Higgs basis exists if and only if $\Im(Z_6^2 Z_5^*)=\Im(Z_7^2 Z_5^*)=\Im(Z_6 Z_7^* )=0$, in which case one can rephase the Higgs basis field $H_2$ appropriately to achieve the real Higgs basis.  In the 2HDM, the existence of a real Higgs basis is a necessary and sufficient condition for a CP-conserving scalar potential and vacuum.}    Thus, $f_3\neq 0$ and $\mathcal{F}=0$ is a basis-independent signal of
spontaneous CP violation.\footnote{Basis-independent conditions for spontaneous CP violation have also been obtained in the bilinear formalism of the 2HDM in  
Refs.~\cite{Nishi:2006tg,Maniatis:2007vn}.}  

If $\mathcal{F}=0$ then \eq{fancy} provides a quadratic equation for $Y_2$ that yields $Y_2\sim\mathcal{O}(Z_i)$.   In contrast, the decoupling limit of the 2HDM corresponds to $Y_2\gg v$~\cite{Haber:2006ue}.
Since $|Z_i|/4\pi\lsim\mathcal{O}(1)$ as a consequence of tree-level unitarity~\cite{Huffel:1980sk,Weldon:1984wt,Akeroyd:2000wc,Ginzburg:2005dt,Horejsi:2005da,Haber:2010bw,Kanemura:2015ska},
it follows that the 2HDM with a softly broken $\mathbb{Z}_2$ symmetry and spontaneous CP violation possesses no decoupling limit~\cite{Nebot:2019lzf}.

To complete the analysis of this subsection, we must address the special cases in which either $Z_1=Z_2$ and/or $Z_{67}=0$.  As noted below \eq{finalcond}, if $Z_1=Z_2$ and $Z_{67}\neq 0$, then 
\eqs{plusb}{minusb} imply that $s_{2\beta}=1$ and $c_{2\beta}=0$, and it follows that $\Im(Z_5^* Z^2_{67})=0$ and $|Z_6|=|Z_7|$ in light of \eqs{cond5}{cond6}.   Then, \eq{imexp}  yields,
\beqa 
\Im\bigl(\lambda_5^*[m^2_{12}]^2\bigr)&=&\mp\frac{v^4 f_3}{8f_1^{3/2}}\Biggl\{f_1\left[4\left(\frac{Y_2}{v^2}\right)^2+\frac{2Y_2}{v^2}\bigl(Z_1+Z_{34}\bigr)+Z_1 Z_{34}\right]
\nonumber\\[-6pt]
&& \qquad\qquad -4f_3^2-\left(Z_1+\frac{2Y_2}{v^2}\right)\Re(Z_5^* Z_{67}^2)\Biggr\}. \label{imm12sq2}
\eeqa
Note that in contrast to \eq{e1}, $\Re(Z_5^* Z_{67}^2)$ is not determined in terms of the $Z_i$, $f_1$ and $f_2$, since in the case of $Z_1=Z_2$, this quantity is not constrained by
\eqs{cond5}{cond6}.   Indeed, another way to derive \eq{imm12sq2} is to use \eq{e1} to solve for $f_2$ in terms of $\Re(Z_5^* Z_6^2)$ and substitute this result back into \eq{fancy}.  In this way, the factor of $Z_1-Z_2$ in the denominator of \eq{imm12sq} is canceled.   The resulting expression is significantly more complicated than the one given in \eq{fancy}.  Nevertheless, by setting $Z_1=Z_2$ in this latter expression, we have checked that one recovers the result of \eq{imm12sq2}.
Thus, we again conclude that spontaneous CP violation occurs if $f_3 \neq 0$ and the following basis-independent condition is satisfied:
\beq
f_1\left[4\left(\frac{Y_2}{v^2}\right)^2+\frac{2Y_2}{v^2}\bigl(Z_1+Z_{34}\bigr)+Z_1 Z_{34}\right]-4f_3^2-\left(Z_1+\frac{2Y_2}{v^2}\right)\Re(Z_5^* Z_{67}^2)=0\,.
\eeq

Next, as noted below \eq{finalcond}, if $Z_{67}=0$ and $Z_1\neq Z_2$, then \eqs{plusa}{minusa} imply that
$Z_6=Z_7=0$.  Thus, an unbroken $\mathbb{Z}_2$ symmetry is manifestly realized in the Higgs basis.   That is, in this case one identifies $m_{12}^2=0$ and thus $\Im(\lambda_5^*[m_{12}^2])=0$
is trivially satisfied.   Moreover, one can rephase the Higgs basis field $H_2$ such that $Z_5$ is real.  Hence, a real Higgs basis
exists which implies that both the scalar potential and the vacuum are CP conserving.  

So far, in all cases considered above, the conditions $\lambda_6=\lambda_7=0$ and $\Im\bigl(\lambda_5^*[m^2_{12}]^2\bigr)=0$ in the $\Phi$ basis were necessary and sufficient for an explicitly CP-conserving scalar potential.
One encounters a surprising result when considering the final case of the exceptional region of parameter space, where $Z_1=Z_2$ and $Z_7=-Z_6\neq 0$, where the only potentially CP-violating invariant is $\Im(Z_5^* Z_6^2)$.   Suppose that the Higgs basis parameters satisfy $\Im(Z_5^* Z_6^2)=0$, $Z_1=Z_2$ and
$Z_7=-Z_6\neq 0$.   Then, there exists a $\Phi$ basis that satisfies $\lambda_6=\lambda_7=0$, $\beta=\tfrac14\pi$ and $\cos(\xi+\theta_6)=0$, where $\theta_6=\arg Z_6$.
It follows that
\beqa
\quad
\Im(Z_5 e^{2i\xi})&=&\frac{\Im(Z_5^* Z_6^2)}{|Z_6|^2}=0\,,\qquad\quad \Re(Z_6 e^{i\xi})=\Re(Z_7 e^{i\xi})=0\,,\label{erpsa} \\
\Re(Z_5 e^{2i\xi})&=&-\frac{\Re(Z_5^* Z_6^2)}{|Z_6|^2}\,,\qquad\quad \Im(Z_6 e^{i\xi})=-\Im(Z_7 e^{i\xi})=\pm |Z_6|\,,\label{erpsb}
\eeqa
where the sign choice in \eq{erpsb} is correlated with $\sin(\xi+\theta_6)=\pm 1$.   In light of \eqs{lam6def} {lam7def}, it follows that $\lambda_6=\lambda_7=0$.
If we now insert the above results into \eqs{m12}{lam5def} and employ the scalar potential minimum conditions [\eq{minconds}], then 
\beqa
m_{12}^2 e^{i\xi}&=&\tfrac14 v^2\left[\left(Z_1+\frac{2Y_2}{v^2}\right)\pm 2 i|Z_6|\right]\,,\nonumber \\
\lambda_5 e^{2i\xi} &=&\half\left[Z_1-Z_{34}-\frac{\Re(Z_5^* Z_6^2)}{|Z_6|^2}\right] \pm 2i|Z_6|\,.
\eeqa
Hence, for generic choices of the remaining scalar potential parameters, one 
can conclude that a parameter regime within the exceptional region of the parameter space exits where 
\beqa
\Im\bigl(\lambda_5^*[m^2_{12}]^2\bigr) &=&\pm \frac{v^4}{8|Z_6|}\left\{|Z_6^2|\left[4|Z_6|^2-\left(Z_1+\frac{2Y_2}{v^2}\right)^2\right]\right. \nonumber \\[4pt]
&&\qquad\qquad \left. +\left(Z_1+\frac{2Y_2}{v^2}\right)\bigl[|Z_6|^2(Z_1-Z_{34})-\Re(Z_5^* Z_6^2)\bigr]\right\}\neq 0\,, \label{nonzero}
\eeqa
in which the scalar potential is explicitly CP conserving, and moreover CP is \textit{not} spontaneously broken!  In this case, CP is conserved despite the fact that no $\mathbb{Z}_2$ basis
exists in which all the scalar potential parameters are real (for further details, see Ref.~\cite{inprep}).

In the exceptional region of parameter space where $\lambda_6=\lambda_7=0$ is achieved for $\beta\neq\tfrac14\pi$, one finds once again that $\Im\bigl(\lambda_5^*[m^2_{12}]^2\bigr)=0$ is both a necessary and sufficient condition for an explicit CP-conserving scalar potential.  Moreover, if $\Im\bigl(\lambda_5^*[m^2_{12}]^2\bigr)=0$ and $\Im(Z_5^* Z_6^2)\neq 0$, then CP is spontaneously broken.
Further details are provided at the end of Appendix~\ref{erps}.

\subsection{Imposing the convention of non-negative real vevs in the $\mathbb{Z}_2$ basis}
\label{xizero}

In some applications, it is convenient to adopt a convention in which $\xi=0$ in the basis where $\lambda_6=\lambda_7=0$.   If this condition is not satisfied initially, it is straightforward to impose this condition by an appropriate rephasing of the Higgs-doublet field $\Phi_2$.  In this convention, 
the real and imaginary parts of \eqs{plusa}{minusa} yield
\beqa
\half\stwob\left(Z_1-Z_2\right)+c_{2\beta}\Re~\!Z_{67}&=& 0\,, \label{plusamod}\\
\Im~\!Z_{67}&=& 0\,,\label{plusamod2} \\
\half\stwob\ctwob\left[Z_1+Z_2-2Z_{34}-2\Re Z_5\right]+c_{4\beta}\Re(Z_6-Z_7)
&=&0\,, \label{minusamod1} \\
s_{2\beta}\,\Im~\!Z_5-c_{2\beta}\,\Im(Z_6-Z_7)&=&0\,. \label{minusamod2} 
\eeqa
\Eqst{plusamod}{minusamod2} are equivalent to eq.~(3.16) of Ref.~\cite{Belusca-Maito:2017iob}.
Because we have fixed $\xi=0$ in the $\Phi$ basis, we must choose $\xi=\zeta=0$ in \eq{U} in defining the $\Phi^\prime$ basis in order to maintain our convention in which the vevs $v_1$ and $v_2$ are real and non-negative.   That is, $\Phi^\prime_a=U_{a\bbar}\Phi_b$ where $U=\left(\begin{smallmatrix} 0 & 1 \\ 1 & 0\end{smallmatrix}\right)$.
Since $\det U=-1$, it follows that pseudoinvariant quantities will change sign between the $\Phi$ and $\Phi^\prime$ bases.
Indeed, the effect of transforming from the $\Phi$ basis to the $\Phi^\prime$ basis is to modify the $\Phi$ basis parameters such that,
\beq
m_{11}^2\leftrightarrow m_{22}^2\,,\quad m_{12}^2\to m_{12}^{2\,\ast}\,,\quad \lambda_1\leftrightarrow\lambda_2\,,\quad \lambda_5\to\lambda_5^\ast\,,\quad
v_1\leftrightarrow v_2\,,
\eeq
whereas $\lambda_3$, $\lambda_4$ and $\lambda_6=\lambda_7=0$ are unchanged.   In light of \eq{rephasing},
the Higgs basis parameters obtained starting from the $\Phi^\prime$ basis differ from those obtained starting from the $\Phi$ basis by the following sign changes:
\beq
\{Y_3\,,\,Z_6\,,\,Z_7\}\to \{-Y_3\,,\,-Z_6\,,\,-Z_7\}\,,
\eeq
In particular, the Higgs basis parameter $Z_5$ is \textit{unchanged} since $(\det U)^2=1$.

As previously noted, $\tan\beta$ is not yet a physical parameter, since the effect of transforming from the $\Phi$ basis to the $\Phi^\prime$ basis is to modify $\beta\to\half\pi-\beta$.   In light of these remarks, one can check that \eqst{plusamod}{minusamod2} are invariant with respect to the transformation $\Phi^\prime_a=U_{a\bbar}\Phi_b$, and thus define the invariant conditions for the existence of a scalar field basis with $\lambda_6=\lambda_7=0$ and non-negative real scalar vevs (i.e., $\xi=0$).

Consider first the case of $Z_{67}\neq 0$.  By virtue of \eq{plusamod2}, it follows that the pseudoinvariant quantity $Z_{67}$ is real.   This condition fixes the
Higgs basis up to a twofold ambiguity that depends on the sign of $Z_{67}$.  This ambiguity is simply a consequence of the freedom to change from
the $\Phi$ basis to the $\Phi^\prime$ basis, while maintaining the $\mathbb{Z}_2$ basis conditions, \mbox{$\lambda_6=\lambda_7=0$}, as discussed above.
Likewise, the pseudoinvariant quantity $e^{i\theta_{23}}$ is determined up to a twofold ambiguity, as its sign can be flipped by transforming from  
the $\Phi$ basis to the $\Phi^\prime$ basis.  

One can obtain an explicit expression for $e^{i\theta_{23}}$ in terms of pseudoinvariant quantities by setting $\xi=0$ in \eq{xit23},
\beq \label{xizt23}
e^{i\theta_{23}}=\left(\frac{Z_2-Z_1}{2Z_{67}e^{-i\theta_{23}}}\right)\frac{s_{2\beta}}{c_{2\beta}}\,.
\eeq
Under $\Phi_1\leftrightarrow\Phi_2$, $c_{2\beta}$ changes sign, and we conclude that $\theta_{23}$ is determined modulo $\pi$.
However, a more practical expression can be obtained as follows.
Writing $Z_6\equiv |Z_6|e^{i\theta_6}$ and $Z_7\equiv |Z_7|e^{i\theta_7}$, \eq{plusamod2} is equivalent to the equation,
$|Z_6|\sin\theta_6+|Z_7|\sin\theta_7=0$.  One can eliminate $\theta_7$ and solve for $\theta_6$ to obtain
\beq \label{th6}
\tan\theta_6=\frac{\Im(Z_6 Z_7^*)}{|Z_6|^2+\Re(Z_6 Z_7^*)}\,,
\eeq
which implies that $\theta_6$ is determined modulo $\pi$.
Under the assumption that $Z_6\neq 0$, one can obtain an explicit formula for $e^{i\theta_{23}}$,
\beq \label{e23}
e^{i\theta_{23}}=\frac{|Z_6|e^{i\theta_6}}{Z_6e^{-i\theta_{23}}}\,,
\eeq
where the numerator and denominator on the right-hand side of \eq{e23} are evaluated by employing \eqs{th6}{zee6id}, respectively.   As expected, $\theta_{23}$ is thus determined modulo $\pi$.

If $Z_6=0$, then \eq{plusamod2} yields $\sin\theta_7=0$, which implies that $Z_7^2=|Z_7|^2$.   In this case, assuming $Z_5\equiv |Z_5|e^{i\theta_5}\neq 0$,it follows that
\beq \label{th5}
\cos\theta_5=\frac{\Re(Z_5^* Z_7^2)}{|Z_5||Z_7|^2}\,,\qquad \sin\theta_5=-\frac{\Im(Z_5^* Z_7^2)}{|Z_5||Z_7|^2}\,,\quad\text{in the case of $Z_6=0$}\,.
\eeq
Hence, 
\beq \label{e223}
e^{2i\theta_{23}}=\frac{ |Z_5|e^{i\theta_5}}{Z_5 e^{-2i\theta_{23}}}\,,
\eeq
where the numerator and denominator on the right-hand side of \eq{e223} are evaluated by employing \eqs{th5}{zee5id}, respectively.  Taking the square root of
\eq{e223} determines $\theta_{23}$ modulo $\pi$.

If $Z_5=Z_6=0$, then the squared-mass matrix of the neutral Higgs scalars is diagonal.  In this case, 
the mass basis and the Higgs basis (with $Z_7$ real) coincide and the scalar potential and vacuum are CP conserving.

The case of $Z_{67}=0$ must be separately considered.
If $Z_{67}=0$ and $Z_1\neq Z_2$, then as discussed below \eq{finalcond} it follows that $Z_6=Z_7=0$ corresponding to the IDM.  
The exceptional region of parameter space corresponding to $Z_{67}=0$, $Z_6\neq 0$ and $Z_1=Z_2$, is treated in Appendix~\ref{erps}.  In this case, \eq{signchoice} is replaced by
\beq
e^{i\xi}=e^{i\xi^\prime}e^{-i\theta_6}\,,
\eeq 
where $Z_6\equiv |Z_6|e^{i\theta_{6}}$ and $\xi^\prime\equiv\xi+\theta_6$ is a pseudoinvariant quantity that is determined modulo $\pi$ in Appendix~\ref{erps}.  Once again, we see that in a convention where $\xi=0$, the $\mathbb{Z}_2$ basis is uniquely defined up to a twofold ambiguity corresponding to the fact that $\xi^\prime$, and hence $\theta_6$ and $\theta_{23}$, have been determined modulo $\pi$. 

Finally, in light of the remarks at the end of Section~\ref{sec:four}, we can conclude that in a convention in which $\xi=0$, 
once a specific discrete symmetry is chosen (among the four specified in Table~\ref{Tab:type}), both $\tan\beta$ and $\theta_{23}$ are promoted to physical parameters of the model.  

\subsection{An exact $\mathbb{Z}_2$ symmetry}
\label{exact}

In Section~\ref{soft}, we defined the $\mathbb{Z}_2$ basis to be the scalar basis in which $\lambda_6=\lambda_7=0$.  If in addition $m_{12}^2=0$ in the same basis, then the scalar potential possesses an exact $\mathbb{Z}_2$ symmetry; i.e., it is invariant under $\Phi_1\to\Phi_1$ and $\Phi_2\to -\Phi_2$. 
In this case, the condition $m_{12}^2=0$ yields additional constraints.  In light of \eq{m12},
\beq \label{yy}
\half(Y_2-Y_1)s_{2\beta}-\Re(Y_3 e^{i\xi})c_{2\beta}-i\Im(Y_3 e^{i\xi})=0\,,
\eeq
where $\xi$ and $\beta$ have been determined previously by \eqs{signchoice}{tan2b}, respectively, under the assumption that $Z_{67}\neq 0$.
Hence, employing $e^{i\xi}=\pm e^{-i\theta_{67}}=\pm Z_{67}^*/|Z_{67}|$ in \eq{yy}, it follows that
\beqa
(Y_2-Y_1)|Z_{67}|^2-(Z_2-Z_1)\Re(Y_3Z^*_{67})&=&0\,,\label{cond3e} \\
\Im(Y_3 Z_{67}^*)&=&0\,.\label{cond4e}
\eeqa
Due to \eq{cond4e}, one can replace $\Re(Y_3 Z_{67}^*)$ in \eq{cond3e} by $Y_3 Z_{67}^*$ and then divide the resulting equation by $Z_{67}^*$.  It follows that for $Z_{67}\neq 0$, 
one can replace \eq{cond3e} by
\beq \label{Ycond}
(Y_2-Y_1)Z_{67}-Y_3(Z_2-Z_1)=0\,.
\eeq
The analysis above relied on the assumption that $Z_{67}\neq 0.$   Thus, we now examine the relevant conditions for an exactly $\mathbb{Z}_2$-symmetric scalar potential when $Z_{67}=0$. 

If $Z_{67}=0$ and $Z_6=0$, then we also have $Z_7=Y_3=0$ [the latter of \eq{minconds}], in which case the exact $\mathbb{Z}_2$ symmetry is manifest in the Higgs basis. 
Consequently, in what follows we shall assume that $Z_{67}=0$ and $Z_6\neq 0$. 

If $Z_{67}=0$ and $Z_1\neq Z_2$ then \eq{plusa} implies that $s_{2\beta}=0$, in which case \eq{yy} yields $\Re(Y_3 e^{i\xi})=\Im(Y_3 e^{i\xi})=\nobreak 0$.  That is, $Y_3=0$, and we again conclude that $Z_6=Z_7=0$ in light of \eq{minconds}, which reduces to the previous case considered.

If $Z_{67}=0$,  $Y_1=Y_2$ and $Z_1=Z_2$, then it follows from \eqs{minconds}{yy} that $\beta=\tfrac14\pi$ and $\Im(Z_6 e^{i\xi})=0$.
The real part of \eq{minusa} then yields $\Re(Z_6 e^{i\xi})=0$, which implies $Z_6=0$, which again reduces to the previous case considered.

In the three subcases considered above, \eq{Ycond} remains valid.    However, there is one last case where \eq{Ycond} is trivially satisfied and yet an additional constraint must be imposed in order to achieve a $\mathbb{Z}_2$-symmetric scalar potential.   
Consider the case of $Z_{67}=0$,  $Y_1\neq Y_2$, $Z_1=Z_2$ and $Z_6\neq 0$.    In this case, 
$\xi$ and $\beta$ are determined from \eq{yy} [since \eq{plusa} is no longer relevant].  We first note that the imaginary part of \eq{yy} yields $\Im(Z_{6}e^{i\xi})=0$
after employing \eq{minconds}.  Denoting
$Z_{6}\equiv  |Z_{6}|e^{i\theta_{6}}$,
it follows that $\xi+\theta_{6}=n\pi$, for some integer $n$.  Hence,
$e^{i\xi}=\pm e^{-i\theta_{6}}=\pm Z_6^*/|Z_6|$, which when applied in \eqs{minusa}{yy} yields,
\beqa
&&
\half s_{2\beta}(Y_2-Y_1)|Z_6|\mp\Re(Y_3 Z_6^*)c_{2\beta}\mp i\Im(Y_3 Z_6^*)=0\,,\label{cond9e} \\
&&
\stwob\ctwob\left[(Z_1-Z_{34})|Z_6|^2-\Re(Z^*_5 Z_6^2)\right]+is_{2\beta}\Im(Z^*_5 Z_6^2)\pm 2c_{4\beta}|Z_6|^3=0\,.
\label{cond10e}
\eeqa
In light of $Y_3=-\half Z_6 v^2$, \eq{cond9e} yields
\beq \label{tan2}
\tan 2\beta=\frac{s_{2\beta}}{c_{2\beta}}=\pm\frac{v^2|Z_6|}{Y_1-Y_2}\,.
\eeq
Since $Z_6\neq 0$, it follows that $s_{2\beta}\neq 0$.  Hence, the imaginary part of \eq{cond10e} yields 
\beq \label{Ycond1}
\Im(Z_5^* Z_6^2)=0\,.
\eeq
Dividing the real part of \eq{cond10e} by $s_{2\beta}^2$ and using the result of \eq{tan2}, we end up with,
\beq \label{Ycond2}
v^2(Y_1-Y_2)\left[(Z_1-Z_{34})|Z_6|^2-\Re(Z^*_5 Z_6^2)\right]+2|Z_6|^2\bigl[(Y_1-Y_2)^2-v^4|Z_6|^2\bigr]=0\,.
\eeq
We can replace \eqs{Ycond1}{Ycond2} by a single complex equation by multiplying \eq{Ycond1} by $-iv^2(Y_1-Y_2)$ and adding the result to \eq{Ycond2}.
Additional simplification ensues by using \eq{minconds} to put $|Z_6|^2(Z_1v^2 +2Y_1)=0$.   It then follows that
\beq \label{lastcond}
(Y_1-Y_2)\left[|Z_6|^2\left(Z_{34}+\frac{2Y_2}{v^2}\right)+Z_5^* Z_6^2\right]+2|Z_6|^4 v^2=0\,.
\eeq

In conclusion, \eqs{finalcond}{Ycond} are necessary conditions for the presence of an exact $\mathbb{Z}_2$ symmetry.   These are also sufficient conditions in all cases with two exceptions.   As previously noted, if $Z_1=Z_2$, $Z_{67}\neq 0$ and $Z_5\neq 0$, then \eq{finalcond} must be supplemented with the additional constraint of $\Im(Z_5^* Z^2_{67})=0$ .  In addition, if $Z_1=Z_2$, $Z_{67}=0$, $Y_1\neq Y_2$ and $Z_6\neq 0$, then \eq{Ycond} must be supplemented by \eq{lastcond}.

In this paper, we are primarily interested in the case where either the scalar potential or the vacuum is CP violating.  However, it is easy to see that if the $\mathbb{Z}_2$ symmetry is
exact, then both the scalar potential and vacuum are CP conserving.   In the $\mathbb{Z}_2$ basis, since $m_{12}^2=\lambda_6=\lambda_7=0$, the only potentially complex scalar potential parameter is $\lambda_5$, whose phase can be removed by an appropriate rephasing of the Higgs fields.  
Moreover, if $\vev{\Phi_1^\dagger\Phi_2}=\half v_1 v_2 e^{i\xi}$, then the $\xi$-dependent term of the scalar potential is of the form $\mathcal{V} \ni \tfrac14\lambda_5 v_1^2 v_2^2\cos 2\xi$, which is minimized when $\xi=0$, $\half\pi$ or $\pi$ (depending on the sign of~$\lambda$).   If $\xi=\half\pi$, then one can rephase $\Phi_2\to i\Phi_2$, which simply changes the sign $\lambda_5$ while rendering the two vevs relatively real.  Hence, the vacuum is CP conserving.  Having achieved a scalar potential with only real parameters and real vevs, it immediately follows that a real Higgs basis exists.  That is a Higgs basis exists such that $Z_5$, $Z_6$ and $Z_7$ (and $Y_3=-\half Z_6 v^2$ via the scalar potential minimum condition) are simultaneously real.

Nevertheless, it is instructive to show directly that the existence of a real Higgs basis can be deduced solely from the relations satisfied by the Higgs basis parameters when an exact $\mathbb{Z}_2$ symmetry is present.  First, consider the case where the exact $\mathbb{Z}_2$ symmetry is manifest in the Higgs basis, i.e. $Y_3=Z_6=Z_7=0$. 
In this case the only potentially complex parameter in the Higgs basis is $Z_5$.  The phase of $Z_5$ can be removed by a rephasing of the Higgs basis field $H_2$.  Hence,
if the $\mathbb{Z}_2$ symmetry is manifest in the Higgs basis, then a real Higgs basis exists and the scalar potential and the vacuum are CP conserving. 

Next, suppose that $Z_{67}\neq 0$.  Then, if we
combine \eqs{cond6}{cond4e} and employ the scalar potential minimum condition, it follows that if the $\mathbb{Z}_2$ symmetry is exact, then
\beq \label{cpconds2}
\Im(Z_5^* Z_{67}^2)=\Im(Z_6^* Z_7)=0\,.
\eeq  
Given that $Z_{67}\neq 0$, the two conditions exhibited in \eq{cpconds2}
are sufficient to guarantee the existence of a real Higgs basis in which $Z_5$, $Z_6$ and $Z_7$ are simultaneously real.   
If $Z_{67}=0$ and $Z_1\neq Z_2$, then \eq{cond5} implies that $Z_6=Z_7=0$ in which case the $\mathbb{Z}_2$ symmetry is manifest in the Higgs basis and the previous considerations apply.  Finally, if $Z_{67}=0$, $Z_6\neq 0$ and $Z_1=Z_2$, then \eq{Ycond1} implies the existence of a real Higgs basis.   
Thus, in all possible cases,
if an exact $\mathbb{Z}_2$ symmetry is present in some scalar field basis, then a real Higgs basis exists and the scalar potential and vacuum in any scalar basis is CP conserving.

If the  $\mathbb{Z}_2$ symmetry is exact, then a real Higgs basis exists, and the Higgs basis parameters in \eq{cond3e} can be taken to be real. 
Employing \eq{minconds} then yields
\beq \label{real2}
\frac{2Y_2}{v^2}(Z_6+Z_7)+Z_1 Z_7+Z_2 Z_6=0\,.
\eeq
\Eqs{real1}{real2} are equivalent to eqs.~(18) and (19) of Ref.~\cite{Lavoura:1994yu}.   Note that \eq{real2} is trivially satisfied if $Z_{67}=0$ and $Z_1=Z_2$.   In this latter case, one must also impose \eq{lastcond} to guarantee the presence of an exact $\mathbb{Z}_2$ symmetry.  This last observation was missed in Ref.~\cite{Lavoura:1994yu}. 

\section{The C2HDM in the $\mathbb{Z}_2$ basis}
\label{sec:six}

The C2HDM is a two-Higgs-doublet model in which either the scalar potential or the vacuum is CP violating.    To avoid tree-level Higgs-mediated FCNCs, one imposes a $\mathbb{Z}_2$ symmetry on the dimension-four terms of the Higgs Lagrangian.  The symmetry is manifest in the $\Phi$ basis by setting $\lambda_6=\lambda_7=0$ in \eq{pot}.
The $\mathbb{Z}_2$ symmetry is assumed to be softly broken by taking $m_{12}^2\neq 0$.  If 
the CP violation in the scalar potential is explicit, then $\Im(\lambda_5^*[m_{12}^2]^2)\neq 0$.
Imposing the $\mathbb{Z}_2$ symmetry on \eq{yuklag}
implies that the Higgs-quark Yukawa couplings are either  of Type I or Type II as discussed in Section~\ref{sec:four}.

In \sect{xizero}, we noted that 
one is always free to rephase the Higgs-doublet fields such that the vevs are real. (The corresponding results prior to rephasing the vevs are given in Appendix~\ref{appE}.)  Henceforth,
we define the C2HDM in the $\mathbb{Z}_2$ basis such that $\xi=\arg(v_1^* v_2)=0$.  That is,
\beq \label{cbsb}
\sqrt{2} \langle \Phi^0_1 \rangle\equiv
v_1 = v\, c_\beta\,,\qquad\quad
\sqrt{2} \langle \Phi^0_2 \rangle
\equiv
v_2 = v\, s_\beta\,,
\eeq
in the notation of \eqs{potmin}{v246}, 
where $c_\beta \equiv \cos{\beta}$ and
$s_\beta \equiv \sin{\beta}$, with $0\leq\beta\leq\half\pi$.
In this convention, one may parametrize the scalar doublets in the $\Phi$ basis as
\begin{equation}
\Phi_1 =
\begin{pmatrix}
\varphi_1^+ \\
\frac{1}{\sqrt{2}} (v_1 + \eta_1 + i \chi_1)
\end{pmatrix}\,,
\qquad \quad
\Phi_2 =
\begin{pmatrix}
\varphi_2^+  \\
\frac{1}{\sqrt{2}} (v_2 + \eta_2 + i \chi_2)
\end{pmatrix}\, .
\label{Phi1Phi2}
\end{equation}
Setting $\lambda_6=\lambda_7=\xi=0$ in \eq{z2softmin} yields the C2HDM scalar potential minimum conditions,
\beqa
m_{11}^2&=&\Re m_{12}^2\tan\beta-\half v^2\bigl[ \lambda_1 c_\beta^2+(\lambda_3+\lambda_4+\Re~\!\lambda_5)s_\beta^2\bigr]\,, \\
m_{22}^2&=&\Re m_{12}^2\cot\beta-\half v^2\bigl[ \lambda_2 s_\beta^2+(\lambda_3+\lambda_4+\Re~\! \lambda_5)c_\beta^2\bigr]\,, \\
\Im m_{12}^2&=&\half v^2 s_\beta c_\beta\Im~\!\lambda_5\,.\label{m12min}
\eeqa
After eliminating $m_{12}^2$, $m_{22}^2$ and $\Im~\!m_{12}^2$, we are left with nine real parameters that govern the C2HDM: $v$, $\tan\beta$, $\Re m_{12}^2$, $\lambda_1$,
$\lambda_2$, $\lambda_3$, $\lambda_4$, $\Re~\!\lambda_5$ and $\Im~\!\lambda_5$.  By adopting the convention where both vevs are real and positive, it follows that if  $s_{2\beta}\neq 0$ and $\Im~\!\lambda_5\neq 0$  [which implies that $\Im~\!m_{12}^2\neq 0$ via \eq{m12min}], then CP is violated in the scalar sector.

If CP is violated in the scalar sector, then the violation is either explicit or spontaneous.  A scalar potential of the 2HDM is explicitly CP conserving if and only if 
a real basis exists~\cite{Gunion:2005ja} (i.e., a basis of scalar fields exists in which all the scalar potential parameters are real).
However, in transforming to a real basis, the vevs (which were real in the original basis by convention) may acquire a relative complex phase that is unremovable
by any further basis change that maintains the reality of the scalar field basis.
This latter scenario corresponds to the case of spontaneous CP violation.
Consequently, both spontaneous and explicit CP violation are treated simultaneously in the convention adopted in \eq{cbsb}. 

It is instructive to perform the counting of parameters using the invariants quantities discussed in previous sections.  After employing \eq{minconds}, one is left initially with
six real parameters, $v$, $Y_2$, $Z_1$, $Z_2$, $Z_3$ and $Z_4$, and three complex parameters, $Z_5$, $Z_6$ and $Z_7$, for a total of 12 parameters.   Since one can rephase the pseudoinvariant Higgs basis field~$H_2$, this freedom removes one phase from the three complex parameters.   Finally, since a softly broken~$\mathbb{Z}_2$ symmetry is present, one obtains one complex constraint equation (derived in Section~\ref{sec:five}) that removes two additional parameters.  This leaves nine independent real parameters in agreement with our previous counting.

If $s_{2\beta}=0$, then the model corresponds to the IDM which is CP conserving.  Consequently, in our considerations of the C2HDM we shall henceforth assume that $s_{2\beta}\neq 0$, which is a necessary ingredient for the presence of CP violation, as noted below \eq{m12min}.
Since $\lambda_6=\lambda_7=0$ (in the $\mathbb{Z}_2$ basis), it then follows from \eq{sum} that if $\lambda_1\neq\lambda_2$ then $\lambda_6+\lambda_7$ is nonzero when evaluated in any other scalar field basis.   In particular, $\lambda_1\neq\lambda_2$ implies that $Z_{67}\neq 0$.   In contrast, if $\lambda_1=\lambda_2$ in the $\mathbb{Z}_2$ basis, then it follows that 
$Z_1=Z_2$ and $Z_{67}=0$, which corresponds to the exceptional region of the parameter space (see Appendix~\ref{erps}).

In light of \eqthree{hbasisdef}{hbasisdef2}{hbasisfields}, one can identify the
massless would-be neutral Goldstone boson with
$G^0 = c_\beta \chi_1 + s_\beta \chi_2$.
Thus, the neutral scalar state orthogonal to $G^0$ is 
\beq
\eta_3 = - s_\beta \chi_1 + c_\beta \chi_2\,.
\eeq
After diagonalizing the squared-mass matrix of the neutral scalar fields,
$\eta_1$, $\eta_2$, and $\eta_3$, the
three neutral mass-eigenstate scalar fields, $h_1$, $h_2$, and $h_3$, can be identified as
\beq
\left(
\begin{array}{c}
h_1\\
h_2\\
h_3
\end{array}
\right)
= \mathcal{R}
\left(
\begin{array}{c}
\eta_1\\
\eta_2\\
\eta_3
\end{array}
\right).
\label{h_as_eta}
\eeq
In the C2HDM literature, the $3\times 3$ orthogonal mixing matrix $\mathcal{R}$ is parametrized as~\cite{ElKaffas:2007rq}
\beq
\mathcal{R} =
\left(
\begin{array}{ccc}
c_1 c_2 & \quad  s_1 c_2 &\quad  s_2\\
-c_1 s_2 s_3 - s_1 c_3 & \quad \phm c_1 c_3 - s_1 s_2 s_3  & \quad c_2 s_3\\
- c_1 s_2 c_3 + s_1 s_3 & \quad -c_1 s_3 - s_1 s_2 c_3 & \quad c_2 c_3
\end{array}
\right)\,,
\label{matrixR}
\eeq
where $s_i = \sin{\alpha_i}$ and
$c_i = \cos{\alpha_i}$ ($i = 1, 2, 3$).

It is now straightforward to relate the angles $\alpha_1$, $\alpha_2$, and $\alpha_3$ of the C2HDM literature to basis-independent quantities introduced in Section~\ref{sec:two}.  
In Appendix~\ref{appE}, we have examined the mixing of the neutral scalars in the $\mathbb{Z}_2$ basis.  Setting $\xi=0$ in \eqst{app:rk1}{app:rk3} yields
\beqa
\mathcal{R}_{k1}&=&q_{k1}c_\beta-\Re(q_{k2}e^{-i\theta_{23}})s_\beta\,,\label{rk1}\\ 
\mathcal{R}_{k2}&=&q_{k1}s_\beta+\Re(q_{k2}e^{-i\theta_{23}})c_\beta\,, \label{rk2}\\ 
\mathcal{R}_{k3}&=&\Im(q_{k2}e^{-i\theta_{23}})\,.\label{rk3}
\eeqa
One can relate the mixing angles $\alpha_1$, $\alpha_2$, and $\alpha_3$  to invariant (or pseudoinvariant) quantities by setting $\xi=0$ in \eqs{app:combo1}{app:combo2}.  It is convenient to define
$\overline\alpha_1\equiv\alpha_1-\beta$.
We then obtain the results exhibited in Table~\ref{alphaq}.
 \begin{table}[ht!]
\centering
\caption{The relation between the neutral Higgs mixing angles $\alpha_i$ of the C2HDM defined in the $\mathbb{Z}_2$ basis and
(pseudo)-invariant combinations of mixing angles defined in the Higgs basis.  In the notation used below,
$\overline{c}_1\equiv\cos\overline{\alpha}_1$ and $\overline{s}_1\equiv\sin\overline{\alpha}_1$, with $\overline\alpha_1\equiv\alpha_1-\beta$.  \\
\label{alphaq}}
\begin{tabular}{|c||c|c|}\hline
$\phaa k\phaa $ &\phaa $q_{k1}\phaa $ & \phaa $q_{k2}e^{-i\theta_{23}} \phaa $ \\ \hline
$1$ & $\overline{c}_1 c_2$ & $\overline{s}_1 c_2+is_2$ \\
$2$ & $-\overline{c}_1 s_2 s_3 -\overline{s}_1 c_3$ & $\overline{c}_1 c_3-\overline{s}_1 s_2 s_3+ic_2 s_3$ \\
$3$ & $-\overline{c}_1 s_2 c_3 +\overline{s}_1 s_3$ & $-\overline{c}_1 s_3-\overline{s}_1 s_2 c_3+ic_2 c_3$ \\
\hline
\end{tabular}
\end{table}

In the presence of a softly broken $\mathbb{Z}_2$ symmetry, \eq{xi23} 
implies that the quantity $e^{i(\xi+\theta_{23})}$ is determined 
up to a twofold ambiguity associated with a residual basis dependence corresponding to the interchange of the two scalar doublets while maintaining $\lambda_6=\lambda_7=0$. 
Having adopted the C2HDM convention where $\xi=0$, it therefore follows that $e^{i\theta_{23}}$ is determined up to a twofold ambiguity.  In particular, 
one no longer has the freedom to rephase the Higgs basis field $H_2$, which would result in an additive shift of the parameter $\theta_{23}$ [cf.~\eq{shift}].  In light of \eqst{U}{xi23}, it follows that under the basis transformation that simply interchanges $\Phi_1$ and $\Phi_2$ (with no rephasing), $s_\beta\leftrightarrow c_\beta$ and $e^{i\theta_{23}}\to -e^{i\theta_{23}}$.  Moreover,
\beq
s_1\to c_1\,,\,c_1\to s_1\,,\,s_2\to -s_2\,,\,c_2\to c_2\,,\,s_3\to -s_3\,,\,c_3\to -c_3\,,
\eeq
which yields $\mathcal{R}_{k1}\leftrightarrow \mathcal{R}_{k2}$ and $\mathcal{R}_{k3}\to -\mathcal{R}_{k3}$.
These results are consistent with \eqst{rk1}{rk3} since the $q_{k1}$ and $q_{k2}$ are basis-invariant
quantities.
 
Finally, we note that the free parameter $\Re m_{12}^2$ can also be related to basis-invariant quantities by employing \eq{m12} with $\xi=0$ and \eq{minconds}, and making use of the results of \sect{xizero}.  
If $\lambda_1\neq \lambda_2$ then $Z_{67}\neq 0$, in which case \eqs{plusamod}{plusamod2} yield
\beq \label{monetwo}
\Re m_{12}^2=\tfrac14 v^2 s_{2\beta}\left[Z_1+\frac{2Y_2}{v^2}-\left(\frac{|Z_6|^2+\Re(Z_6 Z_7^*)}{|Z_{67}|^2}\right)(Z_1-Z_2)\right]\,,
\eeq
where $s_{2\beta}$ is given by \eq{sincos}.  The case of $\lambda_1=\lambda_2$ in the $\mathbb{Z}_2$ basis corresponds to the exceptional region of parameter space, where $Z_1=Z_2$ and $Z_{67}=0$, as previously noted.  In this case, \eq{monetwo} does not apply and one must employ the results of Appendix~\ref{erps}.   The resulting expression for $\Re m_{12}^2$ is unwieldy and we do not present it here.

It is instructive to identify the nine real parameters of the C2HDM in terms of the scalar 
masses and mixing angles.   In order to perform the correct counting, we note the following sum rule, 
\beq \label{sumrule}
\sum_k m_k^2\,\mathcal{R}_{k3}(\mathcal{R}_{k1}c_\beta-\mathcal{R}_{k2}s_\beta)=0\,,
\eeq
which is derived at the end of Appendix~\ref{appE}.
This sum rule imposes one relation among the ten real quantities,
$v$, $\tan\beta$, $\alpha_1$, $\alpha_2$, $\alpha_3$, $m_1$, $m_2$, $m_3$, $\Re~\!m_{12}^2$ and $m_{H^\pm}$, 
resulting in nine independent parameters.
One can repeat the counting of parameters using basis-invariant quantities.  
In light of \eq{plusmass} and \eqst{zee1id}{zee6id}, one can eliminate $Z_1$, $Z_3$, $Z_4$, $Z_5 e^{-2i\theta_{23}}$ and $Z_6 e^{-i\theta_{23}}$ in terms of scalar masses and the invariant mixing angles $\theta_{12}$ and $\theta_{13}$.
This leaves three invariant parameters, $Z_2$, $\Re(Z_7 e^{-i\theta_{23}})$ and $\Im(Z_7 e^{-i\theta_{23}})$, of which two are determined from the one complex constraint equation arising from the condition of a softly broken $\mathbb{Z}_2$ symmetry.   For example, if we eliminate the complex parameter $Z_7$ using \eq{finalcond}, we are left with
the following nine real parameters: $v$, $Y_2$, $Z_2$, $\theta_{12}$, $\theta_{13}$, $m_1$, $m_2$, $m_3$, and $m_{H^\pm}$. 

The complete set of Feynman rules for the C2HDM in terms of the $\mathbb{Z}_2$-basis parameters can be found in Refs.~\cite{Fontes:2017zfn,WebPageC2HDM}.
One can check that all the Higgs couplings obtained this way (after using \eq{chhiggs} to define an invariant charged Higgs field) are invariant with respect
to basis transformations.   As previously noted, all the bosonic couplings of the most general 2HDM (without any imposed discrete symmetries) can be found in Ref.~\cite{Haber:2006ue} 
expressed directly in terms of invariant quantities $q_{k1}$, $q_{k2}$ and the Higgs basis scalar potential coefficients (including appropriate factors of $e^{-i\theta_{23}}$ to ensure basis-independent combinations).   The bosonic couplings of the most general 2HDM also apply to the C2HDM, since as emphasized
in Section~\ref{sec:five}, $\tan\beta$ does not appear explicitly in any of these couplings.   It is a straightforward to verify that the cubic and quartic Higgs self-couplings, which appear in Ref.~\cite{Haber:2006ue} match precisely the corresponding C2HDM couplings given in Ref.~\cite{WebPageC2HDM}.

Finally, the Type-Ia and Type-IIa Higgs-quark couplings are obtained from \eq{Yukawas} by employing \eqs{rhoType1a}{rhoType2a} with $\xi=0$ [in the convention of \eq{cbsb}].   For example,\footnote{As discussed in \sect{sec:four}, the Yukawa couplings for Type Ib and IIb can be obtained from \eqs{onea}{twoa}, respectively,  by replacing $\cot\beta\leftrightarrow \tan\beta$ and changing the sign of $e^{-i\theta_{23}}$. \label{ab}}
\beqa
-\mathscr{L}_{\rm Type-Ia}&=&\frac{1}{v}\biggl\{\overline{U} M_U\bigl[q_{k1}+\Re(q_{k2}e^{-i\theta_{23}})\cot\beta-i\gamma\ls{5}\Im(q_{k2}e^{-i\theta_{23}})\cot\beta\bigr]Uh_k \nonumber \\
&& \quad 
+\overline{D} M_D\bigl[q_{k1}+\Re(q_{k2}e^{-i\theta_{23}})\cot\beta+i\gamma\ls{5}\Im(q_{k2}e^{-i\theta_{23}})\cot\beta\bigr]Dh_k\biggr\},\label{onea} \\
-\mathscr{L}_{\rm Type-IIa}&=&\frac{1}{v}\biggl\{\overline{U} M_U\bigl[q_{k1}+\Re(q_{k2}e^{-i\theta_{23}})\cot\beta-i\gamma\ls{5}\Im(q_{k2}e^{-i\theta_{23}})\cot\beta\bigr]Uh_k \nonumber \\
&& \quad 
+\overline{D} M_D\bigl[q_{k1}-\Re(q_{k2}e^{-i\theta_{23}})\tan\beta-i\gamma\ls{5}\Im(q_{k2}e^{-i\theta_{23}})\tan\beta\bigr]Dh_k\biggr\},\label{twoa}
\eeqa
where there is an implicit sum over the three neutral Higgs mass-eigenstates $h_k$.  Using the results of Table~\ref{alphaq}, one can reproduce the results of Ref.~\cite{Fontes:2017zfn}.
Indeed, as previously noted, $\tan\beta$ and $e^{-i\theta_{23}}$ now appear explicitly in the Yukawa couplings.   However, these quantities are not quite physical parameters, since under the basis change $\Phi_1\leftrightarrow\Phi_2$, it follows that $\cot\beta\leftrightarrow \tan\beta$ and $e^{-i\theta_{23}}$ change sign.  This has the effect of interchanging the $a$ and $b$ versions of the Type-I and Type-II Yukawa couplings (cf.~footnote~\ref{ab}).

In order to promote $\tan\beta$ and $e^{i\theta_{23}}$ to physical parameters, one must remove the remaining freedom to interchange $\Phi_1\leftrightarrow\Phi_2$ in the C2HDM.
This corresponds to making a specific choice of the discrete symmetry among the four specified in Table~\ref{Tab:type}.
In practice, this can be achieved by declaring, e.g.,  that $\tan\beta<1$ corresponds to an \textit{enhanced} coupling of the neutral Higgs bosons to up-time quarks.   Given this additional proviso, it follows that the signs of $c_{2\beta}$ and $e^{i\theta_{23}}$ are then fixed and can now be considered as physical parameters of the model.  Indeed, $c_{2\beta}$ can be expressed in terms of basis-invariant parameters as specified in \eq{sincos}, where the sign ambiguity is fixed by the sign of $\lambda_1-\lambda_2$ [cf.~\eq{spcase}], under the assumption that $\lambda_1\neq \lambda_2$.
Likewise, $e^{i\theta_{23}}$ is uniquely determined by the formal basis-independent expression given by \eq{xizt23} [after employing \eq{sincos} for $s_{2\beta}/c_{2\beta}$ with the sign ambiguity fixed as indicated above].  Finally, the exceptional region of the parameter space where
$\lambda_1=\lambda_2$ in the $\mathbb{Z}_2$ basis is treated in Appendix~\ref{erps}.

\section{Detecting discrete symmetries}
\label{compare}

In Ref.~\cite{Lavoura:1994yu},  Lavoura described ways to detect
the presence of discrete symmetries exhibited by the scalar potential of the 2HDM.
Four cases of discrete symmetries were examined:
(i) exact $\mathbb{Z}_2$ symmetry;
(ii) explicit CP breaking by a complex soft $\mathbb{Z}_2$-breaking squared-mass term
(which defines the C2HDM);
(iii) softly broken $\mathbb{Z}_2$ and spontaneously broken CP symmetries~\cite{Branco:1985aq}; and
(iv) the Lee model of spontaneous CP violation~\cite{Lee:1973iz}, where no (unbroken or softly broken) $\mathbb{Z}_2$ symmetry is present.
For the reader's convenience, we
provide a translation between
Lavoura's notation and the notation of this paper:
\beqa \label{translate}
&&\lambda_1,\lambda_2,\lambda_5\longrightarrow\half Z_1,\half Z_2,\half Z_5\,,\qquad\quad
\lambda_3,\lambda_4,\lambda_6,\lambda_7\longrightarrow Z_3,Z_4,Z_6,Z_7\,,\nonumber \\
&&\mu_1,\mu_2,\mu_3\longrightarrow Y_1,Y_2,Y_3\,,\qquad\qquad\quad\,
v \longrightarrow v/\sqrt{2}\,.
\eeqa

In case~(i), Lavoura asserts that eqs.~(18) and (19) of Ref.~\cite{Lavoura:1994yu}
are the conditions for an exact  $\mathbb{Z}_2$-symmetric scalar potential.   We have confirmed
that these conditions are both necessary and sufficient in Section~\ref{exact}, as indicated below \eq{real2}.

In case~(ii), Lavoura asserts that eqs.~(20) and (21) of Ref.~\cite{Lavoura:1994yu} are
the conditions for explicit CP breaking by a complex soft $\mathbb{Z}_2$-breaking term.
We have confirmed that these results are a consequence of \eqs{cond5}{cond6} 
Indeed, \eq{cond6} is equivalent to eq.~(20) of Ref.~\cite{Lavoura:1994yu}.
In addition, by multiplying  
\eq{finalcond} by $Z_6-Z_7$ and then taking the imaginary part of the resulting expression, one reproduces eq.~(21) of Ref.~\cite{Lavoura:1994yu},
\beq \label{newcond}
(Z_1-Z_2)\Im\bigl[Z_5^*(Z_6^2-Z_7^2)\bigr]-\bigl[(Z_1-Z_2)(Z_1+Z_2-2Z_{34})+4(|Z_6|^2-|Z_7|^2)\bigr]\Im(Z_6 Z^*_7)=0\,.
\eeq

In case~(iii), Lavoura asserts that eqs.~(20)--(22) of Ref.~\cite{Lavoura:1994yu} are
the conditions for a softly broken $\mathbb{Z}_2$-symmetric scalar potential and spontaneously broken CP symmetry.
We have confirmed Lavoura's results in Section~\ref{scpv}, while noting a typographical error in eq.~(22) of Ref.~\cite{Lavoura:1994yu} (see footnote~\ref{fnlav}).
The corresponding corrected equation (with a different overall normalization) was given in \eq{fancy}.  Moreover, 
Lavoura's results are not applicable in cases of $Z_1=Z_2$ and/or $Z_{67}=0$.  The correct expressions that replace
\eq{fancy} in these special cases have been obtained in Section~\ref{scpv} and Appendix~\ref{erps}.   Note that if $Z_6\neq \pm Z_7$, then only two of the three equations among \eqthree{cond5}{cond6}{newcond} are independent.\footnote{Note that $\Im[(Z_6+Z_7)E]=0$ yields \eq{cond6}
and $\Im[(Z_6-Z_7)E]=0$ yields \eq{newcond}, where $E$ denotes the left-hand side of \eq{finalcond}.  It then follows that $\Re[(Z_6+Z_7)E]=0$, which is equivalent to \eq{cond5}.}

In case (iv), Lavoura attempts to discover the conditions on the 2HDM Higgs basis parameters that govern the 
Lee model of spontaneous CP violation~\cite{Lee:1973iz}.  In this model the $\mathbb{Z}_2$ symmetry is absent, i.e., there is no
basis of scalar fields in which $\lambda_6=\lambda_7=0$.  A scalar field basis exists in the Lee model in which all the scalar potential
parameters are simultaneously real, implying that the scalar potential is explicitly CP conserving.   However, 
there is an unremovable relative complex phase between the two
vevs $\vev{\Phi_1^0}$ and $\vev{\Phi_2^0}$.   Moreover, no real Higgs basis exists.  In terms of the
Higgs basis parameters, the nonexistence of a real Higgs basis implies that at least one of the following three quantities,
$\Im(Z_6^2 Z_5^*)$, $\Im(Z_7^2 Z_5^*)$ and $\Im(Z_6 Z_7^*)$ must be nonvanishing [cf.~\eq{cpconds}].
Hence, the vacuum is CP violating; that is, the Lee model exhibits spontaneous CP violation.

When considering the Lee model, Lavoura noted in Ref.~\cite{Lavoura:1994yu}
that there should be two relations among the parameters of the Lee model,
corresponding to the two independent CP-odd invariants.   Lavoura found one relation, that appears in eq.~(27) of Ref.~\cite{Lavoura:1994yu}.
But he was unable to identify the second invariant condition.   We now proceed to confirm Lavoura's invariant quantity and to complete his
mission by finding the second invariant quantity that was missed.  Moreover, we shall demonstrate that in certain regions of the parameter space of
the Lee model, Lavoura's invariant vanishes, in which case two additional invariant quantities must be introduced in order to cover all possible
special cases.

Consider the scalar potential of the general 2HDM given in \eq{pot}, with no constraints initially imposed on the scalar potential parameters.
To check for the presence of explicit CP violation in all possible regions of the 2HDM parameter space, it is
necessary and sufficient to consider four CP-odd basis-invariant quantities,
identified in Ref.~\cite{Gunion:2005ja}, as follows\footnote{Three CP-odd invariants that are equivalent to \eqst{y3zdef}{6zdef} were also identified in Ref.~\cite{Branco:2005em}.
Subsequently, a group-theoretic formulation of the 2HDM scalar potential was developed in Refs.~\cite{Ivanov:2005hg,Nishi:2006tg}  that provided an elegant
form for the basis-independent conditions governing explicit CP conservation in the  2HDM.  The bilinear formalism exploited in
the latter two references has also been employed in the study of the CP properties of the 2HDM scalar potential in Refs.~\cite{Maniatis:2007vn,Ferreira:2010hy,Ferreira:2010yh,Ivanov:2019kyh}.} 
\begin{eqnarray}
I_{Y3Z} & \equiv &
\textrm{Im}(Z^{(1)}_{a\bar{c}}Z^{(1)}_{e\bar{b}}Z_{b\bar{e} c\bar{d}}Y_{d\bar{a}})
\,, \label{y3zdef}\\
I_{2Y2Z} & \equiv &
\textrm{Im}(Y_{a\bar{b}}Y_{c\bar{d}}Z_{b\bar{a} d\bar{f}}Z^{(1)}_{f\bar{c}})
\,,\label{2y2zdef}\\
I_{6Z} &\equiv &
\textrm{Im}(Z_{a \bar{b} c \bar{d} } Z^{(1)}_{b \bar{f}} Z^{(1)}_{d \bar{h}}
Z_{f\bar{a} j \bar{k}} Z_{k \bar{j} m \bar{n}} Z_{n \bar{m} h \bar{c}})\,,
\label{6zdef}\\
I_{3Y3Z}&\equiv &
\textrm{Im}(Z_{a\bar cb\bar
  d}Z_{c\bar ed\bar g}Z_{e\bar h f\bar q}Y_{g\bar a}Y_{h\bar b}Y_{q\bar f}
)\,.
\label{3y3zdef}
\end{eqnarray}
If all four of these CP-odd invariants vanish, then there exists a real $\Phi$ basis, in which case the scalar potential is \textit{explicitly} CP conserving.
Aside from special regions in parameter space, at most two of these
invariants are independent, as we will demonstrate below.

Explicit forms for the above four CP-odd invariants can be found in Ref.~\cite{Gunion:2005ja}.  We proceed to evaluate them in the Higgs basis.
After employing \eq{minconds} it follows that,
\beqa
I_{Y3Z}&=& \half v^2\biggl\{2f_2 f_3+(Z_1-Z_2)\bigl[\Im(Z_5^* Z_6Z_{67})-(Z_1-Z_{34})f_3\bigr]\nonumber \\
&& \qquad\quad\,\,\, -\left(Z_1+\frac{2 Y_2}{v^2}\right)\bigl[\Im(Z_5^* Z_{67}^2)-(Z_1-Z_2)f_3\bigr]\biggr\}\,,\label{Y3Z} \\
I_{2Y2Z}&=& \tfrac14 v^4\Biggl\{(Z_1-Z_2)\Im(Z_5^* Z_6^2)-\left(Z_1+\frac{2 Y_2}{v^2}\right)\bigl[(Z_1-Z_{34})f_3+\Im(Z_5^* Z_6 Z_{67})\bigr]\,,\nonumber \\
&& \qquad\quad +\left[\left(Z_1+\frac{2 Y_2}{v^2}\right)^2-2|Z_6|^2+2\Re(Z_6 Z_7^*)\right]f_3\Biggr\}\,,\label{2Y2Z}
\eeqa
where the $f_i$ are defined in \eq{effs}.
One can check that $-I_{Y3Z}/v^2$ corresponds precisely to the left-hand side of eq.~(27) of Ref.~\cite{Lavoura:1994yu}.   Thus, $I_{2Y2Z}$ is the second invariant quantity that
governs the Lee model, which is the one that Lavoura was unable to find.  

Apart from special regions of the Lee model parameter space, $I_{Y3Z}=I_{2Y2Z}=0$ provide nontrivial relations among the parameters that must hold for a spontaneously CP-violating scalar sector.   However, there exist special regions of the Lee model parameter space where one or both of the invariants exhibited in \eqs{Y3Z}{2Y2Z} automatically vanish.   
One such example arises in the case of a softly broken $\mathbb{Z}_2$ symmetry, corresponding to $\lambda_6=\lambda_7=0$ in the $\Phi$ basis in which the Lee model is initially defined.  This case was studied in detail in Section~\ref{scpv}, where it was shown that $I_{Y3Z}$ automatically vanishes and thus provides no constraint.  Lavoura was well aware of this in Ref~\cite{Lavoura:1994yu}.  Indeed, he noted that $I_{Y3Z}$ is a linear combination of the left-hand sides of \eqs{cond6}{newcond}.  Since both of these quantities vanish if $\lambda_6=\lambda_7=0$ in some scalar field basis, it follows that $I_{Y3Z}=0$ is automatic in a model with a softly broken $\mathbb{Z}_2$ symmetry.   
One can check this explicitly as follows.  First,  if $Z_{67}=0$ then $f_3=0$, and \eq{Y3Z} immediately yields $I_{Y3Z}=0$.  Next, if $Z_{67}\neq 0$, then 
\beqa
\Im(Z_5^* Z_6 Z_{67})&=&\frac{\Im(Z_5^* Z_{67}^2 Z_{67}^* Z_6)}{|Z_{67}|^2}=\frac{\Im(Z_5^* Z_{67}^2)\bigl[|Z_6|^2+\Re(Z_6 Z_7^*)\bigr]+\Re(Z_5^* Z_{67}^2)\Im(Z_6 Z_7^*)}{|Z_{67}|^2}
\nonumber \\[5pt]
&=&\frac{(f_1-f_2)\Im(Z_5^* Z_{67}^2)+2f_3\Re(Z_5^* Z_{67}^2)}{2f_1}\,.\label{zid}
\eeqa
Employing \eqs{e1}{e2} in \eqs{Y3Z}{zid}, one can easily verify that $I_{Y3Z}=0$.  

In \eqs{imm12sq}{fancy}, an invariant condition was identified that guarantees that the scalar sector of the 2HDM with a softly broken $\mathbb{Z}_2$ exhibits spontaneous CP violation.   
We now demonstrate that this invariant condition is equivalent to the requirement that $f_3\neq 0$ and \mbox{$I_{2Y2Z}=0$.}  Assuming that $Z_{67}\neq 0$, we shall make use of the following formulae,
\beqa
\Im(Z_5^* Z_6^2)&=&\frac{\bigl[(f_1-f_2)^2-4f_3^2\bigr]\Im(Z_5^* Z_{67}^2)+4f_3(f_1-f_2)\Re(Z_5^* Z_{67}^2)}{4f_1^2}\,,\label{check1}\\
\Re(Z_6 Z_7^*)-|Z_6|^2&=&\frac{f_2(f_1-f_2)-4f_3^2}{2f_1}\,,\label{check2}
\eeqa
which are derived in the same manner as \eq{zid}.  One can now evaluate $I_{2Y2Z}$ given in \eq{2Y2Z} with the help of \eqst{zid}{check2}.  Imposing the conditions of a softly broken $\mathbb{Z}_2$ symmetry by employing \eqs{e1}{e2}, the end result of this computation is
\beq \label{bigrel}
I_{2Y2Z}=\frac{v^4 f_3\mathcal{F}}{16 f_1^2(Z_1-Z_2)}\,,
\eeq
where $\mathcal{F}$ is given explicitly in \eq{fancy}.
This result confirms that $f_3\neq 0$  and $I_{2Y2Z}=0$ are the invariant conditions for spontaneous CP violation in the softly broken $\mathbb{Z}_2$-symmetric 2HDM.  
As discussed in Section~\ref{scpv}, \eq{bigrel} can be used in the case of $Z_1=Z_2$ by employing \eq{e1} to eliminate $f_2$ in favor of $\Re(Z_5^* Z_{67}^2)$.  This procedure will  remove the
potential singularity due to the factor of $Z_1-Z_2$ in the denominator of \eq{bigrel}.

Because $\lambda_6=\lambda_7=0$ in the $\Phi$ basis, the only potentially nontrivial phase is the relative phase between $m_{12}^2$ and $\lambda_5$.  Thus, only one invariant condition is needed to determine whether or not the model exhibits spontaneous CP violation.   In the special case of $Z_{67}=0$ and $Z_1\neq Z_2$, the conditions for a softly broken $\mathbb{Z}_2$ symmetry given in \eqs{plusa}{minusa} yield $Y_3=Z_6=Z_7=0$ [after using \eq{minconds}], corresponding to the (CP-conserving) IDM treated in \sect{idm}.  
In the exceptional region of parameter space defined by $Z_{67}=0$ and $Z_1=Z_2$, it follows that $I_{2Y2Z}=0$, and one must discover another invariant condition to determine whether the model exhibits spontaneous CP violation.

In order to exhibit cases where Eqs.~(\ref{Y3Z}) and (\ref{2Y2Z}) are not sufficient to determine whether or not the scalar potential is explicitly CP conserving, we shall make use of the observation of
Ref.~\cite{Gunion:2005ja} that it is always possible to perform a basis transformation such that in the transformed basis of 
scalar fields, $\lambda_7 = -\lambda_6$ (a simple proof of this result is 
presented 
in Appendix~\ref{appD}).   Since basis-invariant quantities can be evaluated in any basis without changing their values, 
we shall
evaluate the four CP-odd invariants listed in \eqst{y3zdef}{3y3zdef} in a basis where $\lambda_7=-\lambda_6$, where these invariants take on the following 
simpler forms:
\beqa
I_{Y3Z} &=&
(\lambda_1-\lambda_2)^2\,\textrm{Im}(m^2_{12} \lambda_6^*)\,,
\label{whythreezee}
\\*[1mm]
I_{2Y2Z} &=&
(\lambda_1 - \lambda_2) \left[
   \textrm{Im}(\lambda_5^*[m_{12}^2]^2) -
   (m^2_{11} - m^2_{22}) \textrm{Im}(m^2_{12} \lambda_6^*)
\right]\,,
\label{twowhytwozee}
\\*[1mm]
I_{6Z} &=&
- (\lambda_1 - \lambda_2)^3\, \textrm{Im}(\lambda_6^2 \lambda_5^*)\,,
\label{sixzee}
\\*[1mm]
I_{3Y3Z} &=&
4\,\textrm{Im}([m^2_{12}]^3(\lambda_6^*)^3)-
2\, \textrm{Im}([m^2_{12}]^3\lambda_6(\lambda_5^*)^2) 
\nonumber\\*[1mm]
&&
+[(m^2_{11} - m^2_{22})^2-6|m^2_{12}|^2](m^2_{11} - m^2_{22}) \textrm{Im}(\lambda_5^*\lambda_6^2)
\nonumber\\*[1mm]
&& 
+ \left[(\lambda_1-\lambda_{34})(\lambda_2-\lambda_{34})
+ 2|\lambda_6|^2-|\lambda_5|^2\right](m^2_{11} - m^2_{22}) \textrm{Im}(\lambda_5^*[m_{12}^2]^2)
\nonumber\\*[1mm]
&&
- \Bigl\{ (\lambda_1-\lambda_2)^2 m^2_{11}m^2_{22} +
2(2|\lambda_6|^2-|\lambda_5|^2)\left[(m^2_{11} -m^2_{22})^2-|m^2_{12}|^2\right]
\Bigr\} \textrm{Im}(m^2_{12}\lambda_6^*)
\nonumber\\*[1mm]
&&
-(\lambda_1+\lambda_2-2\lambda_{34})\Bigl\{(m^2_{11} - m^2_{22})
\textrm{Im}([m^2_{12}]^2(\lambda_6^*)^2)+ \textrm{Im}([m^2_{12}]^3 \lambda_5^* \lambda_6^*)
\nonumber\\*[1mm]
&&
\hspace{19ex}
- \left[(m^2_{11} - m^2_{22})^2-|m^2_{12}|^2\right]
\textrm{Im}(m^2_{12}\lambda_6 \lambda_5^*) \Bigr\}\,,\label{3y3zl6eq-l7}
\eeqa
where $\lambda_{34}\equiv\lambda_3 + \lambda_4$.  If $I_{Y3Z}=0$, then additional CP-odd invariants may need 
to be considered.

In a $\Phi$ basis of scalar fields where $\lambda_6 = - \lambda_7$, the invariant
$I_{Y3Z}=0$ if any one of the following four conditions hold:
(i) $\lambda_6=0$,
(ii) $\lambda_1 = \lambda_2$,
(iii) $m^2_{12}=0$,
or (iv) $\textrm{Im} (m_{12}^2 \lambda_6^*)=0$.
We now examine each of these four cases in turn.  Subsequently, we shall examine two additional special cases of interest in which
$I_{Y3Z}$ does not vanish.

\vskip 0.1in

\noindent
\underline{\bf Case 1:} \,$\lambda_6=0$ and $\lambda_1\neq \lambda_2$.

This case corresponds to a scalar potential with a softly broken $\mathbb{Z}_2$ symmetry, since $\lambda_6=\lambda_7=0$ in the $\Phi$ basis.
\Eqst{whythreezee}{3y3zl6eq-l7} yield $I_{Y3Z}=I_{6Z}=0$ and
\beqa
I_{2Y2Z}&=&(\lambda_1 - \lambda_2) 
   \textrm{Im}(\lambda_5^*[m_{12}^2]^2)\,,\label{2Y2ZZ2} \\[8pt]
I_{3Y3Z}&=&\left(\frac{\bigl[(\lambda_1-\lambda_{34})(\lambda_2-\lambda_{34})-|\lambda_5|^2\bigr](m_{11}^2-m_{22}^2)}{\lambda_1-\lambda_2}\right) I_{2Y2Z}\,.
\label{3Y3ZZ2}
\eeqa   
The above results imply that in this case only one invariant quantity, $I_{2Y2Z}$, is needed to determine whether the scalar potential is explicitly CP conserving.   Indeed, \eq{2Y2ZZ2} 
immediately shows that \eq{imm12sq} is proportional to $I_{2Y2Z}$, a result that was obtained above by a rather tedious computation that yielded \eq{bigrel}.
Moreover, \eq{2Y2ZZ2} provides a very simple method for computing $I_{2Y2Z}$ in terms of Higgs basis parameters.  
Using \eqs{lam1def}{lam2def}, it follows that 
\beq \label{onetwo}
\lambda_1-\lambda_2=(Z_1-Z_2)c_{2\beta}-2s_{2\beta}\Re(Z_{67}e^{i\xi})=\mp\sqrt{(Z_1-Z_2)^2+4|Z_{67}|^2}\,,
\eeq
after using \eq{sincos} and noting that $\Re(Z_{67}e^{i\xi})=\pm |Z_{67}|$ [cf.~\eq{signchoice}].
Hence, by using \eqthree{imm12sq}{fancy}{onetwo} in \eq{2Y2ZZ2}, one  
immediately reproduces the result of \eq{bigrel}.
\vskip 0.1in

\noindent
\underline{\bf Case 2:} \,$\lambda_1=\lambda_2$.

In light of eqs.~(\ref{Lam1def}), (\ref{Lam2def}), (\ref{Lam6def}) and (\ref{Lam7def}), it follows that if $\lambda_1=\lambda_2$ and $\lambda_6=-\lambda_7$,
then these relations hold in \textit{any} basis of scalar fields.
Hence, it follows that $Z_1=Z_2$ and
$Z_6=-Z_7$.  This is the exceptional region of the 2HDM parameter space, which is treated in more detail in Appendix~\ref{erps}.  In this case, \eqst{whythreezee}{3y3zl6eq-l7} yield $I_{Y3Z}=I_{2Y2Z}=I_{6Z}=0$ and
\beqa
I_{3Y3Z}&=& -\tfrac18 v^6\Im(Z_5^* Z_6^2)\biggl\{\left(Z_1+\frac{2Y_2}{v^2}\right)^3+2(Z_1-Z_{34})\left(Z_1+\frac{2Y_2}{v^2}\right)^2 
\\*[1mm]
&& 
-\bigl[4|Z_6|^2+|Z_5|^2-(Z_1-Z_{34})^2\bigr]\left(Z_1+\frac{2Y_2}{v^2}\right)-4\bigl[(Z_1-Z_{34})|Z_6|^2+\Re(Z_5^* Z_6^2)\bigr]\Biggr\}\,,\nonumber
\eeqa
after evaluating $I_{3Y3Z}$ in the Higgs basis and employing \eq{minconds}.
If $\Im(Z_5^* Z_6)=0$, then a real Higgs basis exists and both the scalar potential and vacuum are CP conserving.  
If $\Im(Z_5^* Z_6)\neq 0$ and $I_{3Y3Z}=0$, then
the model exhibits spontaneous CP violation.   This result provides the previously missing invariant condition for spontaneous CP violation in the exceptional region of the 2HDM
parameter space.

\vskip 0.1in
\noindent
\underline{\bf Case 3:} \,$m_{12}^2=0$, 
and $\lambda_1\neq \lambda_2$.

 In this case, \eqst{whythreezee}{3y3zl6eq-l7} yield $I_{Y3Z}=I_{2Y2Z}=0$ and
\beqa
I_{6Z} &=&
- (\lambda_1 - \lambda_2)^3\, \textrm{Im}( \lambda_5^*\lambda_6^2)\,,\\[8pt]
I_{3Y3Z} &=& -\left(\frac{m_{11}^2-m_{22}^2}{\lambda_1-\lambda_2}\right)\lasup{\!3} I_{6Z}\,.
\eeqa
The above results imply that in this case only one invariant quantity, $I_{6Z}$, is needed to determine whether the scalar potential is explicitly CP conserving.   
For completeness, we provide an expression for $I_{6Z}$ when evaluated in the Higgs basis~\cite{Gunion:2005ja}, 
\beqa
\!\!\!\!\!\!\!\!
I_{6Z}&=& -4f_2^2 f_3+|Z_5|^2\biggl\{2f_3\bigr[f_1-(Z_1-Z_2)^2\bigr]+(Z_1-Z_2)\Im(Z_5^* Z_{67}^2)\biggr\} \nonumber \\
&& +2f_2\biggl\{(Z_1-Z_{34})\bigl[f_3(Z_1-Z_2)+\Im(Z_5^* Z_{67}^2)\bigr]-(Z_1-Z_2)\Im(Z_5^* Z_6 Z_{67})\biggr\}\nonumber \\
&& -2f_3(Z_1-Z_2)\biggl\{\Re(Z_5^* Z_{67}^2)-2\Re(Z_5^* Z_6 Z_{67})+(Z_1-Z_2)\bigl[\Re(Z_6 Z_7^*)-|Z_6|^2\bigr]\biggr\}\nonumber \\[3pt]
&& +2\,\!\Im(Z_5^* Z_{67}^2)\Re(Z_5^* Z^2_{67})-2\,\Im(Z_5^* Z^2_{67})\Re(Z_5^* Z_6 Z_{67})-2\,\Re(Z_5^* Z_{67}^2)\Im(Z_5^* Z_6 Z_{67})\nonumber \\[3pt]
&& -(Z_1-Z_2)(Z_1-Z_{34})^2\,\Im(Z_5^* Z^2_{67})+(Z_1-Z_2)^2(Z_1+Z_2-2Z_{34})\Im(Z_5^* Z_6 Z_{67}) \nonumber \\[3pt]
&& +2(Z_1-Z_2)\bigl[\Re(Z_6 Z_7^*)\Im(Z_5^* Z_{67}^2)-f_1\Im(Z_5^* Z_6 Z_7)\bigr]+(Z_1-Z_2)^3\Im(Z_5^* Z_6 Z_7)\,,
\eeqa
where the $f_i$ are defined in \eq{effs}.

\vskip 0.1in
\noindent
\underline{\bf Case 4:} \,$\Im(m_{12}^2\lambda_6^*)=0$, $m_{12}^2\neq 0$ and $\lambda_1\neq \lambda_2$.

 In this case, \eqst{whythreezee}{3y3zl6eq-l7} yield $I_{Y3Z}=0$ and
 \beqa
I_{2Y2Z} &=&
(\lambda_1 - \lambda_2) \textrm{Im}(\lambda_5^*[m_{12}^2]^2)\,,
\\[8pt]
I_{6Z} &=&
- \left(\frac{(\lambda_1-\lambda_2)^2\Re\bigl[(m_{12}^2\lambda_6^*)^2\bigr]}{|m_{12}^2|^4}\right)I_{2Y2Z}\,.
\eeqa
As in the case of $I_{6Z}$, one sees that $I_{3Y3Z}$ is also proportional to $ \textrm{Im}(\lambda_5^*[m_{12}^2]^2)$.   
Both results can be understood geometrically by noting that
the condition $\textrm{Im}(m^2_{12} \lambda_6^*)=0$ implies that
$m^2_{12}$ and $ \lambda_6$ are aligned in the complex plane,
whereas $\textrm{Im}(\lambda_5^*[m_{12}^2]^2)=0$ implies that
$[m_{12}^2]^2$ and $ \lambda_5$ are aligned in the complex plane.
Hence, if $I_{2Y2Z}=0$ then $[m_{12}^2]^2 \lambda_6$ and~$\lambda_6^2$ are aligned
with $\lambda_5$, and it follows that $I_{6Z}=0$ and $I_{3Y3Z}=0$.
Once again, only one invariant quantity, $I_{2Y2Z}$, is needed to determine whether the scalar potential is explicitly CP conserving.

To be complete, we 
examine two further cases in which $I_{Y3Z}\neq 0$, where only one CP-odd invariant is needed
to determine whether the scalar potential is explicitly CP conserving.

\vskip 0.1in
\noindent
\underline{\bf Case 5:}  \,$\textrm{Im}(\lambda_5^*[m_{12}^2]^2) =
   (m^2_{11} - m^2_{22}) \textrm{Im}(m^2_{12} \lambda_6^*)$, $m_{12}^2\neq 0$ and
 $\lambda_1\neq \lambda_2$.
 
In this case, \eqst{whythreezee}{3y3zl6eq-l7} yield $I_{2Y2Z}=0$ and
\beqa
\hspace{-0.3in}
 I_{Y3Z} &=&
(\lambda_1-\lambda_2)^2\,\textrm{Im}(m^2_{12} \lambda_6^*)\,,
\\[8pt]
I_{6Z} &=&
\left(\frac{(\lambda_1 - \lambda_2)\bigl[2\Re(m_{12}^2\lambda_6^*)\Re(\lambda_5^*[m_{12}^2]^2)-(m_{11}^2-m_{22}^2)\Re[(m_{12}^2\lambda_6^*)^2]\bigr]}{|m_{12}^2|^4}\right)I_{Y3Z}\,.
\eeqa 
One can show that $I_{3Y3Z}$ is also proportional to $\Im(m_{12}^2\lambda_6^*)$.   Hence, if $I_{Y3Z}=0$, then it follows that $I_{6Z}=I_{3Y3Z}=0$.   That is, only one invariant quantity, $I_{Y3Z}$, is needed to determine whether the scalar potential is explicitly CP conserving.

\vskip 0.1in
\noindent
\underline{\bf Case 6:}  $\lambda_5=0$ and $\lambda_1\neq \lambda_2$.

In this case, \eqst{whythreezee}{3y3zl6eq-l7} yield $I_{6Z}=0$ and
\beqa
 I_{Y3Z} &=&
(\lambda_1-\lambda_2)^2\,\textrm{Im}(m^2_{12} \lambda_6^*)\,,
\\[8pt]
I_{2Y2Z} &=&-\left(\frac{m_{11}^2-m_{22}^2}{\lambda_1-\lambda_2}\right)I_{Y3Z}\,.
\eeqa
As in the previous case, one can show that $I_{3Y3Z}$ is also proportional to $\Im(m_{12}^2\lambda_6^*)$. Hence, if $I_{Y3Z}=0$, then it follows that $I_{2Y2Z}=I_{3Y3Z}=0$.   That is, only one invariant quantity, $I_{Y3Z}$, is needed to determine whether the scalar potential is explicitly CP conserving.

In summary, in generic regions of the 2HDM parameter space, it is sufficient to examine two CP-odd invariant quantities, $I_{Y3Z}$ and $I_{2Y2Z}$ given in 
\eqs{Y3Z}{2Y2Z} in order to determine whether or not the scalar potential explicitly breaks the CP symmetry.    In special regions of parameter space examined in the six cases above,
one CP-odd invariant quantity is sufficient, although in some cases a third CP-odd invariant, $I_{6Z}$, or a fourth CP-odd invariant, $I_{3Y3Z}$, is needed
to determine the CP property of the scalar potential.  In the Lee model of spontaneous CP violation, all four CP-odd invariants vanish, and the scalar potential is explicitly CP conserving, but at least one of the invariants, $\Im(Z_6^2 Z_5^*)$, $\Im(Z_7^2 Z_5^*)$ 
and $\Im(Z_6 Z_7^*)$ is nonvanishing, signaling that in the absence of explicit CP violation, the source of the CP violation must be attributed to the properties of the vacuum.

\section{Conclusions}
\label{conclusions}

The C2HDM is the most general two-Higgs-doublet model that possesses a softly broken $\mathbb{Z}_2$ symmetry (the latter
is imposed to eliminate tree-level Higgs-mediated FCNCs).  \mbox{In the} so-called $\mathbb{Z}_2$ basis where the $\mathbb{Z}_2$ symmetry of the quartic terms in the
scalar potential is mani\-festly realized, one can rephase the scalar fields such that the vevs $v_1$ and $v_2$ \mbox{are real and}
non-negative.   After minimizing the scalar potential and fixing $v=(v_1^2+v_2^2)^{1/2}=246$~GeV, the C2HDM is governed by nine additional real parameters: four scalar masses, one additional squared-mass parameter, $\Re m_{12}^2$, $\tan\beta=v_2/v_1$, and three mixing angles arising from the diagonalization of the neutral scalar squared-mass matrix.  One sum rule [cf.~\eq{sumrule}] reduces the total number of independent degrees of freedom (including $v$) to nine.    

In this paper, we have provided a basis-invariant treatment of the C2HDM.  This involves a number of steps.  First, we transformed to the
Higgs basis, which is defined up to an arbitrary rephasing of the Higgs basis field $H_2$ (which by definition possesses no vacuum expectation value).
Consequently, the real parameters of the Higgs basis scalar potential are invariant quantities, whereas the complex parameters are
pseudoinvariant quantities that are rephased under $H_2\to e^{i\chi}H_2$.  This allows us to easily identify basis-independent quantities, which are related
to physical observables of the model.  The softly broken~$\mathbb{Z}_2$ symmetry constrains the Higgs basis parameters and yields
one complex invariant constraint equation.   Our results are consistent with the more formal results of Ref.~\cite{Davidson:2005cw}, and a recent computation of Ref.~\cite{Belusca-Maito:2017iob}
that was carried out in a convention of real vevs in the $\mathbb{Z}_2$ basis.  For completeness, we have also provided the corresponding constraints if the
$\mathbb{Z}_2$ symmetry is extended to incorporate the dimension-two squared-mass terms of the scalar potential.

Having obtained the constraints due to the presence of a softly broken $\mathbb{Z}_2$ symmetry, one can 
check that the C2HM is governed by nine basis-independent parameters in agreement with our previous counting above.   Moreover, one can
now identify the behavior of the parameters of the C2HDM under basis transformations.  Our analysis revealed that some combinations of the mixing angles $\alpha_1$, $\alpha_2$ and $\alpha_3$ and the parameter $\tan\beta$ possess a residual
basis dependence due to the freedom to interchange the two complex scalar doublet fields of the C2HDM.   In practice, this residual basis dependence is removed
by declaring that \mbox{$\tan\beta<1$} corresponds to an enhanced coupling of some of the neutral Higgs bosons to up-type quarks.   Having adopted this convention (which is implicitly assumed in the literature but never stated explicitly), the angle parameters of the C2HDM and the parameter $\tan\beta$ are promoted to basis-independent quantities that can be directly related to physical observables.

Our work also resolves an apparent conflict between the number of physical phases
in the matrices that diagonalize the squared-mass matrix of the neutral Higgs fields that arise in the two approaches.
Indeed,
the basis-invariant calculation exhibited in \sect{sec:three}
involves two basis-invariant angles ($\theta_{12}$ and $\theta_{13}$),
and one unphysical angle ($\theta_{23}$),
whereas the calculations in the C2HDM resulting in 
eq.~\eqref{matrixR} yields three physical angles $\alpha_{1,2,3}$.
The resolution of this conundrum is associated with the observation that the C2HDM is initially defined in a $\mathbb{Z}_2$ basis where 
both vevs are real. The constraint imposed by the reality of the two vevs ultimately allows one to 
ascribe physical significance to the pseudoinvariant quantity,~$\theta_{23}$.
This can be seen directly in \eq{z67p} which relates the relative phase of the two vevs to the phase of the pseudoinvariant quantity $Z_{67}$.
Thus, by fixing the phase of the two vevs to be zero, one fixes the quantity $Z_{67}$ to be real.   This leaves a sign ambiguity that is resolved
once a twofold ambiguity in the definition of $\tan\beta$ is fixed as indicated above. 
 We have also examined special cases in which $Z_{67}=0$,
where the phase of $Z_6$ is similarly fixed in the convention of real vevs.\footnote{If $Z_6=Z_7=0$, then the model reduces to the IDM
discussed in \sect{idm}.  This model is necessarily CP conserving and thus is not of further interest to us in this work.}
The so-called exceptional region of the 2HDM parameter space where $Z_1=Z_2$ and $Z_{67}=0$ requires special attention and is treated in Appendix~\ref{erps}.

Finally, we have reanalyzed  the techniques for detecting the presence of discrete symmetries originally presented by Lavoura in  \Ref{Lavoura:1994yu}. 
We have obtained results that are in agreement with the corresponding results in Lavoura's paper (after correcting one typographical error in  \Ref{Lavoura:1994yu}). 
In addition, we have extended Lavoura's results in two directions.  First, we noted that the invariant
constraints obtained by Lavoura do not apply in all parameter regimes of the C2HDM.  Some special cases require additional analysis, and we have provided the appropriate
modifications in cases that cannot be obtained directly from considerations of the generic regions of the parameter space. 
Second, Lavoura was only able to obtain one of two relations that must be satisfied in the 2HDM with an explicitly CP-conserving scalar potential but with no (unbroken or broken) $\mathbb{Z}_2$ symmetry, that exhibits spontaneous CP violation (i.e., the Lee model~\cite{Lee:1973iz}).  
We have provided the second relation that was missed by Lavoura (using the results obtained in Ref.~\cite{Gunion:2005ja}), and we have clarified a number of special cases in which only one relation is sufficient (although that relation is typically not the one found by Lavoura).   It is also instructive to apply this analysis in the presence of a softly broken $\mathbb{Z}_2$ symmetry.  In doing so, we noted a surprising aspect of a subset of the exceptional region of the parameter space where no $\mathbb{Z}_2$ basis exists where all the scalar potential parameters are real, and yet the corresponding 2HDM is CP conserving.

In conclusion, the basis-independent formalism possesses many advantages.  For example, just like covariance in relativistic theories where an equation can be checked by ensuring that both sides of the equation behave similarly under Lorentz transformations in the same way, the basis-independent formalism affords similar benefits.  Indeed, errors in numerous equations in this paper were avoided by such considerations.   In addition, due to the close connection of basis-independent quantities to physical observables, one obtains confidence in appreciating the significance of the relations among the various 2HDM parametrizations.   We hope that the application of basis-independent methods in the analysis of the C2HDM presented in this paper has contributed to a better understanding of this model and will be useful in future phenomenological studies of CP-violating Higgs phenomena.

\acknowledgments

The work of R.B., T.V.F., J.C.R. and J.P.S. is supported in part
by the Portuguese \textit{Funda\c{c}\~{a}o para a Ci\^{e}ncia e Tecnologia}
(FCT) under Contracts No.~CERN/FIS-NUC/0010/2015, No.~PTDC/FIS-PAR/29436/2017 and No.~UID/FIS/00777/2013.
H.E.H. is supported in part by the U.S. Department of Energy Grant
No.~\uppercase{DE-SC}0010107, and in part by the Grant H2020-MSCA-RISE-2014
No.~645722 (NonMinimalHiggs).  H.E.H. is grateful for the hospitality and support of the Instituto Superior T\'{e}cnico, Universidade de Lisboa,
during his visit where this work was conceived, and he
also acknowledges the inspiring working
atmosphere of the Aspen Center for Physics, supported by the
National Science Foundation Grant No.\ PHY-1066293, where some
of the research reported in this work was carried out.

\clearpage

\appendix

\section{Changing the basis of scalar fields in the 2HDM}
\label{appA}

Since the scalar doublets $\Phi_1$ and $\Phi_2$
have identical SU(2)$\times$U(1) quantum numbers, one
is free to define two orthonormal linear combinations of the original
scalar fields.  The parameters appearing in \eq{pot} depend on a
particular \textit{basis choice} of the two scalar fields.
Relative to an initial (generic) basis choice, the scalar
fields in the new basis are given by
$\Phi^\prime=U\Phi$,
where $U$ is a U(2) matrix:
\beq \label{u2}
U=\begin{pmatrix} \cos\beta & e^{-i\xi}\sin\beta \\
-e^{i(\xi+\eta)}\sin\beta & e^{i\eta}\cos\beta \end{pmatrix}\,,
\eeq
up to an overall complex phase factor
$e^{i\psi}$ that has no effect on the scalar potential parameters, since
this corresponds to a global hypercharge transformation.

With respect to the new $\Phi'$ basis, the scalar potential takes on the same
form given in \eq{pot} but with new coefficients
$m_{ij}^{\prime\, 2}$ and $\lambda^\prime_j$.  
 For the general U(2)
transformation of \eq{u2} with $\Phi^\prime=U\Phi$, the scalar
potential parameters ($m_{ij}^{\prime\,2}$, $\lambda_i^\prime$) are related
to the original parameters ($m_{ij}^2$, $\lambda_i$)  by
\beqa
\Maa&=&m_{11}^2\ct^2+m_{22}^2\st^2-\Re(m_{12}^2 e^{i\xi})\stwob\,,
\label{maa}\\
\Mbb &=&m_{11}^2\st^2+m_{22}^2\ct^2+\Re(m_{12}^2 e^{i\xi})\stwob\,,
\label{mbb}\\
\Mab e^{i(\xi+\eta)}&=&
\half(m_{11}^2-m_{22}^2)\stwob+\Re(m_{12}^2 e^{i\xi})\ctwot
+i\,\Im(m_{12}^2 e^{i\xi})\,.\label{mab} \\
\!\!\!\!\!\!\!\!\!\!\!\!\!\!\Lama&=&
\lam_1\ct^4+\lam_2\st^4+\half\lamtil\stwob^2
+2\stwob\left[\ct^2\Re(\lam_6 e^{i\xi})+\st^2\Re(\lam_7e^{i\xi})\right]\,,
\label{Lam1def}  \\
\!\!\!\!\!\!\!\!\!\!\!\!\!\!\Lamb &=&
\lam_1\st^4+\lam_2\ct^4+\half\lamtil\stwob^2
-2\stwob\left[\st^2\Re(\lam_6 e^{i\xi})+\ct^2\Re(\lam_7e^{i\xi})\right]\,,
\label{Lam2def}      \\
\!\!\!\!\!\!\!\!\!\!\!\!\!\!\Lamc &=&
\quarter\stwob^2\left[\lam_1+\lam_2-2\lamtil\right]
+\lam_3-\stwob\ctwot\Re[(\lam_6-\lam_7)e^{i\xi}]\,,
\label{Lam3def}      \\
\!\!\!\!\!\!\!\!\!\!\!\!\!\!\Lamd &=&
\quarter\stwob^2\left[\lam_1+\lam_2-2\lamtil\right]
+\lam_4-\stwob\ctwot\Re[(\lam_6-\lam_7)e^{i\xi}]\,,
\label{Lam4def}      \\
\!\!\!\!\!\!\!\!\!\!\!\!\!\!\Lame  e^{2i(\xi+\eta)} &=&
\quarter\stwob^2\left[\lam_1+\lam_2-2\lamtil\right]+\Re(\lam_5 e^{2i\xi})
+i\ctwot\Im(\lam_5 e^{2i\xi})-\stwob\ctwot\Re[(\lam_6-\lam_7)e^{i\xi}]
\nonumber \\
&&\qquad\qquad\qquad\qquad\qquad\qquad\,\,\,\,
-i\stwob\Im[(\lam_6-\lam_7)e^{i\xi}]\,,
\label{Lam5def}      \\
\!\!\!\!\!\!\!\!\!\!\!\!\!\!\Lamf e^{i(\xi+\eta)}&=&
-\half\stwob\left[\lam_1\ct^2
-\lam_2\st^2-\lamtil\ctwot-i\Im(\lam_5 e^{2i\xi})\right]
+\ct\cthreet\Re(\lam_6 e^{i\xi})+\st\sthreet\Re(\lam_7 e^{i\xi})\nonumber \\
&&\qquad\qquad\qquad\qquad\qquad\qquad\,\,\,\,
+i\ct^2\Im(\lam_6 e^{i\xi})+i\st^2\Im(\lam_7 e^{i\xi})\,,
\label{Lam6def}      \\
\!\!\!\!\!\!\!\!\!\!\!\!\!\!\Lamg e^{i(\xi+\eta)} &=&
-\half s_{2\beta}\left[\lam_1\st^2-\lam_2\ct^2+
\lamtil c_{2\beta}+i\Im(\lam_5 e^{2i\xi})\right]
+\st\sthreet\Re(\lam_6 e^{i\xi})+\ct\cthreet\Re(\lam_7e^{i\xi})\nonumber \\
&&\qquad\qquad\qquad\qquad\qquad\qquad\,\,\,\,
+i\st^2\Im(\lam_6 e^{i\xi})+i\ct^2\Im(\lam_7e^{i\xi})\,,
\label{Lam7def}
\eeqa
where $s_\beta\equiv\sin\beta$, $c_\beta\equiv\cos\beta$, etc., and
\beq \label{lamtildef}
\lamtil\equiv\lam_3+\lam_4+\Re(\lam_5 e^{2i\xi})\,.
\eeq

We shall make use of \eqst{maa}{Lam7def} to write out the explicit relations between the scalar potential parameters of a generic basis and the Higgs basis.  We can employ the unitary matrix given by \eq{u2}, where
\beq \label{app:tanbdef}
\tan\beta\equiv \frac{v_2}{v_1}\,,
\eeq
and $v_1$ and $v_2$ are the magnitudes of the vevs of the neutral components of the Higgs fields in the generic basis, defined in \eq{potmin}.
In particular, 
\beq
v_1=v\cos\beta\,,\qquad\quad v_2=v\sin\beta\,,
\eeq
are non-negative quantities, which implies that we may assume that $0\leq\beta\leq\half\pi$.
It follows that the invariant Higgs basis fields defined in \eq{hbasisdef2} are given by
\beq
\begin{pmatrix} \mathcal{H}_1 \\ \mathcal{H}_2\end{pmatrix}=\begin{pmatrix} \cos\beta & e^{-i\xi}\sin\beta \\
-e^{i(\xi+\eta)}\sin\beta & e^{i\eta} \cos\beta \end{pmatrix}\begin{pmatrix}  \Phi_1 \\ \Phi_2\end{pmatrix}\,.
\eeq
Consequently, we can identify the primed scalar potential parameters with the scalar potential coefficients of the Higgs basis, $\{\mathcal{H}_1,\mathcal{H}_2\}$, as specified in
\eq{higgspot}.

As an example, if the $\Phi^\prime$ basis is identified with the Higgs basis then, e.g.,  $\lambda^\prime_1=Z_1$, $\lambda^\prime_2=Z_2$, $\lambda^\prime_6=Z_6 e^{-i\eta}$, $\lambda^\prime_7=Z_7 e^{-i\eta}$, etc.
In particular, the $\eta$ dependence on the left-hand side of eqs.~(\ref{mab}) and (\ref{Lam5def})--(\ref{Lam7def}) cancels out.
Hence, if we identify the $\Phi$ basis as a
$\mathbb{Z}_2$ basis where $\lambda_6=\lambda_7=0$, it then follows from eqs.~(\ref{Lam1def}), (\ref{Lam2def}), (\ref{Lam6def}), and (\ref{Lam7def}) that
\beq \label{spcase}  
Z_1-Z_2=(\lambda_1-\lambda_2)c_{2\beta}\,,\qquad\quad Z_{67} e^{i\xi}=-\half s_{2\beta}(\lambda_1-\lambda_2)\,.
\eeq
Consequently,
\beq \label{z1267}
\half(Z_1-Z_2)s_{2\beta}+c_{2\beta} Z_{67}e^{i\xi}=0\,.
\eeq
Noting that \eq{spcase} implies that $\Im( Z_{67}e^{i\xi})=0$; it follows that \eqs{plusa}{z1267} are consistent equations.

It is convenient to invert the resulting equations and express the $m_{ij}^2$ and $\lambda_i$ in terms of the $Y_i$ and $Z_i$.   This is easily done by employing the inverse matrix $U^{-1}=U^\dagger$, which simply corresponds to taking $\beta\to -\beta$, $\eta\to -\eta$ and $\xi\to\xi+\eta$  (the last two replacements are equivalent to the interchange of $\xi\longleftrightarrow\xi+\eta$).   Hence, it follows that\footnote{Note that the sign in front of $Y_3$ in \eq{higgspot} is positive, whereas the sign in front of $m_{12}^2$ in \eq{pot} is negative.  Thus, we have identified $Y_3=-m_{12}^{\prime\,2}$ in obtaining \eqst{m11}{m12} from \eqst{maa}{mab}.}
\beqa
m_{11}^2&=&Y_1\cosbb+Y_2\sinbb-\Re(Y_{3} e^{i\xi})s_{2\beta}\,,
\label{m11}\\
m_{22}^2 &=&Y_1\sinbb+Y_2\cosbb+\Re(Y_{3} e^{i\xi})s_{2\beta}\,,
\label{m22}\\
m_{12}^2 e^{i\xi}&=&
\half(Y_2-Y_1)s_{2\beta}-\Re(Y_{3} e^{i\xi})c_{2\beta}
-i\,\Im(Y_{3} e^{i\xi})\,,\label{m12} 
\eeqa
and
\beqa
\!\!\!\!\!\!\!\!\!\!\!\!\!\!\lam_1&=&
Z_1 c_\beta^4+ Z_2 s_\beta^4+\half Z_{345}s_{2\beta}^2
-2\stwob\left[c_\beta^2\Re(Z_6 e^{i\xi})+s_\beta^2\Re(Z_7e^{i\xi})\right]\,,
\label{lam1def}  \\
\!\!\!\!\!\!\!\!\!\!\!\!\!\!\lam_2 &=&
Z_1 s_\beta^4+Z_2c_\beta^4+\half Z_{345}\stwob^2
+2\stwob\left[s_\beta^2\Re(Z_6 e^{i\xi})+c_\beta^2\Re(Z_7e^{i\xi})\right]\,,
\label{lam2def}      \\
\!\!\!\!\!\!\!\!\!\!\!\!\!\!\lam_3 &=&
\quarter\stwob^2\left[Z_1+Z_2-2Z_{345}\right]
+Z_3+\stwob\ctwob\Re[(Z_6-Z_7)e^{i\xi}]\,,
\label{lam3def}      \\
\!\!\!\!\!\!\!\!\!\!\!\!\!\!\lam_4 &=&
\quarter\stwob^2\left[Z_1+Z_2-2Z_{345}\right]
+Z_4+\stwob\ctwob\Re[(Z_6-Z_7)e^{i\xi}]\,,
\label{lam4def}      \\
\!\!\!\!\!\!\!\!\!\!\!\!\!\!\lam_5  e^{2i\xi} &=&
\quarter\stwob^2\left[Z_1+Z_2-2Z_{345}\right]+\Re(Z_5 e^{2i\xi})
+i\ctwob\Im(Z_5 e^{2i\xi})
\nonumber \\
&&\qquad\qquad\qquad\,\,\,\,
+\stwob\ctwob\Re[(Z_6-Z_7)e^{i\xi}]+i\stwob\Im[(Z_6-Z_7)e^{i\xi}]\,,
\label{lam5def}      \\
\!\!\!\!\!\!\!\!\!\!\!\!\!\!\lam_6 e^{i\xi}&=&
\half\stwob\left[Z_1c_\beta^2
-Z_2 s_\beta^2-Z_{345}\ctwob-i\Im(Z_5 e^{2i\xi})\right]
+c_\beta\cthreeb\Re(Z_6 e^{i\xi})  \nonumber \\
&&\qquad\qquad\qquad\,\,\,\,
+s_\beta\sthreeb\Re(Z_7 e^{i\xi})
+ic_\beta^2\Im(Z_6 e^{i\xi})+is_\beta^2\Im(Z_7 e^{i\xi})\,,
\label{lam6def}      \\
\!\!\!\!\!\!\!\!\!\!\!\!\!\!\lam_7 e^{i\xi} &=&
\half s_{2\beta}\left[Z_1 s_\beta^2-Z_2 c_\beta^2+
Z_{345} c_{2\beta}+i\Im(Z_5 e^{2i\xi})\right]
+s_\beta\sthreeb\Re(Z_6 e^{i\xi})\nonumber \\
&&\qquad\qquad\qquad\,\,\,\,
+c_\beta\cthreeb\Re(Z_7e^{i\xi})
+is_\beta ^2\Im(Z_6 e^{i\xi})+ic_\beta^2\Im(Z_7e^{i\xi})\,,
\label{lam7def}
\eeqa
where
\beq \label{z345def}
Z_{345}\equiv Z_3+Z_4+\Re(Z_5 e^{2i\xi})\,.
\eeq

It is convenient to take the sum and difference of \eqs{lam6def}{lam7def} to obtain
\beqa
(\lam_6+\lam_7)e^{i\xi}&=&\half\stwob\left(Z_1-Z_2\right)+c_{2\beta}\Re\bigl[(Z_6+Z_7)e^{i\xi}\bigr]+i\Im\bigl[(Z_6+Z_7)e^{i\xi}\bigr]\,, \label{plus}\\
(\lam_6-\lam_7)e^{i\xi}&=&\half\stwob\ctwob\left(Z_1+Z_2-2Z_{345}\right)-is_{2\beta}\Im(Z_5 e^{2i\xi}) \nonumber \\
&&\qquad\qquad\,\,\,\,
+c_{4\beta}\Re\bigl[(Z_6-Z_7)e^{i\xi}\bigr]+ic_{2\beta}\Im\bigl[(Z_6-Z_7)e^{i\xi}\bigr]\,.\label{minus}
\eeqa
As previously noted, all factors of $e^{i\eta}$ have canceled out due to the $\eta$ dependence of the coefficients of the Higgs basis scalar potential given in \eq{higgspot}.

\section{The exceptional case of $\boldsymbol{Z_1=Z_2}$ and $\boldsymbol{Z_7=-Z_6}$}
\label{erps}

In the exceptional case of $Z_1=Z_2$ and $Z_7=-Z_6$, it follows from \eqst{lam1def}{lam7def} that $\lambda_1=\lambda_2$ and $\lambda_7=-\lambda_6$ in all scalar field 
bases.\footnote{We note in passing that the exceptional region of parameter space where
$\lambda_1=\lambda_2$ and $\lambda_7=-\lambda_6$
was identified in Ref.~\cite{Ferreira:2009wh} as the conditions for a softly broken CP2-symmetric scalar potential,
where CP2 is the generalized CP transformation, $\Phi_1\to\Phi_2^*$ and $\Phi_2\to -\Phi_1^*$.  In general, dimension-two soft CP2-breaking squared-mass terms are present and
violate the CP2-symmetric conditions, $m_{11}^2=m_{22}^2$ and $m_{12}^2=0$.  However, the CP2 symmetry is also violated by the dimension-four Yukawa interactions, which constitute a hard breaking of the symmetry~\cite{Ferreira:2010bm}. Consequently, the exceptional region of the parameter space is unnatural and must be regarded as finely tuned.}  
In this appendix, we show that in this exceptional case, there exists a $\Phi$ basis in which $\lambda_6=\lambda_7=0$.  That is, there exists a scalar field basis where the $\mathbb{Z}_2$ symmetry of the quartic terms of the scalar potential is manifest. 

It we set $Z_1=Z_2$ and $Z_{67}=0$ in \eqs{lam6def}{lam7def}, then it follows that a scalar basis with $\lambda_6=\lambda_7=0$ exists if and only if values of $\beta$ 
and $\xi$ can be found such that
\beq
\stwob\ctwob\left[Z_1-Z_{34}-\Re(Z_5e^{2i\xi})\right]-is_{2\beta}\Im(Z_5 e^{2i\xi})
+2c_{4\beta}\Re(Z_6e^{i\xi})+2ic_{2\beta}\Im(Z_6e^{i\xi})=0\,.\label{originaleq}
\eeq
Taking the real and imaginary parts of \eq{originaleq} yields,
\beqa
s_{2\beta}\Im(Z_5 e^{2i\xi})&=&2c_{2\beta}\Im(Z_6e^{i\xi})\,, \label{except1} \\
\stwob\ctwob\left[Z_1-Z_{34}-\Re(Z_5e^{2i\xi})\right]&=&-2c_{4\beta}\Re(Z_6e^{i\xi})\,.\label{except2}
\eeqa
If there exists a scalar basis in which $\lambda_6=\lambda_7=0$, then this basis is not unique since the relation $\lambda_6=\lambda_7=0$ is unchanged under the
basis transformation, $\Phi_a\to U_{a\bbar}\Phi_b$, where $U$ is given by \eq{U}.  Indeed, \eqs{except1}{except2} are unchanged under 
the transformations exhibited in \eq{phitophiprime}, as expected.  Thus when solving \eqs{except1}{except2},
we expect at least a twofold ambiguity in the determination of~$\beta$ and $\xi$ (where $0\leq\beta\leq\half\pi$ and $0\leq\xi<2\pi$).

If $Z_6=0$, then the scalar potential in the Higgs basis manifestly exhibits the $\mathbb{Z}_2$ symmetry, so we shall henceforth assume that $Z_6\neq 0$, in which case we may write $Z_6\equiv |Z_6|e^{i\theta_6}$.   It is convenient to introduce
\beq  \label{xiprimedef}
\xi^\prime\equiv\xi+\theta_6\,.
\eeq
Under the basis transformation $\Phi_a\to U_{a\bbar}\Phi_b$, where $U$ is given by \eq{U}, it follows that 
$e^{i\xi^\prime}\to -e^{i\xi^\prime}$, in light of  \eq{phitophiprime}.  That is, $\xi^\prime$ is only determined modulo $\pi$, corresponding to the twofold ambiguity anticipated above. 

Inserting $e^{i\xi}=e^{i\xi^{\prime}}Z_6^*/|Z_6|$ into \eqs{except1}{except2} yields
\beqa
\hspace{-0.4in}
s_{2\beta}\bigl[\Re(Z_5^* Z_6^2)\sin 2\xi^\prime-\Im(Z_5^* Z_6^2)\cos 2\xi^\prime\bigr]&=&2c_{2\beta}|Z_6|^3\sin\xi^\prime\,, \label{except3} \\
\hspace{-0.4in}
\stwob\ctwob\bigl[|Z_6|^2(Z_1-Z_{34})-\Re(Z_5^* Z_6^2)\cos 2\xi^\prime-\Im(Z_5^* Z_6^2)\sin 2\xi^\prime\bigr]&=&-2c_{4\beta}|Z_6|^3\cos\xi^\prime\,.\label{except4}
\eeqa

We now consider two cases.  First, if we assume that $\Im(Z^*_5 Z_6^2)=0$ then $\sin\xi^\prime=0$ is a solution to \eq{except3}, which implies that
$\cos\xi^\prime=\pm 1$; the twofold ambiguity was anticipated in light of the comment following \eq{xiprimedef}.
Inserting $\cos\xi^\prime=\pm 1$ into \eq{except4} then yields 
a quadratic equation for $\cot 2\beta= c_{2\beta}/s_{2\beta}$,
\beq \label{tantwobeta}
2|Z_6|\cot^2 2\beta\pm\left(Z_1-Z_{34}-\frac{\Re(Z_5^* Z_6^2)}{|Z_6|^2}\right)\cot 2\beta-2|Z_6|=0\,.
\eeq
As expected from \eq{phitophiprime}, changing the sign of $\cos\xi^\prime$ from $+1$ to $-1$ simply changes the sign of $\cot 2\beta$.  Moreover, 
\eq{tantwobeta} possesses two real roots whose product is equal to $-1$.   This observation implies that if 
$\beta$ is one solution of \eq{tantwobeta} then the second solution is $\beta\pm\tfrac14\pi$ (where the sign is chosen such that the second solution lies between 0 and $\half\pi$).
Hence, if $Z_1=Z_2$, $Z_{67}=0$ and $\Im(Z^*_5 Z_6^2)=0$ then there are four choices of $(\beta,\xi)$, where $0\leq\beta\leq\half\pi$ and $\cos\xi^\prime=\pm 1$, in which
\eqs{except1}{except2} are satisfied.

If $\Im(Z^*_5 Z_6^2)=0$ and $\sin\xi^\prime\neq 0$, then additional solutions of \eqs{except3}{except4} exist.  Solving 
\eq{except3} for $c_{2\beta}/s_{2\beta}$ and inserting this result into \eq{except4} yield
\beq \label{cosxi}
\cos\xi^\prime\bigl([\Re(Z_5^* Z_6^2)]^2+\Re(Z_5^* Z_6^2)|Z_6|^2(Z_1-Z_{34})-2|Z_6|^6\bigr)=0\,.
\eeq
Since the coefficient of $\cos\xi^\prime$ is generically nonzero, it follows that $\cos\xi^\prime=0$.  Plugging this result back into \eq{except3} yield $\cos 2\beta=0$.
Hence, $(\beta=\tfrac14\pi\,,\,\xi^\prime=\half\pi)$ and $(\beta=\tfrac14\pi\,,\,\xi^\prime=\tfrac32\pi)$ are also solutions to \eqs{except3}{except4} when $\Im(Z^*_5 Z_6^2)=0$.
These two solutions are again related by the basis transformation $\Phi_a\to U_{a\bbar}\Phi_b$, where $U$ is given by \eq{U}.

Second, if we assume instead that
$\Im(Z^*_5 Z_6^2)\neq 0$ then $\sin\xi^\prime\neq 0$.  In this case, we follow the method employed in Appendix~C of Ref.~\cite{Gunion:2005ja}.  Solving 
\eq{except3} for $s_{2\beta}/c_{2\beta}$ and inserting this result into \eq{except4} yield the following equation for $\xi^\prime$:
\beqa
&& \hspace{-0.2in}
F(\xi^\prime)\equiv \sin\xi^\prime\bigl[R\sin 2\xi^\prime-I\cos 2\xi^\prime\bigr]\bigl[|Z_6|^2(Z_1-Z_{34})-R\cos 2\xi^\prime-I\sin 2\xi^\prime\bigr] \nonumber \\
&&\qquad\qquad\qquad +\cos\xi^\prime\bigl[(R\sin 2\xi^\prime-I\cos 2\xi^\prime)^2-4|Z_6|^6\sin^2\xi^\prime\bigr]=0\,,\label{eff}
\eeqa
where $R\equiv \Re(Z_5^* Z_6^2)$ and $I\equiv\Im(Z_5^* Z_6^2)$.  Noting that $F(\xi^\prime+\pi)=-F(\xi^\prime)$, it follows that
\eq{eff} determines $\xi^\prime$ modulo $\pi$, as expected in light of the comment below \eq{xiprimedef}.   Moreover, given that
$F(\xi^\prime=0)=I^2$ and $F(\xi^\prime=\pi)=-I^2$,
there must exist an angle $\xi^\prime_0$ such that $0<\xi^\prime_0<\pi$ and $F(\xi^\prime_0)=0$.   Plugging $\xi^\prime=\xi^\prime_0$ back into \eq{except3} then yields,
\beq \label{cot2beta}
\cot 2\beta=\frac{R\sin 2\xi^\prime_0-I\cos 2\xi_0^\prime}{2|Z_6|^3\sin\xi^\prime_0}\,.
\eeq   
As expected, under a basis transformation, $\Phi_a\to U_{a\bbar}\Phi_b$, where~$U$ is given by \eq{U}, it follows that $\xi^\prime_0\to\xi^\prime_0+\pi$ and
$\cot 2\beta\to -\cot 2\beta$, which is consistent with \eq{cot2beta}.

Thus, we have shown that there are at least two choices of ($\beta$, $\xi$), where $0\leq\beta\leq\half\pi$ and $0\leq\xi<2\pi$, that satisfy \eq{originaleq}.
That is, we have proven that if $Z_1=Z_2$ and $Z_{67}=0$, then a scalar basis exists in which $\lambda_6=\lambda_7=0$, where the softly broken $\mathbb{Z}_2$ symmetry is manifestly realized.

We end this appendix with a discussion of spontaneous CP violation.  Starting from \eq{imexp}, we can eliminate $\Re(Z_5 e^{2i\xi})$ and $\Im(Z_5 e^{2i\xi})$ by employing 
\eqs{except1}{except2}.  If we denote $\mathcal{R}\equiv \Re(Z_6 e^{i\xi})=|Z_6|\cos\xi^\prime$ and  $\mathcal{I}\equiv \Im(Z_6 e^{i\xi})=|Z_6|\sin\xi^\prime$, the end result is
\beqa
\!\!\!\!\!\!\!\!\!\!\!\!\!\!\!
\Im(\lambda_5^*[m_{12}^2]^2)&=&-\frac{v^4}{8c_{2\beta}s_{2\beta}}\,\mathcal{I}\,\Biggl\{4c_{2\beta}s^2_{2\beta}\left(\frac{Y_2}{v^2}\right)^2+4s^2_{2\beta}\left(\frac{Y_2}{v^2}\right)\bigl[s_{2\beta}\mathcal{R}+c_{2\beta}Z_{34}\bigr]-4c_{2\beta}\,\mathcal{I}^2\nonumber \\
&&\!\!\!\!  -4c_{2\beta}c_{4\beta}\mathcal{R}^2-2s_{2\beta}\bigl[c_{4\beta}Z_1+c^2_{2\beta}(Z_1-2Z_{34})\bigr]\mathcal{R}-c_{2\beta}s^2_{2\beta}Z_1(Z_1-2Z_{34})\Biggr\},\label{im5erps}
\eeqa
where $\lambda_5$ and $m_{12}^2$ are parameters of the scalar potential in the $\mathbb{Z}_2$ basis, and $\beta$ and $\xi$ are solutions to \eqs{except1}{except2}.   

Below \eq{except4}, we showed that if $\Im(Z_5^* Z_6^2)=0$, then one solution to \eqs{except1}{except2} is $\sin\xi^\prime=0$.  In this case, $\mathcal{I}=\Im(Z_6 e^{i\xi})=|Z_6|\sin\xi^\prime=0$, and it immediately follows from \eq{im5erps} that  $\Im(\lambda_5^*[m_{12}^2]^2)=0$.  
We also showed above that if $\Im(Z_5^* Z_6^2)=0$, then a second solution exists in which $c_{2\beta}=0$ and $\cos\xi^\prime=0$.   
In order to employ \eq{im5erps} in this case, one must first use \eq{except2} in order to rewrite $\Im(\lambda_5^*[m_{12}^2]^2)$ in terms of $\mathcal{I}$~and  
$\Re(Z_5 e^{2i\xi})$.  Having done so, the factor of $c_{2\beta}$ in the denominator of the prefactor in \eq{im5erps} cancels out, and one can then set $c_{2\beta}=0$.  
Finally, we employ $\Re(Z_5 e^{2i\xi})=-\Re(Z_5^* Z_6^2)/|Z_6|^2$ (after using $e^{2i\xi}=e^{2i\xi^\prime}(Z_6^*)^2/|Z_6|^2$ and $\cos 2\xi^{\prime}=-1$).  
The resulting expression reproduces \eq{nonzero} and yields  
$\Im(\lambda_5^*[m_{12}^2]^2)\neq 0$, which implies that no $\mathbb{Z}_2$ basis exists in which $m_{12}^2$ and $\lambda_5$ are both real. 
Nevertheless, because $\Im(Z_5^* Z_6^2)=0$ and $Z_{67}=0$,
it follows that a real Higgs basis exists, which signifies that the scalar sector is CP conserving.

If $\Im(Z_5^* Z_6^2)\neq 0$, then no real Higgs basis exists, and thus the scalar sector violates CP either explicitly or spontaneously.  In this case, $\sin\xi^\prime=\sin\xi_0^\prime\neq 0$,
where $\xi^\prime_0$ is determined as discussed below \eq{eff}.  Since CP is explicitly conserved if $\Im(\lambda_5^*[m_{12}^2]^2)=0$, it follows from \eq{im5erps} that a basis-invariant condition for spontaneous CP violation is given by,
\beqa
 && 4c_{2\beta}s^2_{2\beta}\left(\frac{Y_2}{v^2}\right)^2+4s_{2\beta}\left(\frac{Y_2}{v^2}\right)\bigl[s_{2\beta}\mathcal{R}+c_{2\beta}Z_{34}\bigr]-4c_{2\beta}(\mathcal{I}^2+c_{4\beta}\mathcal{R}^2)\nonumber \\[3pt]
&& \qquad\qquad -2s_{2\beta}\bigl[c_{4\beta}Z_1+c^2_{2\beta}(Z_1-2Z_{34})\bigr]\mathcal{R}-c_{2\beta}s^2_{2\beta}Z_1(Z_1-2Z_{34})=0\,,
\eeqa
where $\mathcal{R}=|Z_6|\cos\xi_0^\prime$ and $\mathcal{I}=|Z_6|\sin\xi_0^\prime$, and the angle $2\beta$ is given by
\eq{cot2beta}.

\section{Basis-invariant conditions for the $\mathbb{Z}_2$ symmetry revisited}
\label{appC}

In Section~\ref{sec:five}, conditions for the presence of a $\mathbb{Z}_2$ symmetry in the scalar potential (which may or may not be softly broken) were derived.   These conditions were expressed in terms of the Higgs basis scalar potential parameters and were invariant with respect to an arbitrary rephasing of the Higgs basis field $H_2$ that defines the set of all possible Higgs bases.   In Ref.~\cite{Davidson:2005cw}, a set of manifestly basis-invariant expressions were presented which were sensitive to the presence of
a $\mathbb{Z}_2$ symmetry in the 2HDM scalar potential.\footnote{The group theoretic analysis of the 2HDM scalar potential developed in Ref.~\cite{Ivanov:2005hg}  and the geometric picture of Ref.~\cite{Ferreira:2010yh} provide alternative approaches for obtaining a basis-independent condition for the presence of a softly broken $\mathbb{Z}_2$ symmetry.}
In this appendix, we demonstrate that if these expressions are evaluated in the Higgs basis, then the results of Section~\ref{sec:five} are recovered.

We begin by defining two U(2)-flavor tensors constructed from the 2HDM couplings $Z_{a\bbar c\dbar}$ defined in \eq{genericpot},
\beq \label{zabdef}
Z^{(1)}_{a\dbar}\equiv \delta_{b\cbar}Z_{a\bbar c\dbar}=Z_{a\bbar b\dbar}\,,
\qquad\qquad
Z^{(11)}_{c\dbar}\equiv Z^{(1)}_{b\abar}Z_{a\bbar c\dbar}\,.
\eeq
It is straightforward to work out the following explicit expressions in the $\Phi$ basis:
\beq \label{zone}
Z^{(1)}=\begin{pmatrix} \lam_{14} & \quad \lam_{67} \\ \lam_{67}^* & \quad \lam_{24} \end{pmatrix}\,,
\eeq
and
\beq \label{zoneone}
Z^{(11)}=\begin{pmatrix} \lam_{14}\lam_1+\lam_{24}\lam_3+\lam_{67}\lam_6^*+\lam^*_{67}\lam_6 & \quad \lam_{14}\lam_6+\lam_{24}\lam_7+\lam_{67}\lam_4+\lam^*_{67}\lam_5 \\
 \lam_{14}\lam^*_6+\lam_{24}\lam^*_7+\lam^*_{67}\lam_4+\lam_{67}\lam^*_5 & \quad
  \lam_{14}\lam_3+\lam_{24}\lam_2+\lam_{67}\lam_7^*+\lam^*_{67}\lam_7 \end{pmatrix}.
  \eeq
In \eqs{zone}{zoneone}, we employ the shorthand notation, $\lambda_{ij}\equiv \lambda_i+\lambda_j$.
\clearpage

We now make use of the following result: starting from an arbitrary $\Phi$ basis of scalar fields, one can always transform to a $\Phi^\prime$ basis in which $\lambda^\prime_7=-\lambda^\prime_6$~\cite{Gunion:2005ja}.   A simple proof of this result is given in Appendix~\ref{appD}.
It is then straightforward to evaluate in the $\Phi^\prime$ basis,
\beq
\bigl[Z^{(1)}\,,\,Z^{(11)}\bigr]=(\lam^\prime_1-\lam^\prime_2)^2\begin{pmatrix} \phm 0 & \quad \lam^\prime_6 \\ -\lam^{\prime\,*}_6 & \quad 0\end{pmatrix}\,.
\eeq
Assuming that $\lambda_1^\prime\neq \lambda_2^\prime$ in a basis where $\lambda_7^\prime=-\lambda_6^\prime$, it follows that if $\bigl[Z^{(1)}\,,\,Z^{(11)}\bigr]=0$, then
$\lambda_6^\prime=0$.  That is, in the $\Phi^\prime$ basis, the softly broken $\mathbb{Z}_2$ symmetry is manifestly realized.
In the special case of $\lambda_1^\prime= \lambda_2^\prime$ and $\lambda_7^\prime=-\lambda_6^\prime$, one can check from \eqst{Lam1def} {Lam7def} that  $\lambda_1= \lambda_2$ and $\lambda_7=-\lambda_6$.  That is, this condition holds in \textit{all} scalar field bases.  This result is not surprising given that in this special case, $Z^{(1)}_{a\bbar}=\lambda_{14}\delta_{a\bbar}$, which maintains this form under any U(2) transformation.   Moreover, as shown in Appendix~\ref{erps}, if $\lambda_1= \lambda_2$ and $\lambda_7=-\lambda_6$, then there exists scalar basis in which $\lambda_6=\lambda_7=0$.

Hence, it follows that the condition for the existence of a softly broken $\mathbb{Z}_2$ symmetry that is manifest in some scalar field basis is given by~\cite{Davidson:2005cw}
\beq \label{softcond}
\bigl[Z^{(1)}\,,\,Z^{(11)}\bigr]=0\,.
\eeq
\Eq{softcond} is covariant with respect to U(2) transformations.  Hence, it can be evaluated in any scalar field basis.
Thus, the condition we seek can be determined by evaluating \eq{softcond} in the Higgs basis.

With the help of \textit{Mathematica}, we obtain the following results.  In any basis, 
\beq
\bigl[Z^{(1)}\,,\,Z^{(11)}\bigr]_{22}=-\bigl[Z^{(1)}\,,\,Z^{(11)}\bigr]_{11}\,,\quad \text{and}\quad
\bigl[Z^{(1)}\,,\,Z^{(11)}\bigr]_{12}=-\bigl[Z^{(1)}\,,\,Z^{(11)}\bigr]_{21}^*
\eeq
In the Higgs basis,
\beqa
&&\hspace{-0.3in}\bigl[Z^{(1)}\,,\,Z^{(11)}\bigr]_{11}= 2i\bigl\{(Z_1-Z_2)\Im(Z_6^* Z_7)+\Im\bigl(Z_5^* Z_{67}^2\bigr)\bigr\}=0\,, \\
&&\hspace{-0.3in}\bigl[Z^{(1)}\,,\,Z^{(11)}\bigr]_{12}= (Z_1-Z_2)\bigl[Z_{34} Z_{67}-Z_2 Z_6-Z_1 Z_7+Z_5 Z_{67}^*\bigr]-2Z_{67}\bigl(|Z_6|^2-|Z_7|^2\bigr),
\eeqa
where $Z_{34}\equiv Z_3+Z_4$ and $Z_{67}\equiv Z_6+Z_7$.

Thus, we arrive at two conditions for the Higgs basis scalar potential parameters that imply the existence of a softly broken $\mathbb{Z}_2$ symmetry,
\beqa
&&(Z_1-Z_2)\bigl[Z_{34} Z_{67}-Z_2 Z_6-Z_1 Z_7+Z_5 Z_{67}^*\bigr]-2Z_{67}\bigl(|Z_6|^2-|Z_7|^2\bigr)=0\,,\label{cond1}\\
&&(Z_1-Z_2)\Im(Z_6^* Z_7)+\Im\bigl(Z_5^* Z_{67}^2\bigr)=0\,,\label{cond2}
\eeqa
which reproduce the results of \eqs{finalcond}{cond6}, respectively.   

If the $\mathbb{Z}_2$ symmetry is exact, then in addition to \eq{softcond}, one must impose a second condition~\cite{Davidson:2005cw}, 
\beq \label{softcond2}
\bigl[Z^{(1)}\,,\,Y\bigr]=0\,.
\eeq
This result is established by evaluating the commutator in the $\Phi^\prime$ basis where $\lambda_7^\prime=-\lambda_6^\prime$.  Noting that,
\beq
\bigl[Z^{(1)}\,,\,Y\bigr]=(\lambda_2^\prime-\lambda_1^\prime)\begin{pmatrix} 0 & \,\, m_{12}^{\prime\,2} \\  m_{12}^{\prime\,2*}& \,\, 0\end{pmatrix},
\eeq
it follows that if $\lambda^\prime_1\neq \lambda^\prime_2$ and $m_{12}^{\prime\,2}=0$, then $\bigl[Z^{(1)}\,,\,Y\bigr]=0$.  That is, if \eqs{softcond}{softcond2} are both satisfied
and $\lambda^\prime_1\neq \lambda^\prime_2$, then a basis exists where $m_{12}^{\prime\,2}=\lambda^\prime_6=\lambda^\prime_7=0$ and the $\mathbb{Z}_2$ symmetry is manifest.

Evaluating \eq{softcond2} in the Higgs basis,
\beq
\begin{pmatrix} Z_{14} &\quad Z_{67} \\ Z_{67}^*& \quad Z_{24}\end{pmatrix}\begin{pmatrix} Y_1 &\quad Y_3 \\ Y_3^* &\quad Y_2\end{pmatrix}
=\begin{pmatrix} Y_1 &\quad Y_3 \\ Y_3^* &\quad Y_2\end{pmatrix}\begin{pmatrix} Z_{14} &\quad Z_{67} \\ Z_{67}^*& \quad Z_{24}\end{pmatrix}\,,
\eeq
where we have again employed the notation, $Z_{ij}\equiv Z_i+Z_j$.  This yields two conditions,
\beqa
(Y_1-Y_2)Z_{67}^*+Y_3^*(Z_2-Z_1) &=& 0\,, \label{cond1e} \\
\Im(Y_3^* Z_{67})&=&0\,,\label{cond2e}
\eeqa
which reproduces the results of \eqs{Ycond}{cond4e}, respectively.

The exceptional region of parameter space (where $\lambda_7=-\lambda_6$ and $\lambda_1=\lambda_2$ in all scalar field bases) must be treated separately,  Indeed \eqs{cond1e}{cond2e} are automatically satisfied, and additional considerations are warranted.    Following Ref.~\cite{Davidson:2005cw}, we introduce $Y^{(1)}_{c\dbar}\equiv Y_{b\abar}Z_{a\bbar c\dbar}$, which is explicitly given in the Higgs basis by
\beq
Y^{(1)}=\begin{pmatrix} Y_1 Z_1+Y_3 Z_6^*+Y_3^* Z_6+Y_2 Z_3 & \quad Y_1 Z_6+Y_3 Z_4+Y_3^* Z_5+Y_2 Z_7 \\
Y_1 Z_6^*+Y_3 Z_5^*+Y_3^* Z_4+Y_2 Z_7^* & \quad Y_1 Z_3+Y_3 Z_7^*+Y_3^* Z_7 + Y_2 Z_2\end{pmatrix}\,.
\eeq 
If  $Z_{67}=0$ and $Z_1= Z_2$, then we require that~\cite{Davidson:2005cw},\beq \label{ycomm}
\bigl[Y^{(1)}\,,\,Y\bigr]=0\,.
\eeq
In the Higgs basis, \eq{ycomm} yields
\beqa
&& Y_3\bigl[Y_1(Z_3-Z_1)+Y_3^*(Z_7-Z_6)+Y_3(Z_7^*-Z_6^*)+Y_2(Z_2-Z_3)\bigr] \nonumber \\
&& \qquad\qquad\qquad\qquad +(Y_1-Y_2)(Y_1 Z_6+Y_3 Z_4+Y_3^* Z_5+Y_2 Z_7)=0\,,\label{cond5e} \\
&& Y_1\Im(Y_3 Z_6^*)+\Im(Y_3^2 Z_5^*)+Y_2\Im(Y_3 Z_7^*)=0\,.\label{cond6e}
\eeqa
By assumption, $Z_{67}=0$ and $Z_1=Z_2$.  The end result is
\beqa
&& \hspace{-0.2in}  (Y_1-Y_2)\bigl[Z_6^*(Y_2 +\half  Z_{34}v^2) +\half Z_6 Z_5^* v^2\bigr]+Z_6^*|Z_6|^2 v^4=0\,,\label{cond7e} \\
&&  \hspace{-0.2in}  \Im(Z_5^* Z_6^2)=0\,,\label{cond8e}
\eeqa
after imposing the scalar potential minimum condition, $Y_3=-\half Z_6 v^2$ [cf.~\eq{minconds}].\footnote{In obtaining \eq{cond7e}, we made use of $Y_1Z_6=Y_3Z_1$, which is a consequence of \eq{minconds}.}   
Multiplying \eq{cond7e} by $Z_6$ yields,
\beq \label{cond78e}
(Y_1-Y_2)\left[|Z_6|^2\left(Z_{34}+\frac{2Y_2}{v^2}\right)+Z_5^* Z_6^2\right]+2|Z_6|^4 v^2=0\,,
\eeq
which reproduces \eq{lastcond}.  Indeed, the imaginary part of \eq{cond78e} yields \eq{cond8e}, implying that the latter is not an independent condition.\footnote{If $Y_1=Y_2$, $Z_1=Z_2$ and $Z_{67}=0$, then \eq{cond7e} implies that $Z_6=0$ and \eq{cond8e} is trivially satisfied.  Of course, in this case, the exact $\mathbb{Z}_2$ symmetry is manifestly realized in the Higgs basis, and no further analysis is required.}

\section{Proof of the existence of a scalar field basis in which $\boldsymbol{\lambda^\prime_7=-\lambda^\prime_6}$}
\label{appD}

Starting from an arbitrary $\Phi$ basis of scalar fields, \eqst{maa}{Lam7def} list the coefficients of the scalar potential in the $\Phi^\prime$ basis that are related to the corresponding coefficients of the $\Phi$ basis by the U(2) transformation given by \eq{u2}.
It then follows that
\beq \label{sum}
(\Lamf+\Lamg)e^{i\xi}=-\half  s_{2\beta}(\lam_1-\lam_2)+c_{2\beta}\Re\bigl[(\lam_6+\lam_7)e^{i\xi}\bigr]+i\Im\bigl[(\lam_6+\lam_7)e^{i\xi}\bigr]\,.
\eeq
We assume that $\lambda_7\neq -\lambda_6$.   The goal of this appendix is to show that there exists a choice of $\beta$ and $\xi$ such that $\Lamg=-\Lamf$.  

Consider the diagonalization of the matrix $Z^{(1)}_{a\bbar}\equiv \delta_{c\dbar}Z_{a\cbar d\bbar}$, which is explicitly given by
\beq
Z^{(1)}\equiv\begin{pmatrix} \lambda_1+\lambda_4 & \quad \lambda_6+\lambda_7 \\  \lambda_6^*+\lambda_7^* & \quad \lambda_2+\lambda_4\end{pmatrix}\,.
\eeq
Under a basis transformation, $\Phi_a\to \Phi^{\prime}_a=U_{a\bbar}\Phi_b$, it follows that $Z^{(1)}_{a\bbar}\to U_{a\cbar}Z^{(1)}_{c\dbar}U^\dagger_{d\bbar}$, where the unitary matrix $U$ is 
given by \eq{u2}.   It is possible to choose $\eta$, $\beta$, and $\xi$ such that 
\beq \label{diagonalize}
UZ^{(1)}U^{-1}={\rm diag}(\lambda_+\,,\,\lambda_-)\,,
\eeq
where the $\lambda_\pm$ are the eigenvalues of $Z^{(1)}$,
\beq \label{lpm}
\lambda_{\pm}=\half\biggl[\lambda_1+\lambda_2+2\lambda_4\pm\sqrt{(\lambda_1-\lambda_2)^2+4|\lambda_6+\lambda_7|^2}\,\biggr]\,.
\eeq
In determining the diagonalization matrix $U$, one is free to take $\eta=0$
without loss of generality.\footnote{In light of \eq{diagonalize}, it follows that the columns of $U^{-1}=U^\dagger$ are the normalized eigenvectors of $Z^{(1)}$, which are only defined up to an overall complex phase.  Hence, one is free to rephase the second row of \eq{u2} in order to set $\eta=0$.} 
By convention, we shall also take $0\leq\beta\leq\half\pi$ and $0\leq\xi<2\pi$.  

It is convenient to introduce the notation,
\beq \label{lamsixsever}
\lambda_{67}\equiv \lambda_6+\lambda_7\equiv |\lambda_{67}|e^{i\theta_{67}}.
\eeq
It is then straightforward to check that the diagonalization of $Z^{(1)}$ is achieved if $U$ is given by \eq{u2} with $\eta=0$, $\xi=-\theta_{67}$ and
\beq \label{sc}
s_{2\beta}=\frac{2|\lambda_{67}|}{\sqrt{(\lambda_1-\lambda_2)^2+4|\lambda_{67}|^2}}\,,\qquad\quad
c_{2\beta}=\frac{\lambda_1-\lambda_2}{\sqrt{(\lambda_1-\lambda_2)^2+4|\lambda_{67}|^2}}\,,
\eeq
Indeed, by inserting $\xi=-\theta_{67}$ into \eq{sum} and using \eq{sc}, one readily verifies that 
\beq
(\Lamf+\Lamg)e^{i\xi}=-\half s_{2\beta}(\lambda_1-\lambda_2)+c_{2\beta}|\lambda_{67}|=0\,,
\eeq
after making use of \eq{sc} in the final step.  Hence, we conclude that $\Lamf+\Lamg=0$.
That is, it is always possible to find a basis change such that $\Lamf=-\Lamg$.

For the record, we verify the diagonalization of $Z^{(1)}$ by computing
\beqa
&&  \!\!\!\!
UZ^{(1)}U^{-1}=\begin{pmatrix} c_\beta & \quad e^{i\theta_{67}} s_\beta \\ -e^{-i\theta_{67}} s_\beta & \quad c_\beta\end{pmatrix} 
\begin{pmatrix} \lambda_1+\lambda_4 & \quad |\lambda_{67}|e^{i\theta_{67}} \\
 |\lambda_{67}|e^{-i\theta_{67}}  & \quad \lambda_2+\lambda_4\end{pmatrix}
 \begin{pmatrix} c_\beta & \quad -e^{i\theta_{67}} s_\beta \\ e^{-i\theta_{67}} s_\beta & \quad c_\beta \end{pmatrix} 
\nonumber \\[10pt]
&& =\begin{pmatrix} c_\beta & \quad e^{i\theta_{67}} s_\beta \\ -e^{-i\theta_{67}} s_\beta & \quad c_\beta\end{pmatrix} 
\begin{pmatrix} (\lambda_1+\lambda_4)c_\beta+|\lambda_{67}|s_\beta & \quad -e^{i\theta_{67}}\bigl[(\lambda_1+\lambda_4)s_\beta-|\lambda_{67}|c_\beta\bigr] \\
e^{-i\theta_{67}}\bigl[|\lambda_{67}|c_\beta+(\lambda_2+\lambda_4)s_\beta\bigr] & \quad
-|\lambda_{67}|s_\beta+(\lambda_2+\lambda_4)c_\beta\end{pmatrix}\!. \nonumber \\[6pt]
&& \phantom{line}
 \eeqa
 In particular,
 \beq
 \bigl(UZ^{(1)}U^{-1})_{12}=\bigl(UZ^{(1)}U^{-1})^*_{21}=e^{i\theta_{67}}\left[|\lambda_6+\lambda_7|c_{2\beta}-\half (\lambda_1-\lambda_2) s_{2\beta}\right],
 \eeq
 which vanishes if
 \beq
 \tan 2\beta=\frac{2|\lambda_{67}|}{\lambda_1-\lambda_2}\,.
 \eeq
 Note that this result is consistent with \eq{sc}.
 \clearpage
 
 One can also check that $\bigl(UZ^{(1)}U^{-1})_{11}=\lambda_+$ and $\bigl(UZ^{(1)}U^{-1})_{22}=\lambda_-$, where
 \beq \label{eigen}
 \lambda_\pm = \half(\lambda_1+\lambda_2+2\lambda_4)\pm\left[ \half(\lambda_1-\lambda_2) c_{2\beta}+|\lambda_{67}|s_{2\beta}\right]\,.
 \eeq
 Plugging in \eq{sc} for $s_{2\beta}$ and $c_{2\beta}$ into \eq{eigen} then yields \eq{lpm}, as expected.

\section{Mixing of the neutral Higgs scalars in the $\boldsymbol{\Phi}$ basis}
\label{appE}

In \sect{sec:three}, the mixing of the neutral Higgs scalars was obtained in the Higgs basis.  In this appendix, we examine the mixing in the $\Phi$ basis, where the scalar potential is given by \eq{pot}.   In the $\Phi$ basis, the two scalar doublet fields can be parametrized by
\begin{equation}\label{PhiBasis}
\Phi_1 =
\begin{pmatrix}
\varphi_1^+ \\
\frac{1}{\sqrt{2}} (v_1 + \eta_1 + i \chi_1)
\end{pmatrix}\,,
\qquad \quad
\Phi_2 = e^{i\xi}
\begin{pmatrix}
\varphi_2^+  \\
\frac{1}{\sqrt{2}}(v_2 + \eta_2 + i \chi_2)
\end{pmatrix}\,,
\end{equation}
where 
\beq
v_1=vc_\beta\,,\qquad v_2=vs_\beta\,,
\eeq
$c_\beta \equiv \cos{\beta}$,
$s_\beta \equiv \sin{\beta}$,
$v$ is defined in \eq{v246}, and the ranges of the parameters $\beta$ and~$\xi$ are conventionally chosen to be $0\leq\beta\leq\half\pi$ and $0\leq\xi<2\pi$.   
The minimum conditions for the 2HDM scalar potential specified in \eq{pot} are,
\beqa
m_{11}^2&=&\Re(m_{12}^2 e^{i\xi})\tan\beta-\half v^2\bigl[\lambda_1 c_\beta^2+\lambda_{345} s_\beta^2+3\Re(\lambda_6 e^{i\xi})s_\beta c_\beta+\Re(\lambda_7 e^{i\xi}) s_\beta^2\tan\beta\bigr]\,,\nonumber \\
m_{22}^2&=&\Re(m_{12}^2 e^{i\xi})\cot\beta-\half v^2\bigl[\lambda_2 s_\beta^2+\lambda_{345}c_\beta^2 +\Re(\lambda_6 e^{i\xi})c^2_\beta \cot\beta+3\Re(\lambda_7 e^{i\xi}) s_\beta c_\beta\bigr]\,, \nonumber \\
\Im(m_{12}^2 e^{i\xi})&=&\half v^2\bigl[ s_\beta c_\beta \Im(\lambda_5 e^{2i\xi})+\Im(\lambda_6 e^{i\xi})c_\beta^2+\Im(\lambda_7 e^{i\xi})s_\beta^2\bigr]\,,\label{z2softmin}
\eeqa 
where $\lamtil\equiv\lam_3+\lam_4+\Re(\lam_5 e^{2i\xi})$.

In light of \eqthree{hbasisdef}{hbasisdef2}{hbasisfields}, one can identify the
neutral Goldstone boson with
$G^0 = c_\beta \chi_1 + s_\beta \chi_2$ and the charged Goldstone boson with $G^+=c_\beta\varphi_1^+ + s_\beta\varphi_2^+$.
The neutral scalar state orthogonal to $G^0$ is denoted by $\eta_3$ and is given by
\beq \label{etathree}
\eta_3 =  c_\beta \chi_2- s_\beta \chi_1\,.
\eeq

An expression for the neutral Higgs mass-eigenstate fields was obtained in \eq{hmassinv}, which we repeat here for the convenience of the reader,
\beq \label{app:hmassinv}
h_k=\frac{1}{\sqrt{2}}\left[\overline\Phi_{\abar}\lsup{0\,\dagger}
(q_{k1} \widehat v_a+q_{k2}\widehat w_a e^{-i\theta_{23}})
+(q^*_{k1}\widehat v^*_{\abar}+q^*_{k2}\widehat w^*_{\abar}e^{i\theta_{23}})
\overline\Phi_a\lsup{0}\right]\,,
\eeq
where the shifted neutral fields are defined
by $\overline\Phi_a\lsup{0}\equiv \Phi_a^0-v\widehat v_a/\sqrt{2}$ and
the $q_{k\ell}$ are exhibited in Table~\ref{tabqij}.
Plugging \eq{PhiBasis} into \eq{app:hmassinv} yields
\beqa \label{hkay}
h_k&=&(c_\beta\eta_1+s_\beta\eta_2)\Re q_{k1}+(c_\beta\chi_1+s_\beta\chi_2)\Im q_{k1} \nonumber \\
&&\quad
+(c_\beta\eta_2-s_\beta\eta_1)\Re(q_{k2}e^{-i(\xi+\theta_{23})}) +\eta_3\Im(q_{k2}e^{-i(\xi+\theta_{23})})\,,
\eeqa
after employing \eq{etathree}.
Recall that for $k=0$, we have $q_{01}=i$ and $q_{02}=0$, in which case \eq{hkay} yields $h_0=G^0$, as expected.

Making use of \eqs{chhiggs}{app:hmassinv}, the physical charged Higgs field is given by
\beq \label{hplus}
H^+=e^{i(\xi+\theta_{23})}(c_\beta\varphi_2^+ -s_\beta\varphi_1^+)\,.
\eeq
Focusing next on the three physical neutral Higgs bosons, $h_k$  (for $k=1,2,3$), we introduce the neutral Higgs mixing matrix, $\mathcal{R}$
[cf.~\eqs{h_as_eta}{matrixR}],
\beq \label{Rdef}
h_k=\mathcal{R}_{k\ell}\eta_\ell\,,\quad \text{for $k=1,2,3$}\,.
\eeq
where there is an implicit sum over the repeated index $\ell$.   Comparing \eqs{hkay}{Rdef} and recalling that the $q_{k1}$ are real for $k=1,2,3$, it immediately follows that
\beqa
\mathcal{R}_{k1}&=&q_{k1}c_\beta-\Re(q_{k2}e^{-i(\xi+\theta_{23})})s_\beta\,,\label{app:rk1}\\ 
\mathcal{R}_{k2}&=&q_{k1}s_\beta+\Re(q_{k2}e^{-i(\xi+\theta_{23})})c_\beta\,, \label{app:rk2}\\ 
\mathcal{R}_{k3}&=&\Im(q_{k2}e^{-i(\xi+\theta_{23})})\,.\label{app:rk3}
\eeqa
Not surprisingly, the individual elements of the matrix $\mathcal{R}$ are basis dependent, since 
there is no physical meaning to the parameters $\beta$ and $\xi$ if the
2HDM Lagrangian possesses no Higgs family symmetry (beyond a global U(1) hypercharge).
Nevertheless, one can construct combinations of the matrix elements of $\mathcal{R}$ that are invariant or pseudoinvariant with respect to 
U(2)-basis transformations.  For example,
\beqa
q_{k1}&=& \mathcal{R}_{k1}c_\beta+\mathcal{R}_{k2}s_\beta\,, \label{app:combo1}\\
q_{k2}e^{-i(\xi+\theta_{23})}&=& -\mathcal{R}_{k1}s_\beta+\mathcal{R}_{k2}c_\beta+i\mathcal{R}_{k3}\,.\label{app:combo2}
\eeqa
Indeed, the above combinations appear in the gauge boson--Higgs boson couplings~\cite{Fontes:2017zfn}.\footnote{The combination of matrix elements on the right-hand side of \eq{app:combo2} appears in the couplings of the charged Higgs boson.  The factor of $e^{-i(\xi+\theta_{23})}$ that  multiplies the invariant quantity $q_{k2}$ in \eq{app:combo2} cancels against the phase factor appearing in \eq{hplus}, resulting in charged Higgs couplings that are invariant with respect to U(2)-basis transformations, as expected for the physical couplings of invariant fields.}

All the results above also apply in the 2HDM with a softly broken $\mathbb{Z}_2$ symmetry, where $\lambda_6=\lambda_7=0$ in the $\Phi$ basis.  In this case, we may use the results of \sect{soft}.   In particular, both $c_{2\beta}$ and $e^{-i(\xi+\theta_{23})}$ are now determined up to a twofold ambiguity by \eqs{sincos} {xit23}, respectively, in terms of basis-invariant quantities.   This twofold ambiguity corresponds to a residual basis dependence associated with the interchange of the two scalar fields. 
Under the interchange of $\Phi_1\leftrightarrow\Phi_2$, it follows from \eqst{U}{xi23} that $e^{i(\xi+\theta_{23})}\to -e^{i(\xi+\theta_{23})}$ and $s_\beta\leftrightarrow c_\beta$,
in which case \eqst{app:rk1}{app:rk3} imply that $\mathcal{R}_{k1}\leftrightarrow\mathcal{R}_{k2}$ and $\mathcal{R}_{k3}\to -\mathcal{R}_{k3}$.  

However, $\theta_{23}$ and $\xi$ are not separately determined.  This is not surprising
since $\theta_{23}$ and $\xi$ are each additively shifted by a rephasing of the second Higgs doublet of the $\mathbb{Z}_2$ basis and the $\Phi$ basis, respectively.  Nevertheless, the presence of the softly broken $\mathbb{Z}_2$ symmetry ties these two parameter shifts together such that their sum $\xi+\theta_{23}$ is invariant under a rephasing of the corresponding scalar doublets.  Indeed, the fact that only the sum $\xi+\theta_{23}$ appears in \eq{hplus} and in \eqst{app:rk1}{app:rk3} could have been anticipated on these grounds.

Using \eqst{app:rk1}{app:rk3}, one can now derive a useful sum rule,
\beqa
&&\frac{1}{v^2}\sum_{k=1}^3 m_k^2\mathcal{R}_{k3}(\mathcal{R}_{k1}c_\beta-\mathcal{R}_{k2}s_\beta)  \nonumber \\
&& \hspace{1in}  =\frac{1}{v^2}c_{2\beta}\sum_{k=1}^3
m_k^2 q_{k1}\Im(q_{k2} e^{-i(\xi+\theta_{23})}) 
 -\frac{1}{2v^2}s_{\beta} \sum_{k=1}^3m_k^2\Im(q_{k2}^2 e^{-2i(\xi+\theta_{23})})\nonumber \\[6pt]
&& \hspace{1in} = -c_{2\beta}\,\Im(Z_6 e^{i\xi})+\half s_{2\beta}\,\Im (Z_5 e^{2i\xi})\,,\label{app:srule}
\eeqa
after making use of \eqs{zee5id}{zee6id}.   Since $\lambda_6=\lambda_7=0$ in the $\mathbb{Z}_2$ basis, one 
can employ \eqs{Lam5def}{Lam6def} to obtain\footnote{When using \eqs{Lam6def}{Lam7def}, we identify the coefficients of the Higgs basis scalar potential given by \eq{higgspot}; e.g.,
$\lambda_5^\prime= Z_5 e^{-2i\eta}$ and $\lambda_6^\prime = Z_6 e^{-i\eta}$.   Thus, the factors of $\eta$ cancel out in obtaining \eq{app:zfivesix}.}
\beq \label{app:zfivesix}
\Im (Z_5 e^{2i\xi})= c_{2\beta}\,\Im(\lambda_5 e^{2i\xi})\,,\qquad\quad \Im (Z_6 e^{i\xi})=\half s_{2\beta}\,\Im(\lambda_5 e^{2i\xi})\,.
\eeq
Inserting these results into \eq{app:srule} yields\footnote{\Eq{app:sumrule} first appears explicitly in Ref.~\cite{ElKaffas:2007rq}.}
\beq \label{app:sumrule}
\sum_{k=1}^3 m_k^2\,\mathcal{R}_{k3}(\mathcal{R}_{k1}c_\beta-\mathcal{R}_{k2}s_\beta)=0\,.
\eeq
Note that this sum rule is independent of the parameter $\xi$.  

After setting $\lambda_6=\lambda_7=0$ in \eq{z2softmin}, 
$m_{11}^2$, $m_{22}^2$ and $\Im(m_{12}^2 e^{i\xi})$ are then fixed by the scalar potential minimum conditions.  Hence,
it follows that the most general 2HDM subject to a softly broken $\mathbb{Z}_2$ symmetry is governed by nine independent parameters that can be identified by using
\eq{app:sumrule} to impose one relation among the ten real quantities:
$v$, $\tan\beta$, $\Re(m_{12}^2 e^{i\xi})$, three mixing angles, three neutral Higgs masses and one charged Higgs mass.


\begin{thebibliography}{99}

\bibitem{Lee:1973iz}
T.D.~Lee,
Phys.\ Rev.\ {\bf D8} 1226 (1973);
Phys.\ Rep. {\bf 9}, 143 (1974).

\bibitem{Peccei:1977hh}
R.D.~Peccei and H.R.~Quinn,
Phys.\ Rev.\ Lett.\  {\bf 38}, 1440 (1977).

\bibitem{Fayet:1974fj} 
  P.~Fayet,
  Nucl.\ Phys.\ B {\bf 78}, 14 (1974).
  
  \bibitem{Inoue:1982ej} 
  K.~Inoue, A.~Kakuto, H.~Komatsu and S.~Takeshita,
  Prog.\ Theor.\ Phys.\  {\bf 67}, 1889 (1982).
  
  \bibitem{Flores:1982pr} 
  R.A.~Flores and M.~Sher,
  Annals Phys.\  {\bf 148}, 95 (1983).

\bibitem{ghhiggs}
J.F.~Gunion and H.E.~Haber,
Nucl.\ Phys.\ B {\bf 272}, 1 (1986);
{\bf 278}, 449 (1986)
[Erratum:~{\bf 402}, 567 (1993)].

\bibitem{hhg}
J.F.~Gunion, H.E.~Haber, G.~Kane and S.~Dawson,
 \textit{The Higgs Hunter's Guide}
  (Westview Press, Boulder, CO, 2000).

\bibitem{Botella:1994cs} 
  F.J.~Botella and J.P.~Silva,
  Phys.\ Rev.\ D {\bf 51}, 3870 (1995)
   [hep-ph/9411288].

\bibitem{branco}
G.C.~Branco, L.~Lavoura and J.P.~Silva, {\it CP Violation}
(Oxford University Press, Oxford, England, 1999),
chapters 22 and 23.

\bibitem{Carena:2002es}
M.~Carena and H.E.~Haber,
Prog.\ Part.\ Nucl.\ Phys.\ {\bf 50}, 63 (2003)
[arXiv:hep-ph/0208209].

\bibitem{Djouadi:2005gj} 
  A.~Djouadi,
  Phys.\ Rept.\  {\bf 457}, 1 (2008)
   [hep-ph/0503172];
 {\bf 459}, 1 (2008)
   [hep-ph/0503173].


\bibitem{Branco:2011iw} 
  G.C.~Branco, P.M.~Ferreira, L.~Lavoura, M.N.~Rebelo, M.~Sher and J.P.~Silva,
  Phys.\ Rept.\  {\bf 516}, 1 (2012)
  [arXiv:1106.0034 [hep-ph]].


\bibitem{Weinberg}
S.L.~Glashow and S.~Weinberg,
Phys.\ Rev.\ {\bf D15}, 1958 (1977);

\bibitem{Paschos}
E.A.~Paschos,
Phys.\ Rev.\ {\bf D15}, 1966 (1977).

\bibitem{Georgi}
H.~Georgi and D.V.~Nanopoulos,
Phys.\ Lett. {\bf 82B}, 95 (1979).

\bibitem{Gunion:2002zf} 
  J.F.~Gunion and H.E.~Haber,
  Phys.\ Rev.\ D {\bf 67}, 075019 (2003)
  [hep-ph/0207010].
  
\bibitem{Craig:2013hca}
  N.~Craig, J.~Galloway and S.~Thomas,
  arXiv:1305.2424 [hep-ph].
  
\bibitem{Haber:2013mia}
  H.E.~Haber, in Proceedings of the of the Toyama International
  Workshop on Higgs as a Probe of New Physics 2013 (HPNP2013),  
  arXiv:1401.0152 [hep-ph].

\bibitem{Asner:2013psa}
  D.M.~Asner {\it et al.},
  arXiv:1310.0763 [hep-ph].
  
\bibitem{Carena:2013ooa}
  M.~Carena, I.~Low, N.R.~Shah and C.E.M.~Wagner,
  JHEP {\bf 1404} (2014) 015
  [arXiv:1310.2248 [hep-ph]].

\bibitem{Dev:2014yca}
  P.S.~Bhupal Dev and A.~Pilaftsis,
  JHEP {\bf 1412} (2014) 024
   [Erratum: JHEP {\bf 1511} (2015) 147]
  [arXiv:1408.3405 [hep-ph]].

\bibitem{Das:2015mwa}
  D.~Das and I.~Saha,
  Phys.\ Rev.\ D {\bf 91} (2015) 095024
  [arXiv:1503.02135 [hep-ph]].

\bibitem{Grzadkowski:2018ohf}
  B.~Grzadkowski, H.E.~Haber, O.M.~Ogreid and P.~Osland,
  JHEP {\bf 1812} (2018) 056
   [arXiv:1808.01472 [hep-ph]].
  
\bibitem{Ginzburg:2002wt} 
  I.F.~Ginzburg, M.~Krawczyk and P.~Osland, 
  in \textit{Proceedings of the International Workshop on Physics and Experiments with Future Electron-Positron Linear Colliders}, LCWS 2002, Seogwipo, Jeju Island, Korea, August 26--30, 2002, edited by J.S.~Kang and S.K.~Oh (Sorim Press, 2003) pp. 703--706 [hep-ph/0211371].
  
 \bibitem{Ginzburg:2004vp} 
  I.F.~Ginzburg and M.~Krawczyk,
  Phys.\ Rev.\ D {\bf 72}, 115013 (2005)
    [hep-ph/0408011].

\bibitem{ElKaffas:2006gdt} 
  A.W.~El Kaffas, W.~Khater, O.M.~Ogreid and P.~Osland,
  Nucl.\ Phys.\ B {\bf 775}, 45 (2007)
    [hep-ph/0605142].
  %
\bibitem{Arhrib:2010ju} 
  A.~Arhrib, E.~Christova, H.~Eberl and E.~Ginina,
  JHEP {\bf 1104}, 089 (2011)
   [arXiv:1011.6560 [hep-ph]].
  
%
\bibitem{Barroso:2012wz} 
  A.~Barroso, P.M.~Ferreira, R.~Santos and J.P.~Silva,
  Phys.\ Rev.\ D {\bf 86}, 015022 (2012)
  [arXiv:1205.4247 [hep-ph]].
 
\bibitem{Inoue:2014nva} 
  S.~Inoue, M.J.~Ramsey-Musolf and Y.~Zhang,
  Phys.\ Rev.\ D {\bf 89}, no. 11, 115023 (2014)
  [arXiv:1403.4257 [hep-ph]].
  
\bibitem{Fontes:2014xva} 
  D.~Fontes, J.C.~Rom\~{a}o and J.P.~Silva,
  JHEP {\bf 1412}, 043 (2014)
   [arXiv:1408.2534 [hep-ph]].

\bibitem{Grzadkowski:2014ada} 
  B.~Grzadkowski, O.M.~Ogreid and P.~Osland,
  JHEP {\bf 1411}, 084 (2014)
   [arXiv:1409.7265 [hep-ph]].

\bibitem{Fontes:2017zfn} 
  D.~Fontes, M.~M\"uhlleitner, J.C.~Rom\~{a}o, R.~Santos, J.P.~Silva and J.~Wittbrodt,
  JHEP {\bf 1802}, 073 (2018)
   [arXiv:1711.09419 [hep-ph]].

\bibitem{Davidson:2005cw} 
  S.~Davidson and H.E.~Haber,
  Phys.\ Rev.\ D {\bf 72}, 035004 (2005)
  [Erratum:~{\bf 72}, 099902 (2005)]
    [hep-ph/0504050].
    
  
\bibitem{Haber:2006ue} 
  H.E.~Haber and D.~O'Neil,
  Phys.\ Rev.\ D {\bf 74}, 015018 (2006)
  [Erratum:~{\bf 74}, 059905 (2006)]
   [hep-ph/0602242].
  
\bibitem{Grzadkowski:2016szj} 
  B.~Grzadkowski, O.M.~Ogreid and P.~Osland,
  Phys.\ Rev.\ D {\bf 94}, 115002 (2016)
   [arXiv:1609.04764 [hep-ph]].
  
  \bibitem{Ivanov:2005hg} 
  I.P.~Ivanov,
  Phys.\ Lett.\ B {\bf 632}, 360 (2006)
  [hep-ph/0507132].

\bibitem{Nishi:2006tg} 
  C.C.~Nishi,
  Phys.\ Rev.\ D {\bf 74}, 036003 (2006)
  [Erratum: {\bf 76}, 119901 (2007)]
   [hep-ph/0605153].
  
\bibitem{Maniatis:2007vn} 
  M.~Maniatis, A.~von Manteuffel and O.~Nachtmann,
  Eur.\ Phys.\ J.\ C {\bf 57}, 719 (2008)
   [arXiv:0707.3344 [hep-ph]].

\bibitem{Ferreira:2010hy} 
  P.M.~Ferreira, M.~Maniatis, O.~Nachtmann and J.P.~Silva,
  JHEP {\bf 1008}, 125 (2010)
   [arXiv:1004.3207 [hep-ph]].
  
  \bibitem{Ferreira:2010yh} 
  P.M.~Ferreira, H.E.~Haber, M.~Maniatis, O.~Nachtmann and J.P.~Silva,
  Int.\ J.\ Mod.\ Phys.\ A {\bf 26}, 769 (2011)
   [arXiv:1010.0935 [hep-ph]].
   
  \bibitem{Ivanov:2019kyh} 
  I.P.~Ivanov, C.C.~Nishi and A.~Trautner,
  Eur.\ Phys.\ J.\ C {\bf 79}, 315 (2019)
   [arXiv:1901.11472 [hep-ph]].
  
  
  \bibitem{Haber:1978jt} 
  H.E.~Haber, G.L.~Kane and T.~Sterling,
  Nucl.\ Phys.\ B {\bf 161}, 493 (1979).
  
  \bibitem{Donoghue:1978cj} 
  J.F.~Donoghue and L.F.~Li,
  Phys.\ Rev.\ D {\bf 19}, 945 (1979).
  
  
\bibitem{Hall:1981bc} 
  L.J.~Hall and M.B.~Wise,
  Nucl.\ Phys.\ B {\bf 187}, 397 (1981).
  
\bibitem{Haber:2015pua} 
  H.E.~Haber and O.~St\r{a}l,
  Eur.\ Phys.\ J.\ C {\bf 75}, 491 (2015)
  [Erratum:~{\bf 76}, 312 (2016)]
   [arXiv:1507.04281 [hep-ph]].
  
  \bibitem{Lavoura:1994yu} 
  L.~Lavoura,
  Phys.\ Rev.\ D {\bf 50}, 7089 (1994)
   [hep-ph/9405307].

\bibitem{Ferreira:2009wh} 
  P.M.~Ferreira, H.E.~Haber and J.P.~Silva,
  Phys.\ Rev.\ D {\bf 79}, 116004 (2009)
   [arXiv:0902.1537 [hep-ph]].


\bibitem{cpx}
L.~Lavoura and J.P.~Silva,
Phys.\ Rev.\ {\bf D50}, 4619 (1994)
[arXiv:hep-ph/9404276].

\bibitem{Haber:2010bw} 
  H.E.~Haber and D.~O'Neil,
  Phys.\ Rev.\ D {\bf 83}, 055017 (2011)
   [arXiv:1011.6188 [hep-ph]].
      
\bibitem{Huffel:1980sk} 
  H.~Huffel and G.~Pocsik,
  Z.\ Phys.\ C {\bf 8}, 13 (1981).

\bibitem{Weldon:1984wt} 
  H.A.~Weldon,
  Phys.\ Rev.\ D {\bf 30}, 1547 (1984).

\bibitem{Akeroyd:2000wc} 
  A.G.~Akeroyd, A.~Arhrib and E.M.~Naimi,
  Phys.\ Lett.\ B {\bf 490}, 119 (2000)
  [hep-ph/0006035].

\bibitem{Ginzburg:2005dt} 
  I.F.~Ginzburg and I.P.~Ivanov,
  Phys.\ Rev.\ D {\bf 72}, 115010 (2005)
  [hep-ph/0508020].
  
\bibitem{Horejsi:2005da} 
  J.~Horejsi and M.~Kladiva,
  Eur.\ Phys.\ J.\ C {\bf 46}, 81 (2006)
   [hep-ph/0510154].

\bibitem{Kanemura:2015ska} 
  S.~Kanemura and K.~Yagyu,
  Phys.\ Lett.\ B {\bf 751}, 289 (2015)
   [arXiv:1509.06060 [hep-ph]].

\bibitem{Nebot:2019lzf} 
  M.~Nebot,
  arXiv:1911.02266 [hep-ph].

  \bibitem{inprep}
  H.E.~Haber and J.P.~Silva, in preparation.
  
 \bibitem{Belusca-Maito:2017iob} 
  H.~B\'{e}lusca-Ma\"{\i}to, A.~Falkowski, D.~Fontes, J.C.~Rom\~{a}o and J.P.~Silva,
  JHEP {\bf 1804}, 002 (2018)
   [arXiv:1710.05563 [hep-ph]].
  
    
\bibitem{Gunion:2005ja} 
  J.F.~Gunion and H.E.~Haber,
  Phys.\ Rev.\ D {\bf 72}, 095002 (2005)
   [hep-ph/0506227].
  
  \bibitem{ElKaffas:2007rq} 
  A.W.~El Kaffas, P.~Osland and O.M.~Ogreid,
  Nonlin.\ Phenom.\ Complex Syst.\  {\bf 10}, 347 (2007)
  [hep-ph/0702097 [hep-ph]].
   
   \bibitem{WebPageC2HDM}
  D.~Fontes, M.~M\"uhlleitner, J.C.~Rom\~ao, R.~Santos, J.P.~Silva and J.~Wittbrodt, \textit{Couplings of the C2HDM}, \texttt{http://porthos.tecnico.ulisboa.pt/arXiv/C2HDM/}, October, 2017.

\bibitem{Branco:1985aq}
   G.C.~Branco and M.N.~Rebelo,
   Phys.\ Lett.\  {\bf 160B}, 117 (1985).

\bibitem{Branco:2005em}
G.C.~Branco, M.N.~Rebelo and J.I.~Silva-Marcos,
Phys.\ Lett.\ {\bf B614}, 187 (2005)
[arXiv:hep-ph/0502118].

    
\bibitem{Ferreira:2010bm} 
  P.M.~Ferreira and J.P.~Silva,
  Eur.\ Phys.\ J.\ C {\bf 69}, 45 (2010)
   [arXiv:1001.0574 [hep-ph]].
  
 \end{thebibliography}
\end{document}